\def\eg       {{\it e.g.}}
\def\ie       {{\it i.e.}}

\newcommand{\ee}[1]{\cdot10^{#1}}
\newcommand{\mr}[1]{\mathrm{#1}}
\newcommand{\unit}[1]{\,\mathrm{#1}}

\newcommand{\rtHz}{\sqrt{\mr{Hz}}}

\newcommand{\ket}[1]{\ensuremath{\left|#1\right\rangle}}

\newcommand{\ma}{|0\rangle}
\newcommand{\mad}{|0'\rangle}
\newcommand{\ima}{\langle 0|}
\newcommand{\mb}{|1\rangle}
\newcommand{\mbd}{|1'\rangle}
\newcommand{\imb}{\langle 1|}

\newcommand{\psia}{|\psi_0\rangle}
\newcommand{\psib}{|\psi_1\rangle}
\newcommand{\psit}{|\psi(t)\rangle}

\newcommand{\ipsib}{\langle\psi_1|}

\newcommand{\malpha}{|\alpha\rangle}

\newcommand{\xa}{|+\rangle}
\newcommand{\xb}{|-\rangle}
\newcommand{\xab}{|\pm \rangle}

\newcommand{\xma}{x_{|0\rangle}}
\newcommand{\xmb}{x_{|1\rangle}}

\newcommand{\tc}{t_c}
\newcommand{\Tac}{T_\mr{ac}}
\newcommand{\fac}{f_\mr{ac}}

\newcommand{\wo}{\omega_0}
\newcommand{\woo}{\omega_{01}}
\newcommand{\wone}{\omega_1}
\newcommand{\weff}{\omega_\mr{eff}}

\newcommand{\Dw}{\Delta\omega}

\newcommand{\Ttwoast}{T_2^\ast}
\newcommand{\Tx}{T_\chi}
\newcommand{\tm}{t_\mr{m}}
\newcommand{\SNR}{\mr{SNR}}

\newcommand{\dr}{\mr{DR}}

\newcommand{\sigx}{\sigma_x}

\newcommand{\sigp}{\sigma_p}
\newcommand{\sigpq}{\sigma_{p,\mr{quantum}}}
\newcommand{\sigpr}{\sigma_{p,\mr{readout}}}

\newcommand{\dpp}{\delta p}
\newcommand{\dppobs}{\delta p_\mr{obs}}

\newcommand{\UH}{\hat U_H(0,t)}

\newcommand{\HVzero}{\hat H_{V_{||}}}
\newcommand{\HVone}{\hat H_{V_{\perp}}}
\newcommand{\Vzero}{V_{||}}
\newcommand{\Vone}{V_{\perp}}
\newcommand{\Vpk}{V_\mr{pk}}
\newcommand{\Vrms}{V_\mr{rms}}
\newcommand{\Vmin}{V_\mr{min}}
\newcommand{\vmin}{v_\mr{min}}
\newcommand{\Vmax}{V_\mr{max}}

\newcommand{\dV}{\delta V}
\newcommand{\phirms}{\phi_\mr{rms}}

\newcommand{\phivar}{\langle\phi^2\rangle}
\newcommand{\dphi}{\delta\phi}

\newcommand{\Wcp}{W_\mr{CP}}
\newcommand{\Wpdd}{W_\mr{PDD}}
\newcommand{\DR}{\mr{DR}}

\newcommand{\figurewidth}{\columnwidth} 

\documentclass[rmp,aps,twocolumn,floatfix]{revtex4-1}

\usepackage{multirow}
\usepackage[x11names,dvipsnames,table]{xcolor}
\usepackage{colortbl}
\usepackage{array}
\usepackage{graphicx}
\usepackage{bm}
\usepackage{amsmath}
\usepackage{amssymb}
\usepackage{verbatim}
\usepackage{enumitem}
\usepackage[colorlinks=true, pdfstartview=FitV, linkcolor=blue, citecolor=blue, urlcolor=blue]{hyperref} 

\begin{document}

\global\emergencystretch = .1\hsize 

\title{Quantum Sensing}

\author{C. L. Degen}
  \email{degenc@ethz.ch} 
  \affiliation{Department of Physics, ETH Zurich, Otto Stern Weg 1, 8093 Zurich, Switzerland.}
\author{F. Reinhard}
  \email{friedemann@quantum-minigolf.org} 
  \affiliation{Walter Schottky Institut and Physik-Department, Technische Universit\"at M\"unchen, Am Coulombwall 4, 85748 Garching, Germany.}
\author{P. Cappellaro}
  \email{pcappell@mit.edu} 
  \affiliation{Research Laboratory of Electronics and Department of Nuclear Science \& Engineering, Massachusetts Institute of Technology, 77 Massachusetts Ave., Cambridge MA 02139, USA.}
	

\begin{abstract}
``Quantum sensing'' describes the use of a quantum system, quantum properties or quantum phenomena to perform a measurement of a physical quantity.  Historical examples of quantum sensors include magnetometers based on superconducting quantum interference devices and atomic vapors, or atomic clocks.  More recently, quantum sensing has become a distinct and rapidly growing branch of research within the area of quantum science and technology, with the most common platforms being spin qubits, trapped ions and flux qubits.  The field is expected to provide new opportunities -- especially with regard to high sensitivity and precision -- in applied physics and other areas of science.  In this review, we provide an introduction to the basic principles, methods and concepts of quantum sensing from the viewpoint of the interested experimentalist.
\end{abstract}

\date{\today}
\maketitle


\let\clearpage\relax

\tableofcontents

%

\section{Introduction}
\label{sec:sec01}

Can we find a promising real-world application of quantum mechanics that exploit its most counterintuitive properties?  This question has intrigued physicists ever since quantum theory development in the early twentieth century.  Today, quantum computers \cite{Deutsch85,Divincenzo00} and quantum cryptography \cite{Gisin02} are widely believed to be the most promising ones.

Interestingly, however, this belief might turn out to be incomplete.  In recent years a different class of applications has emerged that employs quantum mechanical systems as \textit{sensors} for various physical quantities ranging from magnetic and electric fields, to time and frequency, to rotations, to temperature and pressure.  ``Quantum sensors'' capitalize on the central weakness of quantum systems -- their strong sensitivity to external disturbances.  This trend in quantum technology is curiously reminiscent of the history of semiconductors: here, too, sensors -- for instance light meters based on selenium photocells \cite{Weston31art} -- have found commercial applications decades before computers. 

Although quantum sensing as a distinct field of research in quantum science and engineering is quite recent, many concepts are well-known in the physics community and have resulted from decades of developments in high-resolution spectroscopy, especially in atomic physics and magnetic resonance.  Notable examples include atomic clocks, atomic vapor magnetometers, and superconducting quantum interference devices.  What can be considered as ``new'' is that quantum systems are increasingly investigated at the single-atom level, that entanglement is used as a resource for increasing the sensitivity, and that quantum systems and quantum manipulations are specifically designed and engineered for sensing purposes.

The focus of this review is on the key concepts and methods of quantum sensing, with particular attention to practical aspects that emerge from non-ideal experiments.  As ``quantum sensors'' we will  consider mostly qubits  -- two-level quantum systems.  Although an overview over actual implementations of qubits is given, the review will not cover any of those implementation in specific detail.  It will also not cover related fields including atomic clocks or photon-based sensors.  In addition, theory will only be considered up to the point necessary to introduce the key concepts of quantum sensing.  The motivation behind this review is to offer an introduction to students and researchers new to the field, and to provide a basic reference for researchers already active in the field.


\subsection*{Content}

The review starts by suggesting some basic definitions for ``quantum sensing'' and by noting the elementary criteria for a quantum system to be useful as a quantum sensor (Section \ref{sec02}).  The next section provides an overview of the most important physical implementations (Section \ref{sec03}).
The discussion then moves on to the core concepts of quantum sensing, which include the basic measurement protocol (Section \ref{sec04}) and the sensitivity of a quantum sensor (Section \ref{sec05}).  Sections \ref{sec06} and \ref{sec07} cover the important area of time-dependent signals and quantum spectroscopy.
The remaining sections introduce some advanced quantum sensing techniques.  These include adaptive methods developed to greatly enhance the dynamic range of the sensor (Section \ref{sec08}), and techniques that involve multiple qubits (Sections \ref{sec09} and \ref{sec10}).  In particular, entanglement-enhanced sensing, quantum storage and quantum error correction schemes are discussed.  The review then concludes with a brief outlook on possible future developments (Section \ref{sec11}).


There have already been several reviews that covered different aspects of quantum sensing.  Excellent introductions into the field are the review \cite{Budker07} and book \cite{Budker13} by Budker, Romalis and Kimball on atomic vapor magnetometry, and the paper by \onlinecite{Taylor08}, on magnetometry with nitrogen-vacancy centers in diamond.  Entanglement-assisted sensing, sometimes referred to as ``quantum metrology'', ``quantum-enhanced sensing'' or ``second generation quantum sensors'' are covered by \onlinecite{Bollinger96}, \onlinecite{Giovannetti04}, \onlinecite{Giovannetti06}, and \onlinecite{Giovannetti11}.  In addition, many excellent reviews covering different implementations of quantum sensors are available; these will be noted in Section \ref{sec03}.


\section{Definitions}
\label{sec02}

\subsection{Quantum sensing}
\label{sec02:definitions}

``Quantum sensing'' is typically used to describe one of the following:

\begin{enumerate}[label=\Roman*.]
\item Use of a quantum object to measure a physical quantity (classical or quantum).  The quantum object is characterized by quantized energy levels. Specific examples include electronic, magnetic or vibrational states of superconducting or spin qubits, neutral atoms, or trapped ions.
\item Use of quantum coherence (\ie, wave-like spatial or temporal superposition states) to measure a physical quantity.
\item Use of quantum entanglement to improve the sensitivity or precision of a measurement, beyond what is possible classically.
\end{enumerate}

Of these three definitions, the first two are rather broad and cover many physical systems.  This even includes some systems that are not strictly ``quantum''.  An example is classical wave interference as it appears in optical or mechanical systems \cite{Novotny10,Faust13}.  The third definition is more stringent and a truly ``quantum'' definition.  
However, since quantum sensors according to definitions I and II are often close to applications, we will mostly focus on these definitions and discuss them extensively in this review. While these types of sensors might not exploit the full power of quantum mechanics, as for type-III sensors, they already can provide several advantages, most notably operation at nano-scales that are not accessible to classical sensors. 

Because type-III quantum sensor rely on entanglement, more than one sensing qubit is required.  A well-known example is the use of maximally entangled states to reach a Heisenberg-limited measurement. 
 Type III quantum sensors are discussed in Section \ref{sec10}.

\begin{figure}[t!]
\centering
\includegraphics[width=0.8\figurewidth]{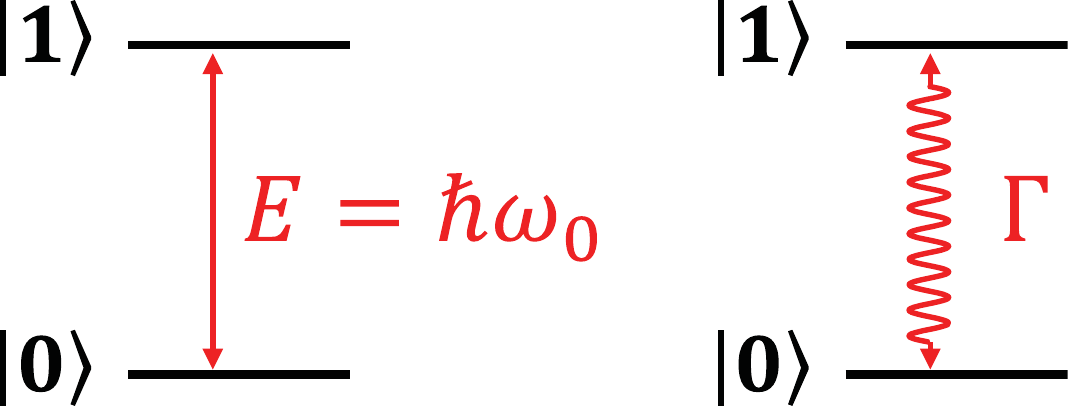}
\caption{Basic features of a two-state quantum system.  $\ma$ is the lower energy state and $\mb$ is the higher energy state.  Quantum sensing exploits changes in the transition frequency $\wo$ or the transition rate $\Gamma$ in response to an external signal $V$.
}
\label{fig:twolevel}
\end{figure}

\subsection{Quantum sensors}

In analogy to the DiVincenzo criteria for quantum computation \cite{Divincenzo00}, a set of four necessary attributes can be listed for a quantum system to function as a quantum sensor.  These attributes include three original DiVincenzo criteria:
\begin{enumerate}[label=(\arabic*)]
\item The quantum system has discrete, resolvable energy levels.  Specifically, we will assume it to be a two-level system (or an ensemble of two-level systems) with a lower energy state $\ma$ and an upper energy state $\mb$ that are separated by a transition energy $E=\hbar\wo$ (see Fig. \ref{fig:twolevel})~\footnote{\label{foot:hbar}Note that this review uses $\hbar = 1$ and expresses all energies in units of angular frequency.}.
\item It must be possible to initialize the quantum system into a well-known state and to read out its state.
\item The quantum system can be coherently manipulated, typically by time-dependent fields. This condition is not strictly required for all protocols; examples that fall outside of this criterion are continuous-wave spectroscopy or relaxation rate measurements.
\end{enumerate}
The focus on two-level systems (1) is not a severe restriction because many properties of more complex quantum systems can be modeled through a qubit sensor \cite{Goldstein10x}. 
The fourth attribute is specific to quantum sensing:
\begin{enumerate}[label=(4)]
\item The quantum system interacts with a relevant physical quantity $V(t)$, like an electric or magnetic field.  The interaction is quantified by a coupling or transduction parameter of the form $\gamma = \partial^q E/\partial V^q$ which relates changes in the transition energy $E$ to changes in the external parameter $V$.  In most situations the coupling is either linear ($q=1$) or quadratic ($q=2$).  The interaction with $V$ leads to a shift of the quantum system's energy levels or to transitions between energy levels (see Fig. \ref{fig:twolevel}).
\end{enumerate}

Experimental realizations of quantum sensors can be compared by some key physical characteristics.  One characteristic is to what kind of external parameter(s) the quantum sensor responds to.  Charged systems, like trapped ions, will be  sensitive to electrical fields, while spin-based systems will mainly respond to magnetic fields.  Some quantum sensors may respond to several physical parameters.

A second important characteristic is a quantum sensor's ``intrinsic sensitivity''.  On the one hand, a quantum sensor is expected to provide a strong response to wanted signals, while on the other hand, it should be minimally affected by unwanted noise.  Clearly, these are conflicting requirements.  In Section \ref{sec05}, we will see that the sensitivity scales as 
\begin{equation}
\text{sensitivity} \propto \frac{1}{\gamma\sqrt{\Tx}} \ ,
\end{equation}
where $\gamma$ is the above transduction parameter and $\Tx$ is a decoherence or relaxation time that reflects the immunity of the quantum sensor against noise.  In order to optimize the sensitivity, $\gamma$ should be large (for example, by choice of an appropriate physical realization of the sensor) and the decoherence time $\Tx$ must be made as long as possible.  Strategies to achieve the latter are discussed at length in the later sections of this review.

\section{Examples of quantum sensors}
\label{sec03}

We now give an overview of the most important experimental implementations of quantum sensors, following the summary in Table \ref{table:quantumsensors}.

\begin{table*}[t!]
\centering
\def\arraystretch{1.2}
\begin{tabular}{p{0.5cm}>{\raggedright}p{3.5cm}>{\raggedright}p{2.5cm}>{\raggedright}p{3cm}p{2cm}cccc}
\hline\hline
\rowcolor{Wheat1}[1.0\tabcolsep]\multicolumn{2}{l}{Implementation} & Qubit(s) & Measured quantity(ies) & Typical frequency & Initalization & Readout & Type\textcolor{blue}{$^a$}\\
\hline
\rowcolor{Cornsilk1}[1.0\tabcolsep]\multicolumn{8}{l}{Neutral atoms}\\
 \rowcolor{white}[1.0\tabcolsep] & Atomic vapor & Atomic spin & Magnetic field, Rotation, Time/Frequency & DC--10\,GHz & Optical & Optical & II--III\\
\rowcolor{Ivory1}[1.0\tabcolsep]  & Cold clouds & Atomic spin & Magnetic field, Acceleration, Time/Frequency & DC--10\,GHz & Optical & Optical & II--III \\

\hline
\rowcolor{Cornsilk1}[1.0\tabcolsep]\multicolumn{8}{l}{Trapped ion(s)}\\
 \rowcolor{white}[1.0\tabcolsep]  & & Long-lived    & Time/Frequency & THz& Optical & Optical & II-III\\
 \rowcolor{white}[1.0\tabcolsep]  & &   electronic state & Rotation && Optical & Optical & II\\
\rowcolor{Ivory1}[1.0\tabcolsep]  & &  Vibrational mode & Electric field, Force& MHz & Optical & Optical & II\\
\hline
\rowcolor{Cornsilk1}[1.0\tabcolsep]\multicolumn{8}{l}{Rydberg atoms}\\
 \rowcolor{white}[1.0\tabcolsep]  & & Rydberg states & Electric field & DC, GHz & Optical & Optical & II-III\\
\hline
\rowcolor{Cornsilk1}[1.0\tabcolsep]\multicolumn{8}{l}{Solid state spins (ensembles)}\\
 \rowcolor{white}[1.0\tabcolsep]  & NMR sensors & Nuclear spins & Magnetic field & DC & Thermal & Pick-up coil & II\\
\rowcolor{Ivory1}[1.0\tabcolsep]  & NV\textcolor{blue}{$^b$} center ensembles & Electron spins & Magnetic field, Electric field, Temperature, Pressure, Rotation & DC--GHz & Optical & Optical & II\\
\hline
\rowcolor{Cornsilk1}[1.0\tabcolsep]\multicolumn{8}{l}{Solid state spins (single spins)}\\
 \rowcolor{white}[1.0\tabcolsep]  & P donor in Si & Electron spin & Magnetic field & DC--GHz & Thermal & Electrical & II\\
\rowcolor{Ivory1}[1.0\tabcolsep]  & Semiconductor \qquad quantum dots & Electron spin & Magnetic field, Electric field & DC--GHz & Electrical, Optical & Electrical, Optical & I--II\\
 \rowcolor{white}[1.0\tabcolsep]  & Single NV\textcolor{blue}{$^b$} center & Electron spin & Magnetic field, Electric field, Temperature, Pressure, Rotation & DC--GHz & Optical & Optical & II\\
\hline
\rowcolor{Cornsilk1}[1.0\tabcolsep]\multicolumn{8}{l}{Superconducting circuits}\\
 \rowcolor{white}[1.0\tabcolsep]  & SQUID\textcolor{blue}{$^c$} & Supercurrent & Magnetic field & DC--10\,GHz & Thermal & Electrical  & I--II\\
\rowcolor{Ivory1}[1.0\tabcolsep]  & Flux qubit & Circulating currents & Magnetic field & DC--10\,GHz & Thermal & Electrical & II \\
 \rowcolor{white}[1.0\tabcolsep]  & Charge qubit & Charge eigenstates & Electric field & DC--10\,GHz & Thermal & Electrical & II \\
\hline
\rowcolor{Cornsilk1}[1.0\tabcolsep]\multicolumn{8}{l}{Elementary particles}\\
 \rowcolor{white}[1.0\tabcolsep]  & Muon & Muonic spin & Magnetic field & DC & Radioactive decay & Radioactive decay & II\\
\rowcolor{Ivory1}[1.0\tabcolsep]  & Neutron & Nuclear spin & Magnetic field, Phonon density, Gravity & DC & Bragg scattering & Bragg scattering & II\\
\hline
\rowcolor{Cornsilk1}[1.0\tabcolsep]\multicolumn{8}{l}{Other sensors}\\
 \rowcolor{white}[1.0\tabcolsep]  & SET\textcolor{blue}{$^d$} & Charge eigenstates & Electric field & DC--100\,MHz & Thermal & Electrical & I\\
\rowcolor{Ivory1}[1.0\tabcolsep]  & Optomechanics & Phonons & Force, Acceleration, Mass, Magnetic field, Voltage & kHz--GHz & Thermal & Optical  & I\\
 \rowcolor{white}[1.0\tabcolsep]  & Interferometer & Photons, (Atoms, Molecules) & Displacement, Refractive Index & -- & & & II-III\\
\hline\hline
\end{tabular}
\caption{Experimental implementations of quantum sensors. $^a$Sensor type refers to the three definitions of quantum sensing on page \pageref{sec02:definitions}. \textcolor{blue}{$^b$} NV: nitrogen-vacancy; \textcolor{blue}{$^c$} SQUID: superconducting quantum interference device; \textcolor{blue}{$^d$}SET: single electron transistor.}
\label{table:quantumsensors}
\end{table*}

\subsection{Neutral atoms as magnetic field sensors} 

Alkali atoms are suitable sensing qubits fulfilling the above definitions \cite{Kitching11}.  Their ground state spin - a coupled angular momentum of electron and nuclear spin - can be both prepared and read out optically by the strong spin-selective optical dipole transition linking their s-wave electronic ground state to the first (p-wave) excited state. 

\subsubsection{Atomic vapors} 
In the simplest implementation, a thermal vapor of atoms serves as a quantum sensor for magnetic fields \cite{Kominis03, Budker07}.  Held in a cell at or above room temperature, atoms are spin-polarized by an optical pump beam.  Magnetic field sensing is based on the Zeeman effect due to a small external field orthogonal to the initial atomic polarization. In a classical picture, this field induces coherent precession of the spin. Equivalently, in a quantum picture, it drives spin transitions from the initial quantum state to a distinct state,  which can be monitored by a probe beam, e.g. via the optical Faraday effect. Despite their superficial simplicity, these sensors achieve sensitivities in the range of $100\unit{aT/\rtHz}$ \cite{Dang10} and approach a theory limit of $<10\unit{aT/\rtHz}$, placing them on par with Superconducting Quantum Interference Device (SQUIDs, see below) as the most sensitive magnetometers to date.  This is owing to the surprising fact that relaxation and coherence times of spins in atomic vapors can be pushed to the second to minute range \cite{Balabas10}.  These long relaxation and coherence times are achieved by coating cell walls to preserve the atomic spin upon collisions, and by operating in the spin exchange relaxation-free (``SERF'') regime of high atomic density and zero magnetic field. Somewhat counterintuitively, a high density suppresses decoherence from atomic interactions, since collisions occur so frequently that their effect averages out, similar to motional narrowing of dipolar interactions in nuclear magnetic resonance \cite{Happer73}. 
Vapor cells have been miniaturized to few mm$^3$ small volumes \cite{Shah07} and have been used to demonstrate entanglement-enhanced sensing \cite{Fernholz08, Wasilewski10}.  The most advanced application of vapor cells is arguably the detection of neural activity \cite{Livanov78, Jensen16}, which has found use in magnetoencephalography \cite{Xia06}. Vapor cells also promise complementary access to high-energy physics, detecting anomalous dipole moments from coupling to exotic elementary particles and background fields beyond the standard model \cite{Pustelny13, Swallows13,Smiciklas11}. 

\subsubsection{Cold atomic clouds}
The advent of laser cooling in the 1980s spawned a revolution in atomic sensing. The reduced velocity spread of cold atoms enabled sensing with longer interrogation times using spatially confined atoms, freely falling along specific trajectories in vacuum or trapped.

Freely falling atoms have enabled the development of atomic gravimeters \cite{Kasevich92, Peters99} and gyrometers \cite{Gustavson97,Gustavson00}. In these devices an atomic cloud measures acceleration by sensing the spatial phase shift of a laser beam along its freely falling trajectory.

Trapped atoms have been employed to detect and image magnetic fields at the microscale, by replicating Larmor precession spectroscopy on a trapped Bose-Einstein condensate \cite{Vengalattore07} and by direct driving of spin-flip transitions by microwave currents \cite{Ockeloen13} or thermal radiofrequency noise in samples \cite{Fortagh02, Jones03}. Sensing with cold atoms has found application in solid state physics by elucidating current transport in microscopic conductors \cite{Aigner08}.

Arguably the most advanced demonstrations of entanglement-enhanced quantum sensing (``Definition III'') have been implemented in trapped cold atoms and vapor cells.  Entanglement -- in the form of spin squeezing \cite{Wineland92} -- has been produced by optical non-destructive measurements of atomic population \cite{Appel09, Leroux10, Schleier-Smith10, Louchet10, Bohnet14, Cox16, Hosten16} and atomic interactions \cite{Esteve08, Riedel10}. It has improved the sensitivity of magnetometry devices beyond the shot noise limit \cite{Sewell12, Ockeloen13} and has increased their bandwidth \cite{Shah10}. 

\subsection{Trapped ions}

Ions, trapped in vacuum by electric or magnetic fields, have equally been explored as quantum sensors. The most advanced applications employ the quantized motional levels as sensing qubits for electric fields and forces. These levels are strongly coupled to the electric field by dipole-allowed transitions and sufficiently (MHz) spaced to be prepared by Raman cooling and read out by laser spectroscopy. The sensor has a predicted sensitivity of $500\unit{nV/m/\rtHz}$ or $1\unit{yN/\rtHz}$ for the force acting on the ion \cite{Maiwald09,Biercuk10}.  Trapped ions have been extensively used to study electric field noise above surfaces \cite{Brownnutt15}, which could  arise from charge fluctuations induced by adsorbents.  Electrical field noise is a severe source of decoherence for ion traps and superconducting quantum processors \cite{Labaziewicz08} and a key limiting factor in ultrasensitive force microscopy \cite{Kuehn06,Tao15nl}. 

Independently, the ground state spin sublevels of ions are magnetic-field-sensitive qubits analogous to neutral atoms discussed above \cite{Maiwald09,Baumgart16,Kotler11}.  Being an extremely clean system, trapped ions have demonstrated sensitivities down to $4.6\unit{pT/\rtHz}$ 
 \cite{Baumgart16} and served as a testbed for advanced sensing protocols such as dynamical decoupling \cite{Biercuk09,Kotler11} and entanglement-enhanced sensing \cite{Leibfried04}.  Recently, trapped ions have also been proposed as rotation sensors, via matter-wave Sagnac interferometry \cite{Campbell17}.
Their use in practical applications, however, has proven difficult.  Practically all sensing demonstrations have focused on single ions, which, in terms of absolute sensitivity, cannot compete with ensemble sensors such as atomic vapors. Their small size could compensate for this downside in applications like microscopy, where high spatial resolution is required. However, operation of ion traps in close proximity to surfaces remains a major challenge.  Recent work on large ion crystals \cite{Arnold15,Bohnet16,Drewsen15} opens however the potential for novel applications to precise clocks and spectroscopy.

\subsection{Rydberg atoms}

Rydberg atoms -- atoms in highly excited electronic states -- are remarkable quantum sensors for electric fields for a similar reason as trapped ions: In a classical picture, the loosely confined electron in a highly excited orbit is easily displaced by electric fields. In a quantum picture, its motional states are coupled by strong electric dipole transitions and experience strong Stark shifts \cite{Herrmann86,Osterwalder99}. Preparation and readout of states is possible by laser excitation and spectroscopy. 

As their most spectacular sensing application, Rydberg atoms in vacuum have been employed as single-photon detectors for microwave photons in a cryogenic cavity in a series of experiments that has been highlighted by the Nobel prize in Physics in 2012 \cite{Nogues99,Gleyzes07,Haroche13}. Their sensitivity has recently been improved by employing Schr\"odinger cat states to reach a level of $300$nV/m/$\sqrt{\text{Hz}}$ \cite{Facon16}.

Recently, Rydberg states have become accessible in atomic vapour cells \cite{Kubler10}. They have been applied to sense weak electric fields, mostly in the GHz frequency range \cite{Sedlacek12, Fan15}, and have been suggested as a candidate for a primary traceable standard of microwave power. 

\subsection{Atomic clocks}

At first sight, atomic clocks -- qubits with transitions so insensitive that their level splitting can be regarded as absolute and serve as a frequency reference -- do not seem to qualify as  quantum sensors since this very definition violates criterion (4).  Their operation as  clocks, however, employs identical protocols as the operation of quantum sensors, in order to repeatedly compare the qubit's transition to the frequency of an unstable local oscillator and subsequently lock the latter to the former.  Therefore, an atomic clock can be equally regarded as a quantum sensor measuring and stabilizing the phase drift of a local oscillator. Vice versa, quantum sensors  discussed above can be regarded as clocks that operate on purpose on a bad, environment-sensitive clock transition in order to measure external fields. 

Today's most advanced atomic clocks employ optical transitions in single ions \cite{Huntemann16} or atomic clouds trapped in an optical lattice \cite{Takamoto05,Hinkley13, Bloom14}. Interestingly, even entanglement-enhanced sensing has found use in actual devices, since some advanced clocks employ multi-qubit quantum logic gates for readout of highly stable but optically inactive clock ions \cite{Schmidt05,Rosenband08}.

\subsection{Solid state spins -- Ensemble sensors}

\subsubsection{NMR ensemble sensors}
Some of the earliest quantum sensors have been based on ensembles of nuclear spins. Magnetic field sensors have been built that infer field strength from their Larmor precession, analogous to neutral atom magnetometers described above \cite{Packard54,Waters58,Kitching11}. Initialization of spins is achieved by thermalization in an externally applied field, readout by induction detection. Although the sensitivity of these devices ($10\unit{pT/\rtHz}$) \cite{Lenz90} is inferior to their atomic counterparts, they have found broad use in geology, archaeology and space missions thanks to their simplicity and robustness. 
More recently, NMR sensor probes have been developed for in-situ and dynamical field mapping in clinical MRI systems \cite{Dezanche08}.

Spin ensembles have equally served as gyroscopes \cite{Woodman87,Fang12}, exploiting the fact that Larmor precession occurs in an independent frame of reference and therefore appears frequency-shifted in a rotating laboratory frame. In the most advanced implementation, nuclear spin precession is read out by an atomic magnetometer, which is equally used for compensation of the Zeeman shift \cite{Kornack05}.  These experiments reached a sensitivity of $5\ee{-7}\unit{rad/s/\rtHz}$, which is comparable to compact implementations of atomic interferometers and optical Sagnac interferometers.

\subsubsection{NV center ensembles}
Much excitement has recently been sparked by ensembles of nitrogen-vacancy centers (NV centers) -- electronic spin defects in diamond that can be optically initialized and read out.  Densely-doped diamond crystals promise to deliver ``frozen vapor cells'' of spin ensembles that combine the strong (electronic) magnetic moment and efficient optical readout of atomic vapor cells with the high spin densities achievable in the solid state.  Although these advantages are partially offset by a reduced coherence time ($T_2 < 1\unit{ms}$ at room temperature, as compared to $T_2 > 1\unit{s}$ for vapor cells), the predicted sensitivity of diamond magnetometers ($250\unit{aT/\rtHz/cm^{-3/2}}$) \cite{Taylor08} or gyroscopes ($10^{-5}\unit{rad/s/\rtHz/mm^{3/2}}$) \cite{Ledbetter12,Ajoy12} would be competitive with their atomic counterparts. 

Translation of this potential into actual devices remains challenging, with two technical hurdles standing out.  First, efficient fluorescence detection of large NV ensembles is difficult, while absorptive and dispersive schemes are not easily implemented \cite{Lesage12,Jensen14,Clevenson15}.  Second, spin coherence times are reduced $100-1000$ times in high-density ensembles owing to interaction of NV spins with parasitic substitutional nitrogen spins incorporated during high-density doping \cite{Acosta09}.  As a consequence, even the most advanced devices are currently limited to $\sim 1\unit{pT/\rtHz}$ \cite{Wolf15} and operate several orders of magnitude above the theory limit. As a technically less demanding application, NV centers in a magnetic field gradient have been employed as spectrum analyzer for high frequency microwave signals \cite{Chipaux15a}.

While large-scale sensing of homogeneous fields remains a challenge, micrometer-sized ensembles of NV centers have found application in imaging applications, serving as detector pixels for microscopic mapping of magnetic fields.  Most prominently, this line of research has enabled imaging of magnetic organelles in magnetotactic bacteria \cite{LeSage13} and microscopic magnetic inclusions in meteorites \cite{Fu14}, as well as contrast-agent-based magnetic resonance microscopy \cite{Steinert13}.

\subsection{Solid state spins - Single spin sensors}  

Readout of single spins in the solid state -- a major milestone on the road towards  quantum computers -- has been achieved both by electrical and optical schemes.  Electrical readout has been demonstrated with phosphorus dopants in silicon \cite{Morello10} and electrostatically-defined semiconductor quantum dots \cite{Elzerman04}.  Optical readout was shown with single organic molecules \cite{Wrachtrup93,Wrachtrup93b}, optically active quantum dots \cite{Kroutvar04,Atature07,Vamivakas10}, and defect centers in crystalline materials including diamond \cite{Gruber97} and silicon carbide \cite{Christle15,Widmann15}.  
In addition, mechanical detection of single paramagnetic defects in silica \cite{Rugar04} and real-time monitoring of few-spin fluctuations \cite{Budakian05} have been demonstrated.

Among all solid state spins, NV centers in diamond have received by far the most attention for sensing purposes. This is in part due to the convenient room-temperature optical detection, and in part due to their stability in very small crystals and nanostructures.  The latter permits use of NV centers as sensors in high-resolution scanning probe microscopy \cite{Chernobrod05,Degen08,Balasubramanian08}, as biomarkers within living organisms \cite{Fu07}, or as stationary probes close to the surface of diamond sensor chips.  Quantum sensing with NV centers has been considered in several recent focused reviews \cite{Rondin14,Schirhagl14}.

Single NV centers have been employed and/or proposed as sensitive magnetometers \cite{Taylor08,Maze08,Balasubramanian08}, electrometers \cite{Dolde11}, pressure sensors \cite{Doherty14} and thermometers \cite{Hodges13,Kucsko13,Toyli13,Neumann13}, using the Zeeman, Stark and temperature shifts of their spin sublevels.  The most advanced nano-sensing experiments in terms of sensitivity have employed near-surface NV centers in bulk diamond crystals.  This approach has enabled sensing of nanometer-sized voxels of nuclear or electronic spins deposited on the diamond surface \cite{Mamin13,Staudacher13,Loretz14apl,Shi15,Lovchinsky16,DeVience15,Sushkov14l}, of distant nuclear spin clusters \cite{Shi14}, and of 2D materials \cite{Lovchinsky17}.
Other applications included the study of ballistic transport in the Johnson noise of nanoscale conductors \cite{Kolkowitz15}, phases and phase transitions of skyrmion materials \cite{Dovzhenko16,Dussaux16}, as well as of spin waves \cite{Wolfe14,Vandersar15}, and relaxation in nanomagnets \cite{Schafer14,Schmidlorch15}.

Integration of NV centers into scanning probes has enabled imaging of magnetic fields with sub-100 nm resolution, with applications to nanoscale magnetic structures and domains \cite{Balasubramanian08,Maletinsky12,Rondin12}, vortices and domain walls \cite{Rondin13,Tetienne14,Tetienne15}, superconducting vortices \cite{Pelliccione16,Thiel16}, and mapping of currents \cite{Chang17}. 

NV centers in $\sim$10-nm-sized nanodiamonds have also been inserted into living cells. They have been employed for particle tracking \cite{McGuinness11} and \textit{in vivo} temperature measurements \cite{Kucsko13,Neumann13,Toyli13} and could enable real-time monitoring of metabolic processes. 

\subsection{Superconducting circuits}

\subsubsection{SQUIDs}

The Superconducting Quantum Interference Device (SQUIDs) is simultaneously one of the oldest and one of the most sensitive type of magnetic sensor \cite{Jaklevic65,Clarke04,Fagaly06}. These devices -- interferometers of superconducting conductors -- measure magnetic fields with a sensitivity down to $10\unit{aT/\rtHz}$ \cite{Simmonds79}. Their sensing mechanism is based on the Aharonov-Bohm phase imprinted on the superconducting wave function by an encircled magnetic field, which is read out by a suitable circuit of phase-sensitive Josephson junctions. 

From a commercial perspective, SQUIDs can be considered the most advanced type of quantum sensor, with applications ranging from materials characterization in solid state physics to clinical magnetoencephalography systems for measuring tiny ($\sim 100\unit{fT}$) stray fields of electric currents in the brain.  In parallel to the development of macroscopic (mm-cm) SQUID devices, miniaturization has given birth to sub-micron sized ``nanoSQUIDs'' with possible applications in nanoscale magnetic, current, and thermal imaging \cite{Vasyukov13,Halbertal16}. Note that because SQUIDs rely on spatial rather than temporal coherence, they are more closely related to optical interferometers than to the spin sensors discussed above.

SQUIDs have been employed to process signals from the DC up to the GHz range \cite{Mueck03, Hatridge11}, the upper limit being set by the Josephson frequency. Conceptually similar circuits, dedicated to amplification of GHz frequency signals, have been explored in great detail in the past decade \cite{Castellanos08, Bergeal10, Ho12, Macklin15}. Arguably the most widely studied design is the Josephson parametric amplifier, which has been pushed to a nearly quantum-limited input noise level of only few photons and is now routinely used for spectroscopic single shot readout of superconducting qubits \cite{Vijay11}.

\subsubsection{Superconducting qubits}
Temporal quantum superpositions of supercurrents or charge eigenstates have become accessible in superconducting qubits \cite{Nakamura99,Vion02,Martinis02,Wallraff04,Clarke08}. Being associated with large magnetic and electric dipole moments, they are attractive candidates for quantum sensing. Many of the established quantum sensing protocols to be discussed in Sections \ref{sec04}--\ref{sec07} have been implemented with superconducting qubits. Specifically, noise in these devices has been thoroughly studied from the sub-Hz to the GHz range, using Ramsey interferometry \cite{Yan12, Yoshihara06}, dynamical decoupling \cite{Nakamura02,Ithier05,Yoshihara06, Bylander11, Yan13}, and $T_1$ relaxometry \cite{Astafiev04, Yoshihara06}. These studies have been extended to discern charge from flux noise by choosing qubits with a predominant electric (charge qubit) or magnetic (flux qubit) dipole moment, or by tuning bias parameters \textit{in situ} \cite{Bialczak07, Yan12}. Operating qubits as magnetic field sensors, very promising sensitivities (3.3 pT$/\sqrt{\textrm{Hz}}$ for operation at 10 MHz) were demonstrated \cite{Bal12}.  Extending these experiments to the study of extrinsic samples appears simultaneously attractive and technically challenging, since superconducting qubits have to be cooled to temperatures of only few tens of millikelvin. 

\subsection{Elementary particle qubits}

Interestingly, elementary particles have been employed as quantum sensors long before the development of atomic and solid state qubits. This somewhat paradoxical fact is owing to their straightforward initialization and readout, as well as their targeted placement in relevant samples by irradiation with a particle beam. 

\subsubsection{Muons}
Muons are frequently described as close cousins of electrons. Both particles are leptons, carry an elementary charge and have a spin that can be employed for quantum sensing.  Sensing with muons has been termed ``muon spin rotation'' ($\mu$SR). It employs antimuons ($\mu^+$) that are deterministically produced by proton-proton collisions, from decay of an intermediate positive pion by the reaction $\pi^+ \rightarrow \mu^+ +\nu_\mu$. 
Here, parity violation of the weak interaction automatically initializes the muon spin to be collinear with the particle's momentum. Readout of the spin is straightforward by measuring the emission direction of positrons from the subsequent decay $\mu^+ \rightarrow e^+ + \nu_e + {\overline{\nu}}_\mu$, which are preferably emitted along the muon spin \cite{Brewer78, Blundell99}. 

Crucially, muons can be implanted into solid state samples and serve as local probes of their nanoscale environment for their few microseconds long lifetime. Larmor precession measurements have been used to infer the intrinsic magnetic field of materials. Despite its exotic nature, the technique of muon spin rotation ($\mu$SR) has become and remained a workhorse tool of solid state physics. In particular, it is a leading technique to measure the London penetration depth of superconductors \cite{Sonier00}. 

\subsubsection{Neutrons}

Slow beams of thermal neutrons can be spin-polarized by Bragg reflection on a suitable magnetic crystal. Spin readout is feasible by a spin-sensitive Bragg analyzer and subsequent detection.  Spin rotations (single qubit gates) are easily implemented by application of localized magnetic fields along parts of the neutron's trajectory. As a consequence, many early demonstrations of quantum effects, such as the direct measurement of Berry's phase \cite{Bitter87}, have employed neutrons. 

Sensing with neutrons has been demonstrated in multiple ways. Larmor precession in the magnetic field of samples has been employed for three-dimensional tomography \cite{Kardjilov08}. Neutron interferometry has put limits on the strongly-coupled chameleon field \cite{Li16}.  Ultracold neutrons have been employed as a probe for gravity on small length scales in a series of experiments termed "qBounce". These experiments exploit the fact that suitable materials perfectly reflect the matter wave of sufficiently slow neutrons so that they can be trapped above a bulk surface by the gravity of earth as a ``quantum bouncing ball'' \cite{Nesvizhevsky02}. The eigenenergies of this anharmonic trap depend on gravity and have been probed by quantum sensing techniques \cite{Jenke11, Jenke14}.

The most established technique, neutron spin echo, can reveal materials properties by measuring  small (down to neV) energy losses of neutrons in inelastic scattering events \cite{Mezei72}.  Here, the phase of the neutron spin, coherently precessing in an external magnetic field, serves as a clock to measure a neutron's time of flight. Inelastic scattering in a sample changes a neutron's velocity, resulting in a different time of flight to and from a sample of interest. This difference is imprinted in the spin phase by a suitable quantum sensing protocol, specifically a Hahn echo sequence whose $\pi$ pulse is synchronized with passage through the sample. 

\subsection{Other sensors}
In addition to the many implementations of quantum sensors already discussed, three further systems deserved special attention for their future potential or for their fundamental role in developing quantum sensing methodology.

\subsubsection{Single electron transistors}
Single electron transistors (SET's) sense electric fields by measuring the tunneling current across a submicron conducting island sandwiched between tunneling source and drain contacts. In the ``Coulomb blockade regime'' of sufficiently small (typically $\approx 100\unit{nm}$) islands, tunneling across the device is only allowed if charge eigenstates of the island lie in the narrow energy window between the Fermi level of source and drain contact.  The energy of these eigenstates is highly sensitive to even weak external electric fields, resulting in a strongly field-dependent tunneling current \cite{Kastner92,Yoo97,Schoelkopf98}. SETs have been employed as scanning probe sensors to image electric fields on the nanoscale, shedding light on a variety of solid-state-phenomena such as the fractional quantum Hall effect or electron-hole puddles in graphene \cite{Ilani04,Martin08}.  In a complementary approach, charge sensing by stationary SETs has enabled readout of optically inaccessible spin qubits such as phosphorus donors in silicon \cite{Morello10} based on counting of electrons \cite{Bylander05}.

\subsubsection{Optomechanics} 
Phonons -- discrete quantized energy levels of vibration -- have recently become accessible at the ``single-particle'' level in the field of optomechanics \cite{Aspelmeyer14,Oconnell10}, which studies high-quality mechanical oscillators that are strongly coupled to light. 

While preparation of phonon number states and their coherent superpositions remains difficult, the  devices built to achieve these goals have shown great promise for sensing applications.  This is mainly due to the fact that mechanical degrees of freedom strongly couple to nearly all external fields, and that strong optical coupling enables efficient actuation and readout of mechanical motion.  Specifically, optomechanical sensors have been employed to detect minute forces ($12\unit{zN/\rtHz}$, \onlinecite{Moser13}), acceleration ($100\unit{ng/\rtHz}$, \onlinecite{Krause12,Cervantes14}), masses ($2\unit{yg/\rtHz}$, \onlinecite{Chaste12}), magnetic fields ($200\unit{pT/\rtHz}$, \onlinecite{Forstner14}), spins \cite{Rugar04,Degen09}, and voltage ($5\unit{pV/\rtHz}$, \onlinecite{Bagci14}).   While these demonstrations have remained at the level of classical sensing in the sense of this review, their future extension to quantum-enhanced measurements appears most promising.

\subsubsection{Photons}
While this review will not discuss quantum sensing with photons, due to the breadth of the subject, several fundamental paradigms have been pioneered with optical sensors including light squeezing and photonic quantum correlations.  These constitute examples of quantum-enhanced sensing according to our ``Definition III''.

Squeezing of light -- the creation of partially-entangled states with phase or amplitude fluctuations below those of a classical coherent state of the light field -- has been proposed \cite{Caves81} and achieved \cite{Slusher85} long before squeezing of spin ensembles \cite{Wineland92,Hald99}. 
 Vacuum squeezed states have meanwhile been employed to improve the sensitivity of gravitational wave detectors. In the GEO gravitational wave detector, squeezing has enhanced the shot-noise limited sensitivity by 3.5 dB \cite{Ligo11}; in a proof-of-principle experiment in the LIGO gravitational wave detector, the injection of 10dB of squeezing lowered the shot-noise in the interferometer output by approximately 2.15dB (28\%)\cite{Aasi13short}, equivalent to an increase by more than 60\% in the power stored in the interferometer arm cavities. Further upgrades associated with Advanced LIGO could bring down the shot noise by 6dB, via frequency dependent squeezing \cite{Oelker16}.

In addition, quantum correlations between photons have proven to be a powerful resource for imaging. This has been noted very early on in the famous Hanbury-Brown-Twiss experiment, where bunching of photons is employed to filter atmospheric aberrations and to perform ``super-resolution'' measurements of stellar diameters smaller than the diffraction limit of the telescope employed \cite{Hanbury56}.  While this effect can still be accounted for classically, a recent class of experiments has exploited non-classical correlations to push the spatial resolution of microscopes below the diffraction limit \cite{Schwartz13}. Vice versa, multi-photon correlations have been proposed and employed to create light patterns below the diffraction limit for superresolution lithography \cite{Boto00, Dangelo01}. They can equally improve image contrast rather than resolution by a scheme known as ``quantum illumination'' \cite{Tan08, Lloyd08, Lopaeva13}. Here, a beam of photons is employed to illuminate an object, reflected light being detected as the imaging signal. Entangled twins of the illumination photons are conserved at the source and compared to reflected photons by a suitable joint measurement. In this way, photons can be certified to be reflected light rather than noise, enhancing imaging contrast. In simpler schemes, intensity correlations between entangled photons have been employed to boost contrast in transmission microscopy of weakly absorbing objects \cite{Brida10} and the reduced quantum fluctuations of squeezed light have been used to improve optical particle tracking \cite{Taylor13}.

The most advanced demonstrations of entanglement-enhanced sensing have been performed with single photons or carefully assembled few-photon Fock states. Most prominently, these include Heisenberg-limited interferometers \cite{Holland93,Mitchell04,Walther04,Higgins07,Nagata07}.  In these devices, entanglement between photons or adaptive measurements are employed to push sensitivity beyond the $1/\sqrt{N}$ scaling of a classical interferometer where $N$ is the number of photons (see Section \ref{sec09}).


\section{The quantum sensing protocol}
\label{sec04}

In this Section, we describe the basic methodology for performing measurements with quantum sensors.  Our discussion will focus on a generic scheme where a measurement consists of three elementary steps: the initialization of the quantum sensor, the interaction with the signal of interest, and the readout of the final state.
Phase estimation \cite{Shor94,Kitaev95} and parameter estimation \cite{Braunstein94,Braunstein96,Goldstein10x} techniques are then used to reconstruct the physical quantity from a series of measurements.
Experimentally, the protocol is typically implemented as an interference measurement using pump-probe spectroscopy, although other schemes are possible.  The key quantity is then the \textit{quantum phase} picked up by the quantum sensor due to the interaction with the signal.  The protocol can be optimized for detecting weak signals or small signal changes with the highest possible sensitivity and precision.

\subsection{Quantum sensor Hamiltonian}

For the following discussion, we will assume that the quantum sensor can be described by the generic Hamiltonian
\begin{equation}
\hat{H}(t) = \hat{H}_0 + \hat{H}_V(t) + \hat{H}_\mr{control}(t) \ ,
\label{eq:hamiltonian}
\end{equation}
where $\hat{H}_0$ is the internal Hamiltonian, $\hat{H}_V(t)$ is the Hamiltonian associated with a signal $V(t)$, and $\hat{H}_\mr{control}(t)$ is the control Hamiltonian.  We will assume that $\hat{H}_0$ is known and that $\hat{H}_\mr{control}(t)$ can be deliberately chosen so as to manipulate or tune the sensor in a controlled way.
The goal of a quantum sensing experiment is then to infer $V(t)$ from the effect it has on the qubit via its Hamiltonian $\hat{H}_V(t)$, usually by a clever choice of $\hat{H}_\mr{control}(t)$.

%
%

\subsubsection{Internal Hamiltonian}

$\hat H_0$ describes the internal Hamiltonian of the quantum sensor in the absence of any signal.  Typically, the internal Hamiltonian is static and defines the energy eigenstates $\ma$ and $\mb$,
\begin{equation}
\hat{H}_0 = E_0 \ma\ima + E_1 \mb\imb \ ,
\end{equation}
where $E_0$ and $E_1$ are the eigenenergies and $\wo=E_1-E_0$ is the transition energy between the states ($\hbar=1$).  Note that the presence of an energy splitting $\wo\neq 0$ is not  necessary, but it represents the typical situation for most implementations of quantum sensors.  The qubit internal Hamiltonian may contain additional interactions that are specific to a quantum sensor, such as couplings to other qubits.  In addition, the internal Hamiltonian contains time-dependent stochastic terms due to a classical environment or  interactions with a quantum bath that are responsible for decoherence and relaxation.

\subsubsection{Signal Hamiltonian}

The signal Hamiltonian $\hat H_V(t)$ represents the coupling between the sensor qubit and a signal $V(t)$ to be measured.  When the signal is weak (which is assumed here) $\hat H_V(t)$ adds a small perturbation to $\hat H_0$.  The signal Hamiltonian can then be separated into two qualitatively different contributions,
\begin{equation}
\hat H_V(t)= \HVzero(t) + \HVone(t) \ ,
\end{equation}
where $\HVzero$ is the parallel (commuting, secular) and $\HVone$ the transverse (non-commuting) component, respectively.  The two components can quite generally be captured by
\begin{eqnarray}
\HVzero(t) & = & \tfrac12 \gamma \Vzero(t) \left\{ \mb\imb - \ma\ima \right\} \ , \nonumber \\
\HVone(t) & = & \tfrac12 \gamma \left\{ \Vone(t) \mb\ima + \Vone^\dagger(t) \ma\imb \right\} \ ,
\label{eq:HVa}
\end{eqnarray}
where $\Vzero(t)$ and $\Vone(t)$ are functions with the same units of $V(t)$.  $\gamma$ is the coupling or \textit{transduction parameter} of the qubit to the signal $V(t)$.  Examples of coupling parameters include the Zeeman shift parameter (gyromagnetic ratio) of spins in a magnetic field, with units of $\mr{Hz/T}$, or the linear Stark shift parameter of electric dipoles in an electric field, with units of $\mr{Hz/(Vm^{-1})}$.  Although the coupling is often linear, this is not required.  In particular, the coupling is quadratic for second-order interactions (such as the quadratic Stark effect) or when operating the quantum sensor in variance detection mode (see Section \ref{sec04:variance}).

The parallel and transverse components of a signal have distinctly different effects on the quantum sensor.  A commuting perturbation $\HVzero$ leads to \textit{shifts} of the energy levels and an associated change of the transition frequency $\wo$.  A non-commuting perturbation $\HVone$, by contrast, can induce \textit{transitions} between levels,  manifesting through an increased transition rate $\Gamma$.  Most often, this requires the signal to be time-dependent (resonant with the transition) in order to have an appreciable effect on the quantum sensor. 

An important class of signals are vector signal $\vec V(t)$, in particular those provided by electric or magnetic fields.  The interaction between a vector signal $\vec V(t) = \{V_x, V_y, V_z \}(t)$ and a qubit can be described by the signal Hamiltonian
\begin{equation}
\hat H_V (t) = \gamma \vec V(t) \cdot \hat{\vec\sigma} \ ,
\end{equation}
where $\vec \sigma=\{\sigma_x,\sigma_y,\sigma_z\}$ is a vector of Pauli matrices.  For a vector signal, the two signal functions $\Vzero(t)$ and $\Vone(t)$ are
\begin{align}
\Vzero(t) &= V_z(t) \nonumber \ , \\
\Vone(t) &= V_x(t) + i V_y(t) ,
\end{align}
where the $z$ direction is defined by the qubit's quantization axis.  The corresponding signal Hamiltonian is
\begin{equation}
\hat H_V(t) = \gamma \mr{Re}[\Vone(t)] \hat\sigma_x + \gamma \mr{Im}[\Vone(t)] \hat\sigma_y + \gamma \Vzero(t) \hat\sigma_z \ .
\end{equation}

\subsubsection{Control Hamiltonian}

For most quantum sensing protocols it is required to manipulate the qubit either before, during, or after the sensing process.  This is achieved via a control Hamiltonian $\hat H_\mr{control}(t)$ that allows implementing a standard set of quantum gates \cite{Nielsen00b}.  The most common gates in quantum sensing include the Hadamard gate and the Pauli-X and Y gates, or equivalently, a set of $\pi/2$ and $\pi$ rotations (pulses) around different axes.  Advanced sensing schemes employing more than one sensor qubit may further require conditional gates, especially controlled-NOT gates to generate entanglement,  Swap gates to exploit memory qubits, and controlled phase shifts in quantum phase estimation.  Finally, the control Hamiltonian can include control fields for systematically tuning the transition frequency $\wo$.  This capability is frequently exploited in noise spectroscopy experiments.

\subsection{The sensing protocol}
\label{sec04:basicprotocol}

\begin{figure}[t!]
\centering
\includegraphics[width=0.7\columnwidth]{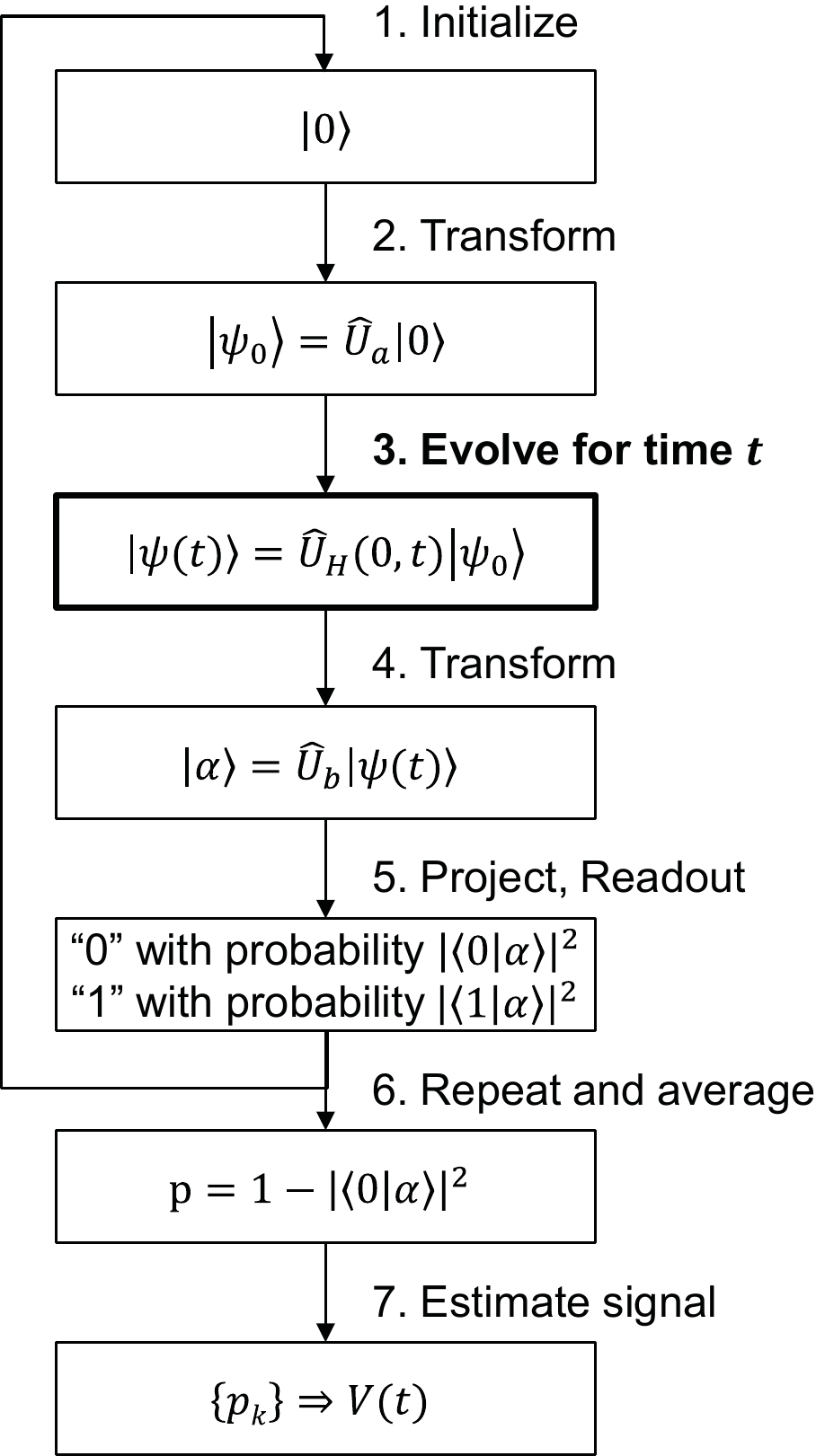}
\caption{Basic steps of the quantum sensing process.}
\label{fig:sensingprocessB}
\end{figure}

Quantum sensing experiments typically follow a generic sequence of sensor initialization, interaction with the signal, sensor readout and signal estimation.
This sequence can be summarized in the following basic protocol, which is also sketched in Fig. \ref{fig:sensingprocessB}:
\begin{enumerate}
\item[1.] The quantum sensor is initialized into a known basis state, for example $\ma$.
\item[2.] The quantum sensor is transformed into the desired initial sensing state $\psia = \hat U_a\ma$.  The transformation can be carried out using a set of control pulses represented by the propagator $\hat U_a$.  In many cases, $\psia$ is a superposition state. 
\item[3.] The quantum sensor evolves under the Hamiltonian $\hat{H}$ [Eq. (\ref{eq:hamiltonian})] for a time $t$.  At the end of the sensing period, the sensor is in the final sensing state
\begin{align}
\psit = \UH \psia = c_0\psia + c_1\psib \ ,
\end{align}
where $\UH$ is the propagator of $\hat{H}$, $\psib$ is the state orthogonal to $\psia$ and $c_0, c_1$ are complex coefficients.
\item[4.] The quantum sensor is transformed into a superposition of observable readout states $\malpha = \hat U_b \psit = c_0'\mad+c_1'\mbd$.  For simplicity we assume that the initialization basis $\{\ma,\mb\}$ and the readout basis $\{\mad,\mbd\}$ are the same and that $\hat U_b= \hat U_a^\dagger$, but this is not required.  Under these assumptions, the coefficients $c_0' \equiv c_0$ and $c_1' \equiv c_1$  represent the overlap between the initial and final sensing states.
\item[5.] The final state of the quantum sensor is read out.  We assume that the readout is projective, although more general positive-operator-valued-measure (POVM) measurements may be possible~\cite{Nielsen00b}.  The projective readout is a Bernoulli process that yields an answer ``0'' with probability $1-p'$ and an answer ``1'' with probability $p'$, where $p'=|c_1'|^2 \propto p$ is proportional to the measurable transition probability,
\begin{align}
p = 1-|c_0|^2 = |c_1|^2
\end{align}
that the qubit changed its state during $t$.
The binary answer is detected by the measurement apparatus as a physical quantity $x$, for example, as a voltage, current, photon count or polarization.
\end{enumerate}
Steps 1-5 represent a single measurement cycle.  Because step 5 gives a binary answer, the measurement cycle needs to be repeated many times in order to gain a precise estimate for $p$:
\begin{enumerate}
\item[6.] Steps 1-5 are repeated and averaged over a large number of cycles $N$ to estimate $p$.  The repetition can be done by running the protocol sequentially on the same quantum system, or in parallel by averaging over an ensemble of $N$ identical (and non-interacting) quantum systems.
\end{enumerate}
Step 6 only provides one value for the transition probability $p$.  While a single value of $p$ may sometimes be sufficient to estimate a signal $V$, it is in many situations convenient or required to record a set of values $\{p_k\}$:
\begin{enumerate}
\item[7.] The transition probability $p$ is measured as a function of time $t$ or of a parameter of the control Hamiltonian, and the desired signal $V$ is inferred from the data record $\{p_k\}$ using a suitable procedure.
\end{enumerate}
More generally, a set of measurements can be optimized to efficiently extract a desired parameter from the signal Hamiltonian (see Section \ref{sec08}).  Most protocols presented in the following implicitly use such a strategy for gaining information about the signal.

Although the above protocol is generic and simple, it is sufficient to describe most sensing experiments.
For example, classical continuous-wave absorption and transmission spectroscopy can be considered as an averaged variety of this protocol.
Also, the time evolution can be replaced by a spatial evolution to describe a classical interferometer.   

To illustrate the protocol, we consider two elementary examples, one for detecting a parallel signal $\Vzero$ and one for detecting a transverse signal $\Vone$.  These examples will serve as the basis for the more refined sequences discussed in later Sections.

\subsection{First example: Ramsey measurement}
\label{sec04:ramsey}

A first example is the measurement of the static energy splitting $\wo$ (or equivalently, a static perturbation $\Vzero$).  The canonical approach for this measurement is a Ramsey interferometry measurement~\cite{Taylor08,Lee02}:
\begin{enumerate}
\item The quantum sensor is initialized into $\ma$.
\item Using a $\pi/2$ pulse, the quantum sensor is transformed into the superposition state
\begin{equation}
\psia = \xa \equiv \frac{1}{\sqrt2}(\ma + \mb) \ .
\end{equation}
\item The superposition state evolves under the Hamiltonian $\hat{H}_0$ for a time $t$.  The superposition state picks up the relative phase $\phi = \wo t$, and the state after the evolution is
\begin{equation}
\psit =  \frac{1}{\sqrt{2}}(\ma + e^{-i\wo t}\mb) \ ,
\label{eq:ramseyphase}
\end{equation}
up to an overall phase factor.
\item Using a second $\pi/2$ pulse, the state $\psit$ is converted back to the measurable state
\begin{equation}
\malpha = \frac12(1+e^{-i\wo t})\ma + \frac12 (1-e^{-i\wo t})\mb \ .
\end{equation}
\item[5,6.] The final state is read out.  The transition probability is
\begin{align}
p &= 1-|\langle 0 \malpha|^2 \nonumber \\
  &= \sin^2(\wo t/2) = \frac{1}{2}[1-\cos(\wo t)] .
\label{eq:pramsey}
\end{align}
\end{enumerate}
By recording $p$ as a function of time $t$, an oscillatory output (``Ramsey fringes'') is observed with a frequency given by $\wo$.  Thus, the Ramsey measurement can directly provide a measurement of the energy splitting $\wo$.

\subsection{Second example: Rabi measurement}
\label{sec04:rabi}

A second elementary example is the measurement of the transition matrix element $|\Vone|$:
\begin{enumerate}
\item[1.] The quantum sensor is initialized into $\psia=\ma$.
\item[3.] In the absence of the internal Hamiltonian, $\hat H_0=0$, the evolution is given by $\HVone=\frac12\gamma V_\perp\sigma_x=\omega_1\sigma_x$, where $\wone$ is the Rabi frequency.  The state after evolution is:
\begin{equation}
\psit = \malpha = \frac12(1+e^{-i\wone t})\ma + \frac12(1-e^{-i\wone t})\mb \ .
\end{equation}
\item[5,6.] The final state is read out.  The transition probability is:
\begin{equation}
p = 1-|\langle 0 \malpha|^2 = \sin^2(\wone t/2).
\label{eq:prabisimple}
\end{equation}
\end{enumerate}

In a general situation where $\hat H_0\neq0$, the transition probability represents the solution to Rabi's original problem~\cite{Sakurai11}, 
\begin{equation}
p = \frac{\wone^2}{\wone^2+\wo^2} \sin^2\left( \sqrt{\wone^2+\wo^2}\, t \right) \ .
\label{eq:prabi}
\end{equation}
Hence, only time-dependent signals with frequency $\omega \approx \wo$ affect the transition probability $p$, a condition known as resonance.
From this condition it is clear that a Rabi measurement can provide information not only on the magnitude $\Vone$, but also on the frequency $\omega$ of a signal \cite{Fedder11,Aiello13}.
\begin{figure}[t!]
\centering
\includegraphics[width=\figurewidth]{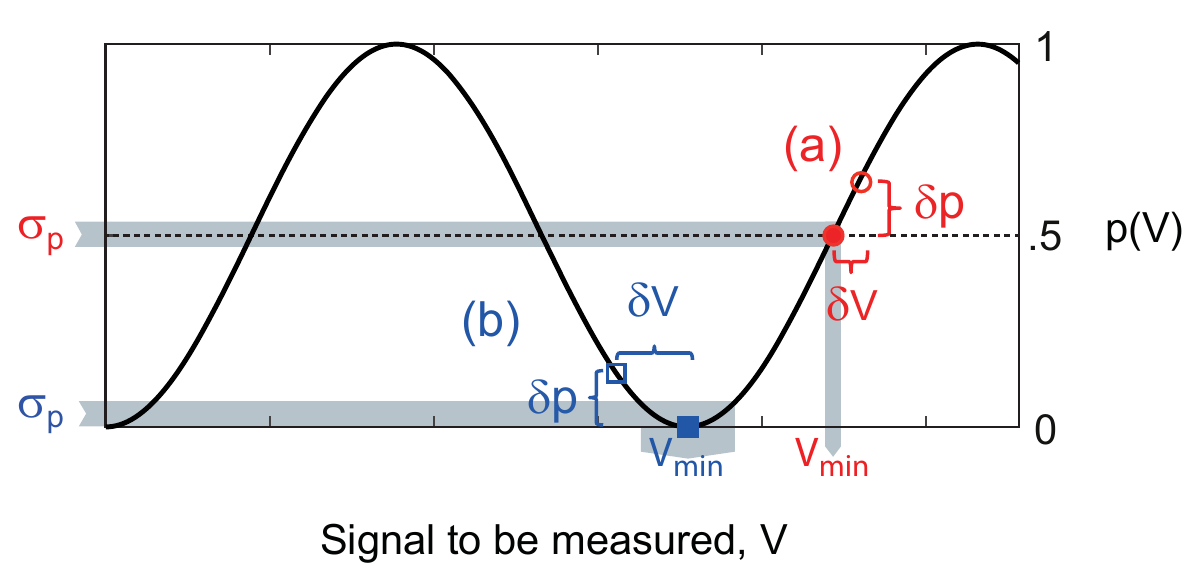}
\caption{
Transition probability $p$ for a Ramsey experiment as a function of the signal $V$ picked up by the sensor.
(a) Slope detection: The quantum sensor is operated at the $p_0=0.5$ bias point (filled red circle).  A small change in the signal $\delta V$ leads to a linear change in the transition probability, $\dpp = \delta\phi/2 = \gamma\delta Vt/2$ (empty red circle). The uncertainty $\sigma_p$ in the measured transition probability, leads to an uncertainty in the estimated signal, $\Vmin$ (grey shade). 
(b) Variance detection: The quantum sensor is operated at the $p_0=0$ bias point (filled blue square).  A small change in the signal $\delta V$ leads to a quadratic change, $\dpp = \dphi^2/4 = \gamma^2\delta V^2 t^2/4$ (empty blue square). The information on the sign of $\delta V$ is lost.  The experimental readout error $\sigp$ translates into an uncertainty in the estimated signal, $\Vmin$, according to the slope or curvature of the Ramsey fringe (grey shade).
}
\label{fig:ramseyrabi}
\end{figure}

\subsection{Slope and variance detection}
\label{sec:slopevariance}

A central objective of quantum sensing is the detection of small signals.  For this purpose, it is advantageous to measure the deviation of the transition probability from a well-chosen reference point $p_0$, which we will refer to as the \textit{bias point} of the measurement, corresponding to a known value of the external signal $V_0$ or reached by setting some additional parameters in the Hamiltonian under the experimenter's control.  The quantity of interest is then the difference $\dpp = p - p_0$ between the probability measured in the presence and absence of the signal, respectively.  Experimentally, the bias point can be adjusted by several means, for example by adding a small detuning to $\wo$ or by measuring the final state $\psit$ along different directions.

\subsubsection{Slope detection (linear detection)}
\label{sec04:slope}

The Ramsey interferometer is most sensitive to small perturbations $\delta V$ around $V_0=\wo/\gamma$ when operated at the point of maximum slope where $p_0 = 0.5$, indicated by the filled red dot in Fig. \ref{fig:ramseyrabi}(a).  This bias point is reached when $\wo t = k\pi/2$, with $k=1,3,5,...$.  Around $p_0=0.5$ , the transition probability is \textit{linear} in $\delta V$ and $t$,
\begin{align}
\dpp
  &= \frac12[1-\cos(\omega_0 t+\gamma\delta V t)] - \frac12 \nonumber \\
  &\approx \pm \frac12 \gamma\delta V t,
\label{eq:pslope}
\end{align}
where the sign is determined by $k$.

Note that slope detection has a limited linear range because phase wrapping occurs for $|\gamma\delta V t| > \pi/2$.  The phase wrapping restricts the dynamic range of the quantum sensor.  Section \ref{sec08} discusses adaptive sensing techniques designed to extend the dynamic range.

\subsubsection{Variance detection (quadratic detection)}
\label{sec04:variance}

If the magnitude of $\delta V$ fluctuates between measurements so that $\langle \delta V\rangle=0$,  readout at $p_0 = 0.5$ will yield no information about $\delta V$, since $\langle p\rangle\approx p_0 = 0.5$.  In this situation, it is advantageous to detect the signal \textit{variance} by biasing the measurement to a point of minimum slope, $\wo t=k\pi$, corresponding to the bias points $p_0=0$ and $p_0=1$ (filled blue square in Fig.  \ref{fig:ramseyrabi}(b)).  If the interferometer is tuned to $p_0=0$, a signal with variance $\langle \delta V^2 \rangle = \Vrms^2$ gives rise to a mean transition probability that is \textit{quadratic} in $\Vrms$ and $t$ \cite{Meriles10},
\begin{align}
\dpp &= p = \left\langle \frac12[1-\cos(\omega_0 t+\gamma\delta V t)]\right\rangle \nonumber \\
  &\approx \frac14\gamma^2\Vrms^2 t^2.
	\label{eq:pvariancesimple}
\end{align}
This relation holds for small $\gamma\Vrms t\ll 1$.  If the fluctuation is Gaussian, the result can be extended to any large value of $\gamma \Vrms t$,
\begin{align}
p &= \frac{1}{2}\left[ 1-\exp( -\gamma^2\Vrms^2 t^2/2 ) \right] \ .
	\label{eq:pvariance}
\end{align}
Variance detection is especially important for detecting ac signals when their synchronization with the sensing protocol is not possible (Section \ref{sec06:acrandom}), or when the signal represents a noise source (Section \ref{sec07}).


\section{Sensitivity}
\label{sec05}
The unprecedented level of sensitivity offered by many quantum sensors has been a key driving force of the field.  In this section, we quantitatively define the sensitivity.  We start by discussing the main sources of noise that enter a quantum sensing experiment, and derive expressions for the signal-to-noise ratio (SNR) and the minimum detectable signal, \ie, the signal magnitude that yields unit SNR.  This will lead us to a key quantity of this paper: the \textit{sensitivity}, $\vmin$, defined as the minimum detectable signal per unit time.  In particular, we will find in Sec.~\ref{sec:vmin} that $\vmin$ is approximately
\begin{equation}
\vmin \approx \frac{\sqrt{2e}}{\gamma C \sqrt{\Tx}}
\label{eq:equation_1_slope}
\end{equation}
for slope detection and
\begin{equation}
\vmin \approx \frac{\sqrt{2e}}{\gamma \sqrt C \sqrt[4]{\Tx^3}}
\label{eq:equation_1_var}
\end{equation}
for variance detection, where $\Tx$ is the sensor's coherence time,  $C\leq 1$ is a dimensionless constant quantifying readout efficiency, and $e$ is Euler's number (see Eqs. \ref{eq:voptslope} and \ref{eq:vminvar}).  In the remainder of the section signal averaging and the Allen variance are briefly discussed, and a formal definition of sensitivity by the quantum Cram\'{e}r-Rao bound (QCRB) is given.

\subsection{Noise}

Experimental detection of the probability $p$ will have a non-zero error $\sigp$.  This error translates into an error for the signal estimate, which is determined by the slope or curvature of the Ramsey fringe (see Fig. \ref{fig:ramseyrabi}).  In order to calculate  SNR and  sensitivity, it is therefore important to analyze the main sources of noise that enter $\sigp$.

\subsubsection{Quantum projection noise}

Quantum projection noise is the most fundamental source of uncertainty in quantum sensing.  The projective readout during ``Step 5'' of the quantum sensing protocol (Section \ref{sec04:basicprotocol}) does not directly yield the fractional probability $p\in[0...1]$, but one of the two values ``0'' or ``1'' with probabilities $1-p$ and $p$, respectively.  In order to precisely estimate $p$, the experiment is repeated $N$ times and the occurrences of ``0'' and ``1'' are binned into a histogram (see Fig. \ref{fig:readout}(a)).  The estimate for $p$ is then
\begin{equation}
p = \frac{N_1}{N} \ ,
\label{eq:p}
\end{equation}
where $N_1$ is the number of measurements that gave a result of ``1''.  The uncertainty in $p$ is given by the variance of the binomial distribution \cite{Itano93},
\begin{equation}
\sigpq^2 = \frac{1}{N}p(1-p) \ .
\end{equation}
The uncertainty in $p$ therefore depends on the bias point $p_0$ of the measurement.  For slope detection, where $p_0=0.5$, the uncertainty is
\begin{equation}
\sigpq^2 = \frac{1}{4N} \quad\quad\text{for $p_0=0.5$}
\label{eq:sigpq}
\end{equation}
Thus, the projective readout adds noise of order $1/(2\sqrt{N})$ to the probability value $p$.  For variance detection, where ideally $p_0=0$, the projection noise would in principle be arbitrarily low.  In any realistic experiment, however, decoherence will shift the fringe minimum to a finite value of $p$ (see below), where Eq. (\ref{eq:sigpq}) holds up to a constant factor.

\subsubsection{Decoherence}
\label{sec:decoherence}

A second source of error includes decoherence and relaxation during the sensing time $t$.  Decoherence and relaxation cause random transitions between states or random phase pick-up during coherent evolution of the qubit (for more detail, see Section \ref{sec07}).  The two processes lead to a reduction of the observed probability $\dpp$ with increasing sensing time $t$,
\begin{equation}
\dppobs(t) = \dpp(t) e^{-\chi(t)} \ ,
\label{eq:dpobs}
\end{equation}
where $\dpp(t)$ is the probability that would be measured in the absence of decoherence (see Eqs. (\ref{eq:pslope},\ref{eq:pvariancesimple})).
$\chi(t)$ is a phenomenological \textit{decoherence function} that depends on the noise processes responsible for  decoherence (see Section \ref{sec07:decoherencefunction}).  Although the underlying noise processes may be very complex, $\chi(t)$ can often be approximated by a simple power law,
\begin{equation}
\chi(t) = (\Gamma t)^a \ ,
\end{equation}
where $\Gamma$ is a  decay rate and typically $a=1...3$.  The decay rate can be associated with a decay time $\Tx = \Gamma^{-1}$ that equals the evolution time $t$ where $\dppobs/\dpp=1/e \approx 37\%$.  The decay time $\Tx$, also known as the \textit{decoherence time} or \textit{relaxation time} depending on the noise process, is an important figure-of-merit of the qubit, as it sets the maximum possible evolution time $t$ available for sensing.

\subsubsection{Errors due to initialization and qubit manipulations}

Errors can also enter through the imperfect initialization or manipulations of the quantum sensor.  An imperfect initialization leads to a similar reduction in the observed probability $\dppobs$ as with decoherence,
\begin{equation}
\dppobs = \beta\,\dpp \ ,
\end{equation}
where $\beta<1$ is a constant factor that describes the reduction of the observed $\dppobs$ as compared to the ideal $\dpp$.  Contrary to the case of decoherence, this reduction does not depend on the sensing time $t$.  Errors in qubit manipulations can cause many effects, but will typically also lead to a reduction of $\dpp$.  A more general approach, considering, \eg, faulty initialization through a density matrix approach, will be briefly discussed in the context of quantum limits to sensitivity (see Sec.~\ref{Sec05:QCRB}).
In addition, the observed probability is sometimes reduced by the control sequence of the sensing protocol, for example if there is no one-to-one mapping between the initialization, sensing and readout basis (``Step 2'' and ``Step 4'' in the protocol).
Since $\beta$ is a constant of time, we will assume $\beta=1$ in the following for reasons of simplicity.

\subsubsection{Classical readout noise}
\begin{figure}[t!]
\centering
\includegraphics[width=\figurewidth]{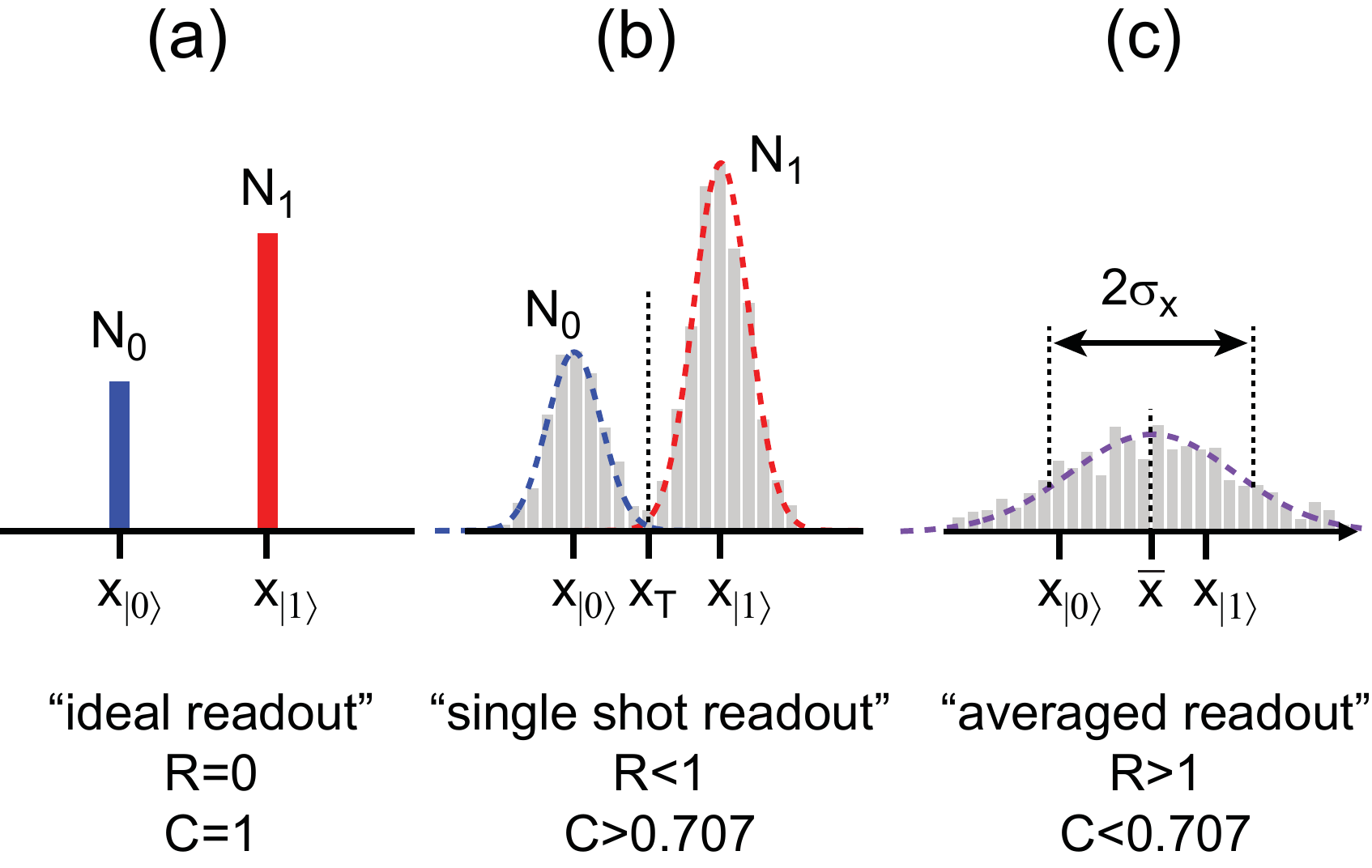}
\caption{Illustration of the sensor readout. $N$ measurements are performed producing $\{x_j\}_{j=1...N}$ readings on the physical measurement apparatus.  The readings $\{x_j\}$ are then binned into a histogram.
(a) Ideal readout.  Only two values are observed in the histogram, $\xma$ and $\xmb$, which correspond to the qubit states $\ma$ and $\mb$. All $\{x_j\}$ can be assigned to ``0'' or ``1'' with 100\% fidelity.
(b) Single shot readout. Most $\{x_j\}$ can be assigned, but there is an overlap between histogram peaks leading to a small error.
(c) Averaged readout. $\{x_j\}$ cannot be assigned. The ratio between ``0'' and ``1'' is given by the relative position of the mean value $\bar{x}$ and the error is determined by the histogram standard deviation $\sigma_x$.  $R$ is the ratio of readout and projection noise, and $C$ is an overall readout efficiency parameter that is explained in the text.
}
\label{fig:readout}
\end{figure}

A final source of error is the classical noise added during the readout of the sensor.  Two situations can be distinguished, depending on whether the readout noise is small or large compared to the projection noise.  We will denote them as the ``single shot'' and ``averaged'' readout regimes, respectively.  Due to the widespread inefficiency of quantum state readout, classical readout noise is often the dominating source of error.

\paragraph{Single shot readout}

In the ``single shot'' regime, classical noise added during the readout process is small.  The physical reading $x$ produced by the measurement apparatus will be very close to one of the two values $\xma$ and $\xmb$, which would have been obtained in the ideal case for the qubit states $\ma$ and $\mb$, respectively.  By binning the physical readings $x_j$ of $j=1...N$ measurements into a histogram, two peaks are observed centered at $\xma$ and $\xmb$, respectively (see Fig. \ref{fig:readout}(b)).  However, compared to the ideal situation (Fig. \ref{fig:readout}(a)), the histogram peaks are broadened and there is a finite overlap of between the tails of the peaks.  To obtain an estimate for $p$, all $x_j$ are assigned to either ``0'' or ``1'' based on a threshold value $x_T$ chosen roughly midway between $\xma$ and $\xmb$,
\begin{eqnarray}
N_0 & = & \text{number of measurements $x_j<x_T$} \\
N_1 & = & \text{number of measurements $x_j>x_T$} \ ,
\end{eqnarray}
where $p = N_1/N$.  Note that the choice of the threshold is not trivial; in particular, for an unbiased measurement, $x_T$ depends itself on the probability $p$.

Because of the overlap between histogram peaks, some values $x_j$ will be assigned to the wrong state.  The error introduced due to wrong assignments is
\begin{equation}
\sigpr^2 = \frac{1}{N}\left[ \kappa_0(1-\kappa_0)p + \kappa_1(1-\kappa_1)(1-p) \right] \ ,
\end{equation}
where $\kappa_{0}$ and $\kappa_{1}$ are the fraction of measurements that are erroneously assigned.
The actual values for $\kappa_{0,1}$ depend on the exact type of measurement noise and are determined by the cumulative distribution function of the two histogram peaks.
Frequently, the peaks have an approximately Gaussian distribution, such that
\begin{equation}
\kappa_0 \approx \frac12 \left[ 1 + \mr{erf}\left(\frac{|\xma - x_T|}{\sigma_x}\right) \right] \ ,
\end{equation}
and likewise for $\kappa_1$, where $\mr{erf}(x)$ is the Gauss error function.
Moreover, if $\kappa \equiv \kappa_{0}\approx\kappa_{1}\ll 1$ are small and of similar magnitude,
\begin{equation}
\sigpr^2  \approx  \frac{\kappa}{N} \ .
\end{equation}

\paragraph{Averaged readout}


When the classical noise added during the quantum state readout is large, only one peak appears in the histogram and the $x_j$ can no longer be assigned to $\xma$ or $\xmb$.  The estimate for $p$ is then simply given by the mean value of $x$,
\begin{eqnarray}
p & = & \frac{\bar{x}-\xma}{\xmb-\xma} = \frac{1}{N}\sum_{j=1}^{N} \frac{x_j-\xma}{\xmb-\xma} \ , 
\end{eqnarray}
where $\bar{x} = \frac1N\sum x_j$ is the mean of $\{ x_j \}$.  The standard error of $p$ is
\begin{eqnarray}
\sigpr^2 & = & \frac{\sigx^2}{(\xmb-\xma)^2} = \frac{R^2}{4N} \ ,
\end{eqnarray}
where $|\xmb-\xma|$ is the \textit{measurement contrast} and
\begin{equation}
R  \equiv \frac{\sigpr}{\sigpq} = \frac{2\sqrt N\sigx}{|\xmb-\xma|}
\end{equation}
is the ratio between classical readout noise and quantum projection noise.

As an example, we consider the optical readout of an atomic vapor magnetometer \cite{Budker07} or of NV centers in diamond \cite{Taylor08}.  For this example, $\xma$ and $\xmb$ denote the average numbers of photons collected per readout for each state.  The standard error is (under suitable experimental conditions) dominated by shot noise, $\sigx \approx \sqrt{\bar{x}}$.  The readout noise parameter becomes
\begin{equation}
R \approx \frac{2\sqrt{\bar{x}}}{|\xmb-\xma|}
	= \frac{2\sqrt{1-\epsilon/2}}{\epsilon\sqrt\xmb}
	\approx \frac{2}{\epsilon\sqrt\xmb} \ ,
\end{equation}
where $\epsilon = |1 - \xma/\xmb|$ is a relative optical contrast between the states, $0<\epsilon<1$, and the last equation represents the approximation for $\epsilon\ll 1$.

\paragraph{Total readout uncertainty}

The classical readout noise $\sigpr$ is often combined with the quantum projection noise $\sigpq$ to obtain a total readout uncertainty,
\begin{align}
\sigp^2
  &= \sigpq^2 +\sigpr^2 \nonumber \\
	&\approx (1+R^2)\sigpq^2
   \approx \frac{\sigpq^2}{C^2} = \frac{1}{4C^2 N} \ ,
\label{eq:sigp}
\end{align}
where $C=1/\sqrt{1+R^2} \approx 1/\sqrt{1+4\kappa}$ is an overall readout efficiency parameter \cite{Taylor08}. $C\leq 1$ describes the reduction of the signal-to-noise ratio compared to an ideal readout ($C=1$), see Fig. \ref{fig:readout}.  We will in the following use Eq. (\ref{eq:sigp}) to derive the SNR and minimum detectable signal.

\subsection{Sensitivity}
\label{sec:SNRvmin}
\subsubsection{Signal-to-noise ratio}

The signal-to-noise ratio (SNR) for a quantum sensing experiment can be defined as 
\begin{equation}
\SNR = \frac{\dppobs}{\sigp} = {\dpp(t)\, e^{-\chi(t)}}{2C\sqrt{N}} \ ,
\label{eq:snr}
\end{equation}
where $\dppobs$ is given by Eq. (\ref{eq:dpobs}) and $\sigp$ is given by Eq. (\ref{eq:sigp}).
To further specify the SNR, the change in probability $\dpp$ can be related to the change in signal $\delta V$ as
$\dpp = \delta V^q |\partial_V^q p(t)| \propto (\gamma t\delta V)^q$,
with $q=1$ for slope detection and $q=2$ for variance detection (see Fig. \ref{fig:ramseyrabi}).
In addition, the number of measurements $N$ is equal to $T/(t+\tm)$, where $T$ is the total available measurement time and $\tm$ is the extra time needed to initialize, manipulate and readout the sensor.
The updated SNR becomes
\begin{equation}
\SNR = {\dV^q |\partial_V^q p(t)|\, e^{-\chi(t)} 2C(\tm)} \frac{\sqrt{T}}{\sqrt{t+\tm}} \ ,
\label{eq:snr2}
\end{equation}
where $C(\tm)$ is a function of $\tm$ because the readout efficiency often improves for longer readout times.

\subsubsection{Minimum detectable signal and sensitivity}
\label{sec:vmin}
The sensitivity is  defined as the {minimum detectable signal} $\vmin$ that yields unit SNR for an integration time of one second ($T = 1\unit{s}$),
\begin{equation}
\vmin^q
  = \frac{e^{\chi(t)}\sqrt{t+\tm}}{2C(\tm) |\partial_V^q p(t)|}
  \propto \frac{e^{\chi(t)}\sqrt{t+\tm}}{2C(\tm) \gamma^q t^q} \ .
\label{eq:dpminB}
\end{equation}
Eq. (\ref{eq:dpminB}) provides clear guidelines for maximizing the sensitivity.
First, the sensing time $t$ should be made as long as possible.
However, because the decay function $\chi(t)$ exponentially penalizes the sensitivity for $t>\Tx$, the optimum sensing time is reached when $t\approx \Tx$.
Second, the sensitivity can be optimized with respect to $\tm$.  In particular, if $C(\tm)$ does improve as $C\propto \sqrt{\tm}$ -- which is a typical situation when operating in the averaged readout regime -- the optimum choice is $\tm \approx t$.  Conversely, if $C$ is independent of $\tm$ -- for example, because the sensor is operated in the single-shot regime or because readout resets the sensor -- $\tm$ should be made as short as possible.
Finally, $C$ can often be increased by optimizing the experimental implementation or using advanced quantum schemes, such as quantum logic readout.

We now evaluate Eq. (\ref{eq:dpminB}) for the most common experimental situations:

\paragraph{Slope detection}

For slope detection, $p_0=0.5$ and $\dpp(t) \approx \tfrac12 \gamma V t$  (Eq. \ref{eq:pslope}).  The sensitivity is
\begin{equation}
\vmin = \frac{e^{\chi(t)}\sqrt{t+\tm}}{\gamma C(\tm) t} \ .
\label{eq:vminslope}
\end{equation}
Note that the units of sensitivity are then typically given by the units of the signal $V$ to be measured times Hz$^{-1/2}$.
Assuming $t_m\ll t$, we can find an exact optimum solution with respect to $t$.  Specifically, for a Ramsey measurement with an exponential dephasing $e^{-\chi(t)}=e^{-t/\Ttwoast}$, the optimum evolution time is $t=\Ttwoast/2$ and
\begin{equation}
\vmin = \frac{\sqrt{2e}}{\gamma C\sqrt{\Ttwoast}} \quad\quad\text{for $t=\frac12\Ttwoast$}\ .
\label{eq:voptslope}
\end{equation}
This corresponds to Eq. (\ref{eq:equation_1_slope}) highlighted in the introduction to this section.


\paragraph{Variance detection}

For variance detection,  $\dpp \approx \frac14 \gamma^2 \Vrms^2 t^2$ (Eq. \ref{eq:pvariancesimple}).  The sensitivity is
\begin{equation}
\vmin = \left[ \frac{2e^{\chi(t)}\sqrt{t+\tm}}{C(\tm) \gamma^2 t^2} \right]^{1/2} \ ,
\end{equation}
In the limit of $t_m\approx0$ and $t\approx \Tx$, this expression simplifies to 
\begin{equation}
\vmin = \frac{\sqrt{2e}}{\gamma \sqrt C \sqrt[4]{\Tx^3}} \ ,
\label{eq:vminvar}
\end{equation}
which corresponds to Eq. (\ref{eq:equation_1_var}) highlighted in the introduction of the section.
Thus, variance detection profits more from a long coherence time $\Tx$ than slope detection (but is, in turn, more vulnerable to decoherence).  Alternatively, for the detection of a noise spectral density $S_V(\omega)$, the transition probability is  $\dpp \approx \frac12\gamma^2 S_V(\omega) \Tx$ (see Eqs. (\ref{eq:chiphi}) and (\ref{eq:phirms:noise})) and
\begin{equation}
S_{\vmin}(\omega) \approx \frac{e}{\gamma^2 C\sqrt{\Tx}} \ .
\end{equation}
%

\subsubsection{Signal integration}

The above expressions for sensitivity all refer to the minimum detectable signal per unit time.  By integrating the signal over longer measurement times $T$, the sensor performance can be improved.  According to Eq. (\ref{eq:snr2}), the minimum detectable signal for an arbitrary time $T$ is $\Vmin^q(T) = \vmin^q/\sqrt{T}$.
Therefore, the minimum detectable signal for slope and variance detection, Eqs. (\ref{eq:vminslope}) and (\ref{eq:vminvar}), respectively, scale as
\begin{align}
\Vmin(T) &= \vmin T^{-1/2} \quad\quad\text{for slope detection,} \label{eq:vminscaling1} \\
\Vmin(T) &= \vmin T^{-1/4} \quad\quad\text{for variance detection.} \label{eq:vminscaling2}
\end{align}
The corresponding scaling for the spectral density is $S_{\Vmin} = S_{\vmin} T^{-1/2}$.
We notice that variance detection improves only $\propto T^{1/4}$ with the integration time, while slope detection improves $\propto T^{1/2}$.  Hence, for weak signals with long averaging times $T\gg\Tx$, variance detection is typically much less sensitive than slope detection. 
As we will discuss in Section~\ref{sec08}, adaptive sensing methods can improve on these limits.

\subsection{Allan variance}

Sensors are typically also characterized by their stability over time. Indeed, while the sensitivity calculation suggests that one will always improve the minimum detectable signal by simply extending the measurement time, slow variations affecting the sensor might make this not possible.
These effects can be quantified by the Allan variance \cite{Allan66} or its square root, the Allan deviation.  While the concept is based on a classical analysis of the sensor output, it is still important for characterizing the performance of quantum sensors.  In particular, the Allan variance is typically reported to evaluate the performance of quantum clocks \cite{Hollberg01,Leroux10l}.

Assuming that the sensor is sampled over time at constant intervals $t_s$ yielding the signal $x_j=x(t_j)=x(j t_s)$, the Allan variance is defined as
\begin{equation}
	\sigma_X^2(\tau)={\frac{1}{2(N-1) t_s^2}\sum_{j=1}^{N-1}(x_{j+1}-x_j)^2} \ ,
\label{eq:Allan1}
\end{equation}
where $N$ is the number of samples $x_j$. 
One is typically interested in knowing how $\sigma_X^2$ varies with time, given the recorded sensor outputs. To calculate $\sigma_X^2(t)$ one can group the data in variable-sized bins and calculate the Allan variance for each grouping. The Allan variance for each grouping time $t=m t_s$ can then be calculated as
\begin{equation}
	\sigma_X^2(m t_s)={\frac{1}{2(N-2m)m^2 t_s^2}\sum_{j=1}^{N-2m}(x_{j+m}-x_j)^2}.
\end{equation}
The Allan variance can also be used to reveal the performance of a sensor beyond the standard quantum limit \cite{Leroux10l}, and its extension to and limits in quantum metrology have been recently explored \cite{Chabuda16}.

\subsection{Quantum Cram\'{e}r Rao Bound for parameter estimation}
\label{Sec05:QCRB}

The sensitivity of a quantum sensing experiment can be more rigorously considered in the context of the Cram\'{e}r-Rao bound applied to parameter estimation.  Quantum parameter estimation aims at measuring the value of a continuous parameter $V$ that is encoded in the state of a quantum system $\rho_V$, via, e.g., its interaction with the external signal we want to characterize.  The estimation process consists of two steps: in a first step, the state $\rho_V$ is measured;  in a second step, the estimate of $V$ is determined by data-processing the measurement outcomes. 

In the most general case, the measurement can be described by a positive-operator-valued measure (POVM) $\mathcal M=\{E_x^{N}\}$ over the $N$ copies of the quantum system.  The measurement yields the outcome $x$ with conditional probability $p_N(x|V) = \mbox{Tr}[ E_x^{(N)} \rho_V^{\otimes N}]$.

With some further data processing, we arrive at the estimate $v$ of the parameter $V$. The estimation uncertainty can be described by the probability $P_N(v|V) := \sum_x p_\mr{est}^{(N)}(v|x) p_N(x|V)$, where $p_\mr{est}^{(N)}(v|x)$ is the probability of estimating $v$ from the measurement outcome $x$. We can then define the estimation uncertainty as  $\Delta V_N:= \sqrt{\sum_{v} [{v}-V]^2  P_N(v|V)}$.
Assuming that the estimation procedure is asymptotically locally unbiased, $\Delta V_N$ obeys the so-called Cram\'{e}r-Rao bound
%
\begin{equation}
\Delta V_N\geq {1}/{\gamma \sqrt{F_N(V)}} \ ,
\label{eq:CCRB}
\end{equation}
where 
\begin{equation}\begin{array}{l}
	F_N(V) := \displaystyle\sum_x \frac1{p_N(x|V)}\left(\frac{\partial p_N(x|V)}{\partial
    V}\right)^2 \\=\displaystyle \sum_x \frac1{\mbox{Tr}[ E_x^{(N)} \rho_V^{\otimes n}]}\left(\frac{\partial\, \mbox{Tr}[ E_x^{(N)} \rho_V^{\otimes n}]}{\partial
    V}\right)^2
\end{array}
\label{eq:Fisher}
\end{equation}
is the Fisher information associated with the given POVM measurement \cite{Braunstein94}.

By optimizing Eq.~(\ref{eq:CCRB}) with respect to all possible POVM's, one obtains the \textit{quantum} Cram\'{e}r-Rao bound (QCRB) \cite{Helstrom67,Braunstein94,Braunstein96b,Holevo82,Paris09, Goldstein10x}
\begin{eqnarray}
   \Delta V_N \geqslant  \frac{1}{\gamma\sqrt{ \max_{{\cal M}^{(N)}}[
       F_N(V)]}} 
   \geqslant \frac{1}{\gamma\sqrt{N \mathcal F(\rho_V)}}, \label{QCRN}
\end{eqnarray}
where the upper bound of $\max_{{\cal M}^{(N)}}[ F_N(V)]$ is expressed in terms of the {quantum Fisher information} $\mathcal F(\rho_V)$, defined as 
\begin{equation}
\mathcal F(\rho_V):={\rm Tr}[{\cal
    R}^{-1}_{\rho_V}(\partial_V\rho_V)\rho_V{\cal R}^{-1}_{\rho_V}(\partial_V\rho_V)],
\end{equation}
with
\begin{equation}
{\cal
    R}^{-1}_{\rho}(A):=\sum_{j,k:\lambda_j+\lambda_k\neq
    0}\frac{2A_{jk}|j\rangle\langle k|}{\lambda_j+\lambda_k} 
\end{equation}
being the symmetric logarithmic derivative written in the basis that diagonalizes the state, $\rho_V=\sum_j\lambda_j|j\rangle\langle j|$. 

A simple case results when $\rho_V$ is a pure state, obtained from the evolution of the reference initial state $|0\rangle$ under the signal Hamiltonian, $|\psi_V\rangle=e^{-i\hat H_V t}|0\rangle$.
Then, the QCRB is a simple uncertainty relation \cite{Helstrom67,Braunstein94,Braunstein96b,Holevo82},
\begin{eqnarray}
  \Delta V_N \geqslant \frac{1}{2\gamma\sqrt{N} \; \Delta H }  \label{QCRN2} \;,
\end{eqnarray}
where $\Delta H := \sqrt{ \langle H^2\rangle-\langle H\rangle^2  }$.
We note that the  scaling of the QCRB with the number of copies, $N^{-1/2}$, is a consequence of the additivity of the quantum Fisher information for tensor states $\rho^{\otimes N}$. This is the well-known standard quantum limit (SQL).  To go beyond the SQL, one then needs to use entangled states (see Section \ref{sec09}) -- in particular, simply using correlated POVMs is not sufficient. Thus, to reach the QCRB, local measurements and at most adaptive estimators are sufficient, without the need for entanglement.
 
While the quantum Fisher information (and the QCRB) provide the ultimate lower bound to the achievable uncertainty for optimized quantum measurments, the simpler Fisher information can be used to evaluate a given measurement protocol, as achievable, \eg, within experimental constraints. 

Consider for example the sensing protocols described in Section~\ref{sec04}.  For the Ramsey protocol, the quantum sensor state after the interaction with the signal $V$ is given by
\begin{equation}
\rho_{11}(V,t)=\frac12  \qquad
\rho_{12}(V,t)=-\frac i2e^{-i  \gamma V t}e^{-\chi(t)} \ .
\label{rho}
\end{equation}
Here, $e^{-\chi(t)}$ describes decoherence and relaxation as discussed with Eq. (\ref{eq:dpobs}).  If we assume to perform a projective measurement in the $\sigma_x$ basis,
$\{ |\pm\rangle \} = \{ \frac{1}{\sqrt{2}} (\ma\pm\mb)$, giving the outcome probabilities $p(x_\pm| V)=\langle\pm|\rho(V)|\pm\rangle $, the Fisher information is  
\begin{equation}
F=\sum_x \frac{1}{p(x| V)}\left[\partial_{ V} p(x| V)\right]^2=\frac{t^2 \cos^2(\gamma Vt)e^{-2\chi}}{1-e^{-2\chi}\sin^2(\gamma Vt)}.
\end{equation}
The Fisher information thus oscillates between its minimum, where $\gamma V t = (k+1/2)\pi$ and $F=0$, and its optimum, where $\gamma V t = k\pi$ and $F = t^2e^{-2\chi}$.  The uncertainty in the estimate $\delta V=1/\gamma \sqrt{N F}$ therefore depends on the sensing protocol bias point.  In the optimum case $F$ corresponds to the quantum Fisher information and we find the QCRB
\begin{equation}
\Delta V_N=\frac{1}{\gamma \sqrt{N \mathcal F}}=\frac{e^\chi}{\gamma t\sqrt{N}} \ .
\label{eq:QCRB}	
\end{equation}
Depending on the functional form of $\chi(t)$, we can further find the optimal sensing time for a given total measurement time.
Note that if we remember that $N$ experiments will take a time $T=N(t+t_m)$, and we add inefficiency due to the sensor readout, we can recover the sensitivity $\vmin$ of Eq.~(\ref{eq:vminslope}). 

  Similarly, we can analyze more general protocols, such as variance detection of random fields, simultaneous estimation of multiple parameters \cite{Baumgratz16} or optimized protocols for signals growing over time \cite{Pang16}.


\section{Sensing of AC signals}
\label{sec06} 

So far we have implicitly assumed that signals are static and deterministic.  For many applications it is important to extend sensing to time-dependent signals.  For example, it may be required to detect the amplitude, frequency or phase of an oscillating signal.  More broadly, one may be interested in knowing the waveform of a time-varying parameter or reconstructing a frequency spectrum.  A diverse set of quantum sensing methods has been developed for this purpose that are summarized in the following two sections.

Before discussing the various sensing protocols in more detail, it is important to consider the type of information that one intends to extract from a time-dependent signal $V(t)$.  In this Section \ref{sec06}, we will assume that the signal is composed of one or a few harmonic tones and our goal will be to determine the signal's amplitude, frequency, phase or overall waveform.  In the following Section \ref{sec07}, we will discuss the measurement of stochastic signals with the intent of reconstructing the noise spectrum or measuring the noise power in a certain bandwidth. 

\subsection{Time-dependent signals}
As measuring arbitrary time-dependent signals is a complex task, we first focus on developing a basic set of AC sensing protocols, assuming a single-tone AC signal given by
\begin{equation}
V(t') = \Vpk \cos(2\pi\fac t' + \alpha) \ .
\label{eq:singletone}
\end{equation}
This signal has three basic parameters, including the signal amplitude $\Vpk$, the frequency $\fac$ and the relative phase $\alpha$.  Our aim will be to measure one or several of these parameters using suitable sensing protocols.

Signal detection can be extended to multi-tone signals by summing over individual single-tone signals,
\begin{equation}
V(t')=\sum_m V_{\mr{pk},m} \cos(2\pi f_{\mr{ac},m} t' + \alpha_m) \ ,
\label{eq:multitone}
\end{equation}
where $V_{\mr{pk},m}$, $f_{\mr{ac},m}$ and $\alpha_m$ are the individual amplitudes, frequencies and phases of the tones, respectively.

\subsection{Ramsey and Echo sequences}

\begin{figure}[t]
\centering
\includegraphics[width=\figurewidth]{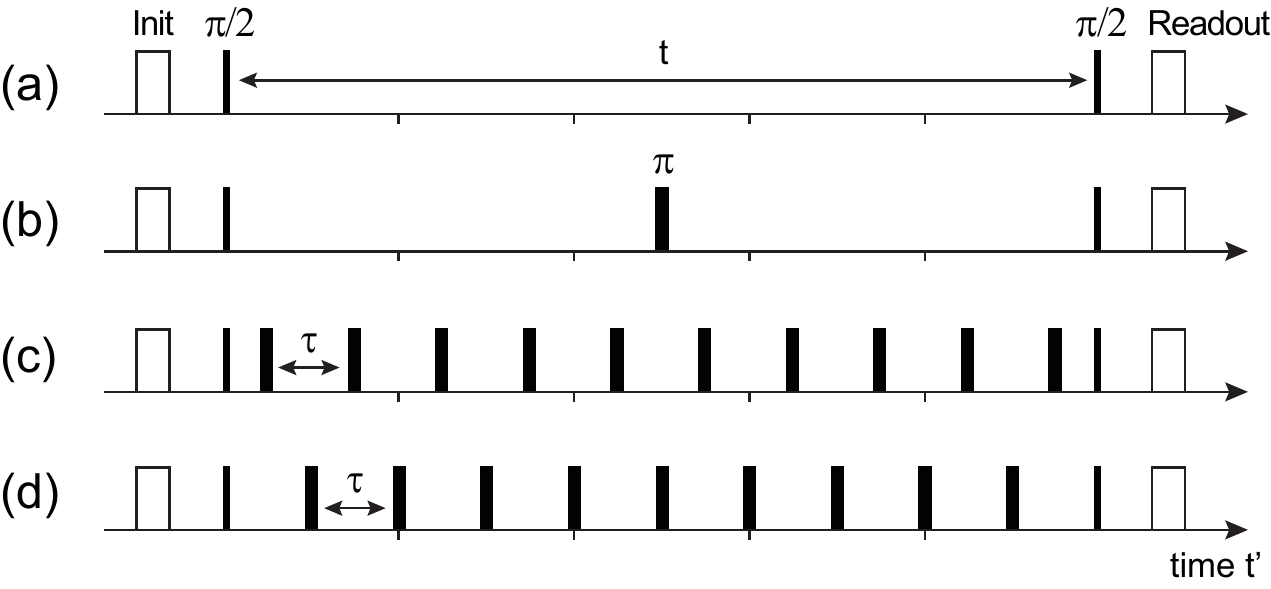}
\caption{Pulse diagrams for DC and AC sensing sequences.
Narrow blocks represent $\pi/2$ pulses and wide blocks represent $\pi$ pulses, respectively.
$t$ is the total sensing time and $\tau$ is the interpulse delay.
(a) Ramsey sequence.
(b) Spin-echo sequence.
(c) CP multipulse sequence.
(d) PDD multipulse sequence.
}
\label{fig:acsequences}
\end{figure}

To illustrate the difference between DC and AC sensing, we re-examine the Ramsey measurement from Section \ref{sec04:ramsey}.  The corresponding pulse diagram is given in Fig. \ref{fig:acsequences}(a).
This protocol is ideally suited to measure static shifts of the transition energy.  But is it also capable of detecting dynamical variations?
In order to answer this question, one can inspect the phase $\phi$ accumulated during the sensing time $t$ due to either a static or a time-dependent signal $V(t)$,
\begin{equation}
\phi = \int_0^{t} \gamma V(t') \, dt' \ .
\end{equation}
For a static perturbation, the accumulated phase is simply $\phi = \gamma V t$.  For a rapidly oscillating perturbation, by contrast, phase accumulation is averaged over the sensing time, and $\phi = \gamma \langle V(t') \rangle t \approx 0$.  To answer our question, the Ramsey sequence will only be sensitive to slowly varying signals up to some cut-off frequency $\approx t^{-1}$.

Sensitivity to alternating signals can be restored by using time-reversal (``spin echo'') techniques \cite{Hahn50}.  To illustrate this, we assume that the AC signal goes through exactly one period of oscillation during the sensing time $t$.  The Ramsey phase $\phi$ due to this signal is zero because the positive phase build-up during the first half of $t$ is exactly canceled by the negative phase build-up during the second half of $t$.  However, if the qubit is inverted at time $t/2$ using a $\pi$ pulse (see Fig. \ref{fig:acsequences}(b)), the time evolution of the second period is reversed, and the accumulated phase is non-zero,
\begin{align}
\phi
  &= \int_0^{t/2} \gamma V(t') \, dt' - \int_{t/2}^{t} \gamma V(t') \, dt' 
	 = \frac{2}{\pi}\gamma\Vpk t\cos\alpha \ .
\end{align}
%

\subsection{Multipulse sensing sequences}
\label{sec06:multipulse}

The spin echo technique can be extended to sequences with many $\pi$ pulses.  These sequences are commonly referred to as \textit{multipulse sensing sequences} or multipulse control sequences, and allow for a detailed shaping of the frequency response of the quantum sensor.  To understand the AC characteristics of a multipulse sensing sequence, we consider the phase accumulated for a general sequence of $n$ $\pi$ pulses applied at times $0<t_j<t$, with $j=1...n$.  The accumulated phase is given by
\begin{equation}
\phi = \int_0^{t} \gamma V(t') y(t') \, dt'\ ,
\end{equation}
where $y(t') = \pm 1$ is the \textit{modulation function} of the sequence that changes sign whenever a $\pi$ pulse is applied.  For a harmonic signal $V(t') = \Vpk \cos(2\pi\fac t' + \alpha)$ the phase is
\begin{align}
  \phi
	&= \frac{\gamma \Vpk}{2\pi\fac}\left[\sin(\alpha)-(-1)^n\sin(2\pi\fac t+\alpha)\right. \nonumber \\
	&  \qquad\qquad +2\sum_{j=1}^n(-1)^j \left. \sin(2\pi\fac t_j+\alpha)\right] \nonumber \\
  &= \gamma \Vpk t \times W(\fac,\alpha) \ .
\label{eq:ACsensing}	
\end{align}
This defines for any multipulse sequence a \textit{weighting function} $W(\fac,\alpha)$.  For composite signals consisting of several harmonics with different frequencies and amplitudes, Eq. (\ref{eq:multitone}), the accumulated phase simply represents the sum of individual tone amplitudes multiplied by the respective weighting functions.

\subsubsection{CP and PDD sequences}

The simplest pulse sequences used for sensing have been initially devised in nuclear magnetic resonance (NMR) \cite{Slichter96} and have been further developed in the context of dynamical decoupling (DD) \cite{Viola98}. They are composed of $n$ equally spaced $\pi$-pulses with an interpulse duration $\tau$.  The most common types are Carr-Purcell (CP) pulse trains \cite{Carr54} (Fig. \ref{fig:acsequences}(c)) and periodic dynamical decoupling (PDD) sequences \cite{Khodjasteh05} (Fig. \ref{fig:acsequences}(d)).

For a basic CP sequence, $t_j = \frac{2j-1}{2}\tau$, and the weighting function is \cite{Taylor08,Hirose12}
\begin{equation}
\Wcp(\fac,\alpha) = 
\frac{ \sin (\pi  \fac n \tau ) }{\pi  \fac n \tau }[1-\sec (\pi  \fac \tau )]\cos (\alpha +\pi  \fac n \tau ) \ .
\end{equation}
Similarly, for a PDD sequence, $t_j=j\tau$ and 
\begin{equation}
\Wpdd(\fac,\alpha) =
\frac{ \sin (\pi  \fac n \tau ) }{\pi  \fac n \tau }\tan (\pi  \fac \tau )\sin (\alpha +\pi  \fac n \tau ) \ .
\end{equation}
%
Because of the first (sinc) term, these weighting functions resemble narrow-band filters around the center frequencies $\fac=f_k = k/(2\tau)$, where $k=1,3,5,...$ is the harmonic order.  In fact, they can be rigorously treated as \textit{filter functions} that filter the frequency spectrum of the signal $V(t)$ (see Section \ref{sec07}).  For large pulse numbers $n$, the sinc term becomes very peaked and the filter bandwidth $\Delta f \approx 1/(n\tau) = 1/t$ (full width at half maximum) becomes very narrow.
The narrow-band filter characteristics can be summed up as follows (see Fig. \ref{fig:acsequences}(c)),
\begin{alignat}{3}
&f_k           = k/(2\tau)                       && \quad \text{center frequencies} \label{eq:filter:frequency} \\
&\Delta f      \approx 1/t = 1/(n\tau)          && \quad \text{bandwidth} \label{eq:filter:bandwidth} \\
&\begin{array}{l}\hspace{-4pt}\Wcp(\alpha)  = \displaystyle \frac{2}{\pi  k }(-1)^{\frac{k-1}2}\!\! \cos(\alpha)\\\hspace{-4pt}\Wpdd(\alpha) = \displaystyle-\frac{2}{\pi  k } \sin(\alpha)\end{array}  					 &&  \quad \text{peak transmission} \label{eq:filter:peak}
\end{alignat}
The advantage of the CP an PDD sequences is that their filter parameters can be easily tuned.  In particular, the pass-band frequency can be selected via the interpulse delay $\tau$, while the filter width can be adjusted via the number of pulses $n=t/\tau$ (up to a maximum value of $n\approx T_2/\tau$).  The resonance order $k$ can also be used to select the pass-band frequency, however, because $k=1$ provides the strongest peak transmission, most reported experiments used this resonance.  The time response of the transition probability is
\begin{align}
p &= \frac{1}{2}\left[ 1 - \cos\left(W(\fac,\alpha) \gamma \Vpk t \right) \right]  \nonumber \\
  & =\frac{1}{2}\left[ 1 - \cos\left( \frac{2 \gamma \Vpk t \cos\alpha}{k \pi } \right) \right],
\label{eq:multipulse:fixedphase}
\end{align}
where the last expression represents the resonant case ($\fac=k/2\tau$) for CP sequences.

\subsubsection{Lock-in detection}

The phase $\phi$ acquired during a CP or PDD sequence depends on the relative phase difference $\alpha$ between the AC signal and the modulation function $y(t)$.  For a signal that is in-phase with $y(t)$, the maximum phase accumulation occurs, while for an out-of-phase signal, $\phi = 0$.

This behavior can be exploited to add further capabilities to AC signal detection.  \onlinecite{Kotler11}, have shown that both quadratures of a signal can be recovered, allowing one to perform lock-in detection of the signal.  Furthermore, it is possible to correlate the phase acquired during two subsequent multipulse sequences to perform high-resolution spectroscopy of AC signals (see Section~\ref{sec06:frequencyestimation}).
\begin{figure}[t]
\centering
\includegraphics[width=0.9\columnwidth]{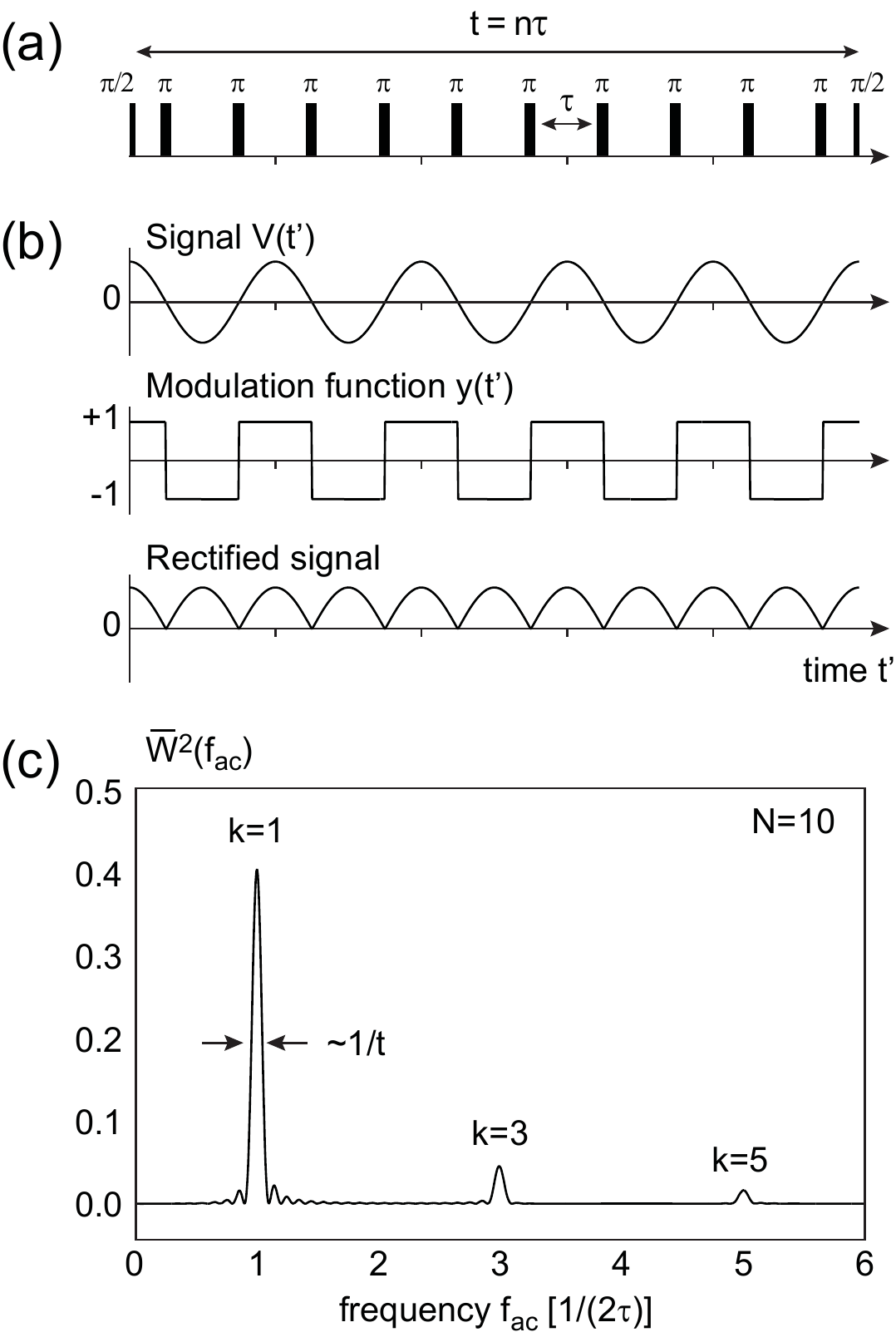}
\caption{Modulation and weight functions of a CP multipulse sequence.
(a) CP multipulse sequence.
(b) Signal $V(t')$, modulation function $y(t')$ and ``rectified'' signal $V(t')\times y(t')$.
The accumulated phase is represented by the area under the curve.
(c) Weight function $\overline W^2(\fac)$ associated with the modulation function.  $k$ is the harmonic order of the filter resonance.
}
\label{fig:acfilter}
\end{figure}

\subsubsection{Other types of multipulse sensing sequences}

Many varieties of multipulse sequences have been conceived with the aim of optimizing the basic CP design, including improved robustness against pulse errors, better decoupling performance, narrower spectral response and sideband suppression, and avoidance of signal harmonics.

A systematic analysis of many common sequences has been given by~\onlinecite{Cywinski08}.  One favorite has been the XY4, XY8 and XY16 series of sequences \cite{Gullion90} owing to their high degree of pulse error compensation.  A downside of XY type sequences are signal harmonics \cite{Loretz15} and the sidebands common to CP sequences with equidistant pulses.  Other recent efforts include sequences with non-equal pulse spacing \cite{Zhao14,Casanova15,Ajoy17} or sequences composed of alternating subsequences \cite{Albrecht15}.  A Floquet spectroscopy approach to multipulse sensing has also been proposed \cite{Lang15}.

\subsubsection{AC signals with random phase}
\label{sec06:acrandom}

Often, the multipulse sequence cannot be synchronized with the signal or the phase $\alpha$ cannot be controlled. Then, incoherent  averaging leads to phase cancellation, $\langle \phi \rangle=0$.   In this case, it is advantageous to measure the variance of the phase $\langle \phi^2 \rangle$ rather than its average $\langle \phi \rangle$.  (Although such a signal technically represents a stochastic signal, which will be considered in more detail in the next section, it is more conveniently described here.)

For a signal with fixed amplitude but random phase, the variance is
\begin{equation}
\langle \phi^2 \rangle = \gamma^2 \Vrms^2 t^2 \overline{W}^2(\fac),
\end{equation}
where $\Vrms = \Vpk/\sqrt{2}$ is the rms amplitude of the signal and $\overline{W}^2(\fac)$ is the average over $\alpha=0\dots2\pi$ of the weighting functions,
\begin{align}
\overline{W}^2(\fac)
	&= \frac{1}{2\pi} \int_0^{2\pi} W^2(\fac,\alpha) \, d\alpha
\end{align}
For the CP and PDD sequences, the averaged functions are given by
\begin{align}
\overline{W}_\mr{CP}^2(\fac)  &= \frac{\sin^2 (\pi  \fac n\tau)}{2(\pi \fac n\tau)^2} \left[1-\sec\left(\pi \fac \tau\right)\right]^2 , \nonumber \\
\overline{W}_\mr{PDD}^2(\fac) &= \frac{\sin^2 (\pi  \fac n\tau)}{2(\pi \fac n\tau)^2} \tan \left({\pi  \fac \tau}\right)^2 ,
\label{eq:filter}	
\end{align}
and the peak transmission at $\fac = k/2\tau$ is $\overline{W}^2 = 2/(k\pi)^2$.  The time response of the transition probability is \cite{Kotler13}
\begin{align}
p(t)
  &= \frac{1}{2}\left[ 1-J_0\left(\sqrt{2}\overline{W}(\fac)\gamma \Vrms t \right) \right]  \nonumber \\
  &= \frac{1}{2}\left[ 1-J_0\left(\frac{2\sqrt{2}\gamma \Vrms t}{k\pi} \right) \right]
\label{eq:multipulse:randomphase}
\end{align}
where $J_0$ is the Bessel function of the first kind and where the second equation reflects the resonant case.

\subsubsection{AC signals with random phase and random amplitude}

If the amplitude $\Vpk$ is not fixed, but slowly fluctuating between different measurements, the variance $\phivar$ must be integrated over the probability distribution of $\Vpk$.  A particularly important situation is a Gaussian amplitude fluctuation with an rms amplitude $V_\mr{rms}$.  In this case, the resonant time response of the transition probability is
\begin{equation}
p(t) = \frac{1}{2}
  \left[ 1 
  - \exp\left(-\frac{\overline{W}^2 \gamma^2 \Vrms^2 t^2}{2k^2} \right)
	   I_0\left( \frac{\overline{W}^2 \gamma^2 \Vrms^2 t^2}{2k^2} \right)
  \right]
\label{eq:multipulse:randomamplitude}
\end{equation}
where $I_0$ is the modified Bessel function of the first kind.
\begin{figure}[t]
\centering
\includegraphics[width=\figurewidth]{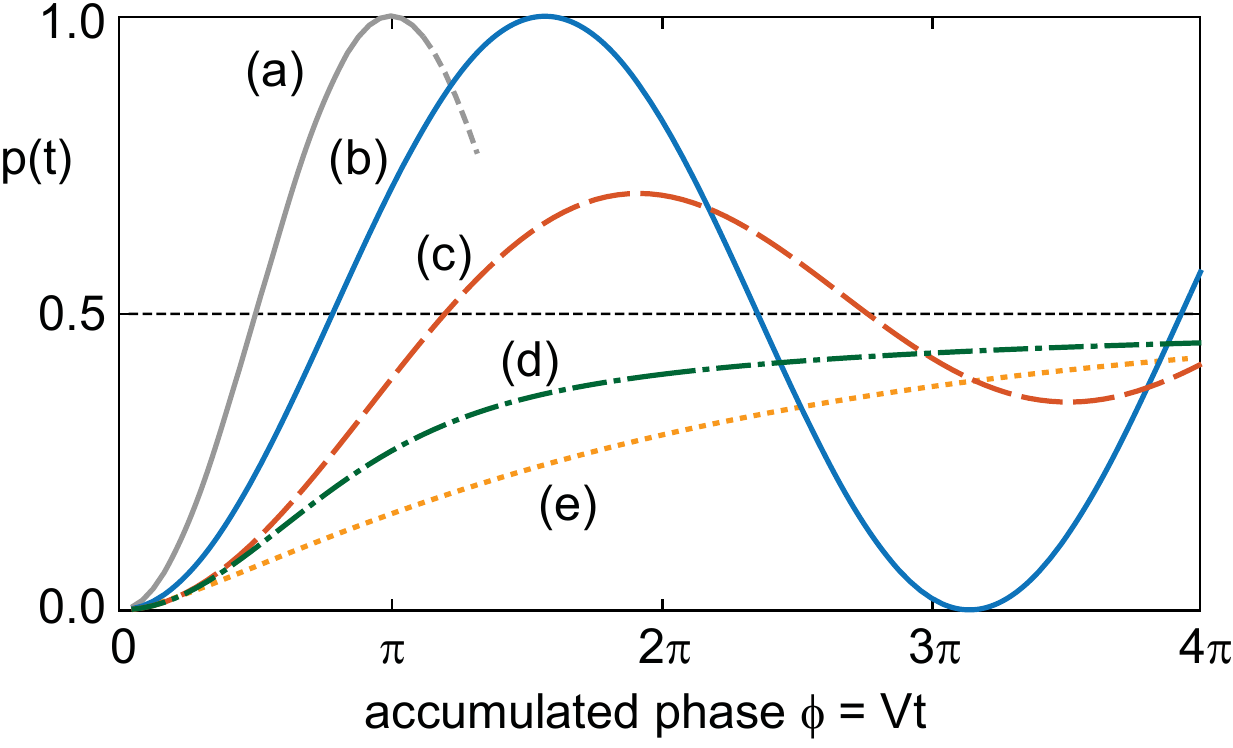}
\caption{Transition probability $p(t)$ as a function of phase accumulation time $t$.
(a) Ramsey oscillation (Eq. \ref{eq:pramsey}).
(b) AC signal with fixed amplitude and optimum phase (Eq. \ref{eq:multipulse:fixedphase}).
(c) AC signal with fixed amplitude and random phase (Eq. \ref{eq:multipulse:randomphase}).
(d) AC signal with random amplitude and random phase (Eq. \ref{eq:multipulse:randomamplitude}).
(e) Broadband noise with $\chi = \Gamma t$ (Eq. \ref{eq:decoherence}).
}
\label{fig:transitionprobabilitycurves}
\end{figure}

\subsection{Waveform reconstruction}
\label{sec06:Walsh}

The detection of AC fields can be extended to the more general task of sensing and reconstructing arbitrary time dependent fields.  A simple approach is to record the time response $p(t)$ under a specific sensing sequence, such as a Ramsey sequence, and to reconstruct the phase $\phi(t)$ and signal $V(t)$ from the time trace \cite{Balasubramanian09}.  This approach is, however, limited to the bandwidth of the sequence and by readout deadtimes.

To more systematically reconstruct the time dependence of an arbitrary signal, one may use a family of pulse sequences that forms a basis for the signal.
A suitable basis are Walsh dynamical decoupling sequences \cite{Hayes11}, which apply a $\pi$ pulse every time the corresponding Walsh function \cite{Walsh23} flips its sign.  Under a control sequence with $n$ $\pi$-pulses applied at the zero-crossings of the $n$-th Walsh function $y(t')=w_n(t'/t)$, the phase acquired after an acquisition period $t$ is
\begin{equation}
\phi(t)=\gamma\int_0^t V(t')y(t')dt'=\gamma V_n t \ ,
\end{equation}
which is proportional to the $n$-th Walsh coefficient $V_n$ of $V(t')$.  By measuring $N$ Walsh coefficients (by applying $N$ different sequences) one can reconstruct an $N$-point functional approximation to the field $V(t')$ from the $N$-th partial sum of the Walsh-Fourier series \cite{Cooper14,Magesan13},
\begin{equation}
	V_N(t')=\sum_{n=0}^{N-1} V_n w_n(t'/t) ,	
\end{equation}
which can be shown to satisfy $\lim_{N\to\infty}V_N(t')=V(t')$.  A similar result can be obtained using different basis functions, such as  Haar wavelets,  as long as they can be easily implemented experimentally \cite{Xu16}. 

An advantage of these methods is that they provide protection of the sensor against dephasing, while extracting the desired information.   In addition, they can be combined with compressive sensing techniques \cite{Magesan13c,Candes06,Puentes14} to reduce the number of acquisition needed to reconstruct the time-dependent signal.   The ultimate metrology limits in waveform reconstruction have also been studied \cite{Tsang11}.

\subsection{Frequency estimation}
\label{sec06:frequencyestimation}

\begin{figure}[t]
\centering
\includegraphics[width=\figurewidth]{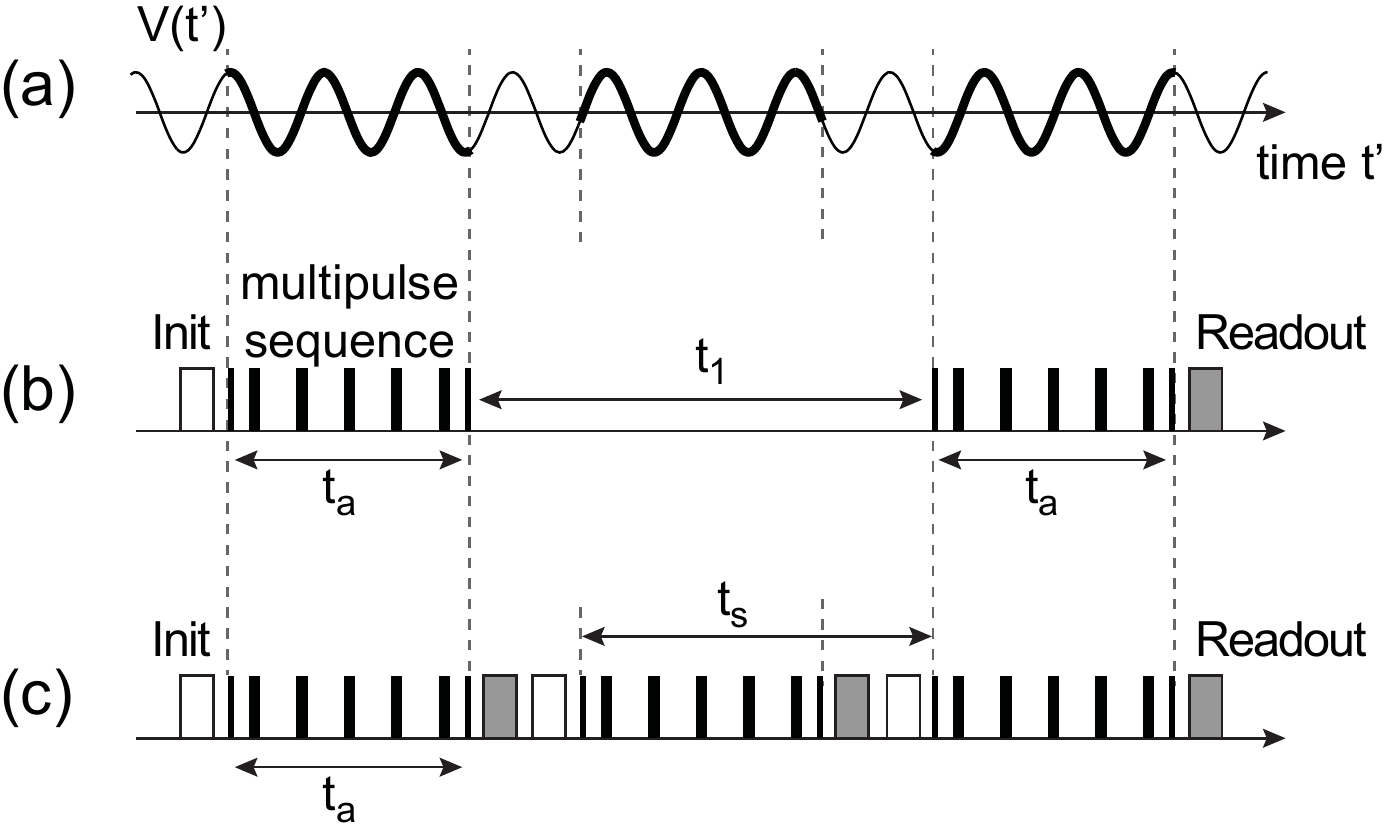}
\caption{Correlation spectroscopy.
(a) AC signal $V(t')$.
(b) Correlation sequence. Two multipulse sequences are interrupted by an incremented delay time $t_1$.  Because the multipulse sequences are phase sensitive, the total phase accumulated after the second multipulse sequence oscillates with $\fac t_1$.  The maximum $t_1$ is limited by the relaxation time $T_1$, rather than the typically short decoherence time $T_2$.
(c) Continouous sampling.  The signal is periodically probed in intervals of the sampling time $t_s$.  The frequency can be estimated from a sample record of arbitrary duration, permitting an arbitrarily fine frequency resolution.
}
\label{fig:correlation}
\end{figure}
An important capability in AC signal detection is the precise estimation of frequencies.  In quantum sensing, most frequency estimation schemes are based on dynamical decoupling sequences.  These are discussed in the following.  Fundamental limits of frequency estimation based on the quantum Fisher information have been considered by \onlinecite{Pang16}.

\subsubsection{Dynamical decoupling spectroscopy}

A simple approach for determining a signal's frequency is to measure the response to pulse sequences with different pulse spacings $\tau$.  This is equivalent to stepping the frequency of a lock-in amplifier across a signal.  The spectral resolution of dynamical decoupling spectroscopy is determined by the bandwidth of the weighting function $W(\fac,\tau)$, which is given by $\Delta f\approx 1/t$ (see Eq. (\ref{eq:filter:bandwidth})).  Because $t$ can only be made as long as the decoherence time $T_2$, the spectral resolution is limited to $\Delta f\approx 1/T_2$.  The precision of the frequency estimation, which also depends on the signal-to-noise ratio, is directly proportional to $\Delta f$.

\subsubsection{Correlation sequences}

Several schemes have been proposed and demonstrated to further narrow the bandwidth and to perform high resolution spectroscopy.  All of them rely on correlation-type measurements where the outcomes of subsequent sensing periods are correlated.

A first method is illustrated in Fig. \ref{fig:correlation}(a) in combination with multipulse detection.  The multipulse sequence is subdivided into two equal sensing periods of duration $t_a=t/2$ that are separated by an incremented free evolution period $t_1$.  Since the multipulse sequence is phase sensitive, constructive or destructive phase build-up occurs between the two sequences depending on whether the free evolution period $t_1$ is a half multiple or full multiple of the AC signal period $\Tac=1/\fac$.
The final transition probability therefore oscillates with $t_1$ as
\begin{align}
p(t_1)
  &= \frac{1}{2} \left\{ 1 - \sin[\Phi\cos(\alpha)] \sin[\Phi\cos(\alpha+2\pi\fac t_1)] \right\}  \nonumber \\
  &\approx \frac{1}{2} \left\{ 1 - \Phi^2\cos(\alpha) \cos(\alpha+2\pi\fac t_1) \right\}
	\label{eq:correlation}
\end{align}
where $\Phi = \gamma \Vpk t/(k \pi)$ is the maximum phase that can be accumulated during either of the two multipulse sequences.  The second equation is for small signals where $\sin\Phi \approx \Phi$.  For signals with random phase, Eq. (\ref{eq:correlation}) can be integrated over $\alpha$ and the transition probability simplifies to
\begin{equation}
p(t_1) \approx \frac{1}{2} \left\{ 1 - \frac{\Phi^2}{2}\cos(2\pi\fac t_1) \right\}
\end{equation}
Since the qubit is parked in $\ma$ and $\mb$ during the free evolution period, relaxation is no longer governed by $T_2$, but by the typically much longer $T_1$ relaxation time \cite{Laraoui13}.  In this way, a Fourier-limited spectral resolution of $\Delta f\sim 1/T_1$ is possible.  The resolution can be further enhanced by long-lived auxiliary memory qubits (see Section \ref{sec10}) and resolution improvements by two-to-three orders of magnitude over dynamical decoupling spectroscopy have been demonstrated \cite{Pfender16,Rosskopf16,Zaiser16}.  The correlation protocol was further shown to eliminate several other shortcomings of multipulse sequences, including signal ambiguities resulting from the multiple frequency windows and spectral selectivity \cite{Boss16}.

\subsubsection{Continuous sampling}

A second approach is the continuous sampling of a signal, illustrated in Fig. \ref{fig:correlation}(b).  The output signal can then be Fourier transformed to extract the undersampled frequency of the original signal.  Because continuous sampling no longer relies on quantum state lifetimes, the Fourier-limited resolution can be extended to arbitrary values and is only limited by total experiment duration $T$, and ultimately the control hardware.  The original signal frequency can then be reconstructed from the undersampled data record using compressive sampling techniques \cite{Nader11}.  Continuous sampling has recently led to the demonstration of $\mu$Hz spectral resolution \cite{Jelezko17,Boss17}.

%

\section{Noise spectroscopy}
\label{sec07} 

In this Section, we discuss methods for reconstructing the frequency spectrum of stochastic signals, a task commonly referred to as noise spectroscopy.  Noise spectroscopy is an important tool in quantum sensing, as it can provide much insight into both external signals and the intrinsic noise of the quantum sensor.  In particular, good knowledge of the noise spectrum can help the adoption of suitable sensing protocols (like dynamical decoupling or quantum error correction schemes) to maximize the sensitivity of the quantum sensor.

Noise spectroscopy relies on the systematic analysis of decoherence and relaxation under different control sequences.  This review will focus on two complementary frameworks for extracting noise spectra.  A first concept is that of ``filter functions'', where the phase pick-up due to stochastic signals is analyzed under different dynamical decoupling sequences.  The second concept, known as ``relaxometry'', has its origins in the field of magnetic resonance spectroscopy and is closely related to Fermi's golden rule.

\subsection{Noise processes}

For the following analysis we will assume that the stochastic signal $V(t)$ is Gaussian.  Such noise can be described by simple noise models, like a semi-classical Gaussian noise or the Gaussian spin-boson bath.  
In addition, we will assume that the autocorrelation function of $V(t)$,
\begin{equation}
G_V(t) = \langle V(t'+t) V(t') \rangle\ ,
\label{eq:autocorrelation}
\end{equation}
decays on a time scale $\tc$ that is  shorter than the sensing time $t$, such that successive averaging measurements are not correlated.  The noise can then be represented by a power spectral density \cite{Biercuk11},
\begin{equation}
S_V(\omega) = \int_{-\infty}^{\infty} e^{-i\omega t} G_V(t) \, dt \ ,
\label{eq:spectraldensity}
\end{equation}
that has no sharp features within the bandwidth $\Delta f \approx 1/t$ of the sensor.  The aim of a noise spectroscopy experiment is to reconstruct $S_V(\omega)$ as a function of $\omega$ over a frequency range of interest.

Although this Section focuses on Gaussian  noise with $\tc\lesssim t$, the analysis can be extended to other noise models and correlated noise.  
When $\tc\gg t$, the frequency and amplitude of $V(t)$ are essentially fixed during one sensing period and the formalism of AC sensing can be applied (see Section \ref{sec06:multipulse}).  A rigorous derivation for all ranges of $\tc$, but especially $\tc \approx t$ is given by~\onlinecite{Cummings62}.  More complex noise models, such as $1/f$ noise with no well-defined $\tc$, or models that require a cumulant expansion beyond a first order approximation on the noise strength can also be considered \cite{Bergli07,Ban09}.  Finally, open-loop control protocols have been introduced~\cite{Cywinski14,PazSilva14,Norris16,Barnes16} to characterize stationary, non-Gaussian dephasing.

\subsection{Decoherence, dynamical decoupling and filter functions}
\label{sec07:decoherence}

There have been many proposals for examining decoherence under different control sequences to investigate noise spectra~\cite{Faoro04,Yuge11,Young12,Almog11}.  In particular, dynamical decoupling sequences based on multipulse protocols (Section \ref{sec06:multipulse}) provide a systematic means for filtering environmental noise~\cite{Biercuk11,Alvarez11,Kotler11}.  These have been implemented in many physical systems~\cite{Kotler13,Bar-Gill12,Bylander11,Dial13,Muhonen14,Yoshihara14,Yan12,Yan13,Romach15}.  A brief introduction to the method of filter functions is presented in the following.

\subsubsection{Decoherence function $\chi(t)$}
\label{sec07:decoherencefunction}

Under the assumption of a Gaussian, stationary noise, the loss of coherence can be captured by an exponential decay of the transition probability with time $t$,
\begin{equation}
p(t)=\frac12\left(1-e^{-\chi(t)}\right) \ .
\label{eq:decoherence}
\end{equation}
where $\chi(t)$ is the associated decay function or \textit{decoherence function} that was already discussed in the context of sensitivity (Section \ref{sec05}).  Quite generally, $\chi(t)$ can be identified with an rms phase accumulated during time $t$,
\begin{equation}
\chi(t) = \frac12 \phirms^2 \ .
\label{eq:chiphi}
\end{equation}
according to the expression for variance detection, Eq. (\ref{eq:pvariance}).

Depending on the type of noise present, the decoherence function shows a different dependence on $t$.  For white noise, the dephasing is Markovian and $\chi(t)=\Gamma t$, where $\Gamma$ is the decay rate.  For Lorentzian noise centered at zero frequency the decoherence function is $\chi(t)=(\Gamma t)^3$.  For a generic $1/f$-like decay, where the noise falls of $\propto \omega^a$ (with $a$ around unity), the decoherence function is $\chi(t)=(\Gamma t)^{a+1}$ \cite{Cywinski08,Medford12} with a logarithmic correction depending on the ratio of total measurement time and evolution time \cite{Dial13}.  Sometimes, decoherence may even need to be described by several decay constants associated with several competing processes.  A thorough discussion of decoherence is presented in the recent review by \onlinecite{Suter16}.

\subsubsection{Filter function $Y(\omega)$}
\label{sec07:filterfunction}

The decoherence function $\chi(t)$ can be analyzed under the effect of different control sequences.  Assuming the control sequence has a modulation function $y(t)$ (see Section \ref{sec06:multipulse}), the decay function is determined by the correlation integral \cite{DeSousa09,Biercuk11}
\begin{align}
\chi(t) 
  &= \frac12 \int_0^t  dt'\, \int_0^{t} dt''\, y(t') y(t'') \gamma^2 G_V(t'-t'') \ ,
\end{align}
where $G_V(t)$ is the autocorrelation function of $V(t)$ (Eq.~\ref{eq:autocorrelation}).  In the frequency domain the decay function can be expressed as
\begin{align}
\chi(t) 
  &= \frac2\pi \int_{0}^{\infty} \gamma^2 S_V(\omega) |Y(\omega)|^2 d\omega  \ ,
\label{eq:chi}	
\end{align}
where $|Y(\omega)|^2$ is the so-called \textit{filter function} of $y(t)$, defined by the finite-time Fourier transform  
\begin{equation}
Y(\omega) = \int_0^{t} y(t')e^{i\omega t'} \, dt' \ . 	
\end{equation}
(Note that this definition differs by a factor of $\omega^2$ from the one by~\onlinecite{Biercuk11}).
Thus, the filter function plays the role of a transfer function, and the decay of coherence is captured by the overlap with the noise spectrum $S_V(\omega)$.

To illustrate the concept of filter functions we reconsider the canonical example of a Ramsey sensing sequence.
Here, the filter function is
\begin{equation}
|Y(\omega)|^2 = \frac{\sin^2(\omega t/2)}{\omega^2} \ .
\end{equation}
The decoherence function $\chi(t)$ then describes the ``free-induction decay'',
\begin{equation}
\chi(t)
= \frac2\pi \int_{0}^{\infty} \gamma^2 S_V(\omega) \frac{\sin^2(\omega t/2)}{\omega^2} d\omega
\approx \frac12 \gamma^2 S_V(0) t \ ,
\label{eq:chi:ramsey}
\end{equation}
where the last equation is valid for a spectrum that is flat around $\omega \lesssim \pi/t$.  The Ramsey sequence hence acts as a simple $\mr{sinc}$ filter for the noise spectrum $S_V(\omega)$, centered at zero frequency and with a lowpass cut-off frequency of approximately $\pi/t$.

\subsubsection{Dynamical decoupling}

To perform a systematic spectral analysis of $S_V(\omega)$, one can examine decoherence under various dynamical decoupling sequences.
Specifically, we inspect the filter functions of periodic modulation functions $y_{n_c,\tau_c}(t)$, where a basic building block $y_1(t)$ of duration $\tau_c$ is repeated $n_c$ times.
The filter function of $y_{n_c,\tau_c}(t)$ is given by
\begin{align}
Y_{n_c,\tau_c}(\omega)
  &= Y_{1,\tau_c}(\omega) \sum_{k=0}^{n_c-1}e^{i \tau_c k}  \nonumber \\
	&= Y_{1,\tau_c}(\omega) e^{-i(n_c-1)\omega \tau_c/2} \frac{\sin(n_c\omega\tau_c/2)}{\sin(\omega\tau_c/2)} \ ,
\end{align}
where $Y_{1,\tau_c}(\omega)$ is the filter function of the basic building block.
For large cycle numbers, $Y_{n_c,\tau_c}(\omega)$ presents sharp peaks at multiples of the inverse cycle time $\tau_c^{-1}$, and it can be approximated by a series of $\delta$ functions.

Two specific examples of periodic modulation functions include the CP and PDD sequences considered in Section \ref{sec06:multipulse}, where $\tau_c = 2\tau$ and $n_c = n/2$.
The filter function for large pulse numbers $n$ is
\begin{align}
|Y_{n,\tau}|^2
  &\approx \sum_{k} \frac{2\pi}{(k\pi)^2} \textrm{sinc}[(\omega-\omega_k)t/2]^2  \nonumber \\
  &\approx \sum_{k} \frac{2\pi}{(k\pi)^2} t \delta(\omega-\omega_k)
\end{align}
where $\omega_k = 2\pi \times k/(2\tau)$ are resonances with $k=1,3,5,...$ (note that these expression are equivalent to the filters Eq.~(\ref{eq:filter}) found for random phase signals).  

The decay function can then be expressed by a simple sum of different spectral density components,
\begin{align}
\chi(t)
  &= \frac2\pi \int_{0}^{\infty} \gamma^2 S_V(\omega) \sum_{k} \frac{2\pi}{(k\pi)^2} t \delta(\omega-\omega_k) \, d\omega \nonumber \\
  &= \frac{4t}{\pi^2} \sum_{k} \frac{\gamma^2 S_V(\omega_k)}{k^2}
\end{align}

This result provides a simple strategy for reconstructing the noise spectrum.  By sweeping the time $\tau$ between pulses the spectrum can be probed at various frequencies.  Since the filter function is dominated by the first harmonic ($k=1$) the frequency corresponding to a certain $\tau$ is $1/(2\tau)$.  For a more detailed analysis the contributions from higher harmonics as well as the exact shape of the filter functions has to be taken into account.  The spectrum can then be recovered by spectral decomposition~\cite{Alvarez11,Bar-Gill12}.

The filter analysis can be extended to more general dynamical decoupling sequences.  In particular, \onlinecite{Zhao14}, consider periodic sequences with more complex building blocks, and~\onlinecite{Cywinski08}, consider aperiodic sequences like the UDD sequence.

\subsection{Relaxometry}
\label{sec07:relaxometry}

An alternative framework for analyzing relaxation and doherence has been developed in the field of magnetic resonance spectroscopy, and is commonly referred to as ``relaxometry'' \cite{Abragam61}.  The concept has later been extended to the context of qubits \cite{Schoelkopf03}.  The aim of relaxometry is to connect the spectral density $S_V(\omega)$ of a noise signal $V(t)$ to the relaxation rate $\Gamma$ in first-order kinetics, $\chi(t) = \Gamma t$.  Relaxometry is based on first-order perturbation theory and Fermi's golden rule.  The basic assumptions are that the noise process is approximately Markovian and that the noise strength is weak, such that first-order perturbation theory is valid.  Relaxometry has found many applications in magnetic resonance and other fields, especially for mapping high-frequency noise based on $T_1$ relaxation time measurements \cite{Kimmich04}.

\subsubsection{Basic theory of relaxometry}
\label{sec:relaxationtheory}

To derive a quantitative relationship between the decay rate $\Gamma$ and a noise signal $V(t)$, we briefly revisit the elementary formalism of relaxometry \cite{Abragam61}.  In a first step, $V(t)$ can be expanded into Fourier components,
\begin{align}
V(t) 
	&= \frac{1}{2\pi} \int_{-\infty}^{\infty} d\omega \,
	   \left\{ V(\omega) e^{-i\omega t} + V^\dagger(\omega) e^{i\omega t} \right\}
\end{align}
where $V(\omega) = V^\dagger(-\omega)$.
Next, we calculate the probability amplitude $c_1$ that a certain frequency component $V(\omega)$ causes a transition between two orthogonal sensing states $\psia$ and $\psib$ during the sensing time $t$.    Since the perturbation is weak, perturbation theory can be applied.  The probability amplitude $c_1$ in first order perturbation theory is
\begin{align}
c_1(t)
  &= -i \int_0^t dt'\, \ipsib\hat{H}_V(\omega)\psia e^{i(\woo-\omega)t'}  \nonumber \\
	&= -i \ipsib\hat{H}_V(\omega)\psia \frac{e^{i(\woo-\omega)t}-1}{i(\woo-\omega)}  
\end{align}
where $\hat H_V(\omega)$ is the Hamiltonian associated with $V(\omega)$ and  $\woo$ is the transition energy between states $\psia$ and $\psib$.  The transition probability is
\begin{align}
|c_1(t)|^2
  &= |\ipsib\hat{H}_V(\omega)\psia|^2 \left( \frac{\sin[(\woo-\omega)t/2]} {(\woo-\omega)/2} \right)^2  \nonumber \\
	&\approx 2\pi |\ipsib\hat{H}_V(\omega)\psia|^2 t \delta(\woo-\omega)
\end{align}
where the second equation reflects that for large $t$, the $\mr{sinc}$ function approaches a $\delta$ function peaked at $\woo$.  The associated transition rate is
\begin{equation}
\frac{\partial |c_1(t)|^2}{\partial t} \approx 2\pi |\ipsib\hat{H}_V(\omega)\psia|^2 \delta(\woo-\omega) \ .
\label{eq:dpdt}
\end{equation}
This is Fermi's golden rule expressed for a two-level system that is coupled to a radiation field with a continuous frequency spectrum \cite{Sakurai11}.

The above transition rate is due to a single frequency component of $\hat{H}_V(\omega)$.  To obtain the overall transition rate $\Gamma$, Eq. (\ref{eq:dpdt}) must be integrated over all frequencies, 
\begin{align}
\Gamma &= \frac{1}{\pi} \int_0^{\infty} d\omega\, 2\pi |\ipsib\hat{H}_V(\omega)\psia|^2 \delta(\woo-\omega) \nonumber \\
  &= 2 |\ipsib\hat{H}_V(\woo)\psia|^2  \nonumber \\
	&= 2 \gamma^2  S_{V_{01}}(\woo)\cdot |\ipsib \sigma_V/2 \psia|^2
	\label{eq:gamma}
\end{align}
where in the last equation, $S_{V_{01}}$ is the spectral density of   the component(s) of $V(t)$ than can drive transitions
between $\psia$ and $\psib$,  multiplied by a transition matrix element $|\ipsib  \sigma_V/2 \psia|^2$ of order unity that represents the operator part of $\hat H_V=V(t)\sigma_V/2$ (see Eq. \ref{eq:HVa}).

The last equation (\ref{eq:gamma}) is an extremely simple, yet powerful and quantitative relationship:  the transition rate equals the spectral density of the noise evaluated at the transition frequency, multiplied by a matrix element of order unity \cite{Abragam61,Schoelkopf03}.
The expression can also be interpreted in terms of the rms phase $\phirms$.
According to Eq. (\ref{eq:chiphi}), $\phirms^2 = 2\chi(t) = 2\Gamma t$, which in turn yields (setting $|\ipsib  \sigma_V/2 \psia|^2 = \frac14$)
\begin{equation}
\phirms^2 = \gamma^2  S_{V_{01}}(\woo) t \ .
\label{eq:phirms:noise}
\end{equation}
The rms phase thus corresponds to the noise integrated over an equivalent noise bandwidth of $1/(2\pi t)$.

\begin{table*}[t!]
\centering
\begin{tabular}{l|>{\centering}p{0.12\linewidth}|>{\centering}p{0.18\linewidth}|>{\centering}p{0.18\linewidth}|l}
\hline\hline
Method & Sensing states $\{\psia,\psib\}$ & Sensitive to $\Vzero$ at frequency & Sensitive to $\Vone$ at frequency & Frequency tunable via \\
\hline
Ramsey                       & $\{\xa,\xb\}$  & 0                                 & ---$^a$       & --- \\
Spin echo                    & $\{\xa,\xb\}$  & $1/t$                             & ---$^a$       & --- \\
Dynamical decoupling         & $\{\xa,\xb\}$  & $\pi k/\tau$, with $k=1,3,..$     & ---$^a$       & Pulse spacing $\tau$, resonance order $k$ \\
$T_1$ relaxometry            & $\{\ma,\mb\}$  & ---                               & $\wo$         & Static control field \\
Dressed states (resonant)    & $\{\xa,\xb\}$  & $\wone$                           & ---$^a$       & Drive field amplitude $\wone$ \\
Dressed states (off-resonant)& $\{\xa,\xb\}$  & $\weff\approx\Dw$                 & ---$^a$       & Detuning $\Dw$ \\
\hline\hline
\end{tabular}
\caption{Summary of noise spectroscopy methods.  $\xab=(\ma\pm\mb)/\sqrt2$.  $^a$also affected by $T_1$ relaxation.
}
\label{table:relaxometry}
\end{table*}

The relation between the transition rate $\Gamma$ and the spectral density can be further specified for vector signals $\vec V$.
In this case the transition rate represents the sum of the three vector components of $V_j$, where $j=x,y,z$,
\begin{align}
\Gamma
  &= 2 \sum_{j=x,y,z} |\langle \psi_1|\hat{H}_{V_j}(\omega_j)|\psi_0\rangle|^2  \nonumber \\
  &= 2 \sum_{j=x,y,z} \gamma^2 S_{V_j}(\omega_j) |\langle \psi_1|\hat\sigma_j|\psi_0\rangle|^2
\end{align}
where $S_{V_j}(\omega_j)$ is the spectral density of $V_j$, $\omega_j$ is a transition frequency that reflects the energy exchange required for changing the state, and $\hat{\sigma}_j$ are Pauli matrices.  Note that if $\{\psia,\psib\}$ are coherent superposition states, $V_x$ and $V_y$ represent the components of $\Vone$ that are in-phase and out-of-phase with the coherence, rather than the static components of the vector signal $\vec V$.

Relaxation rates can be measured between any set of sensing states $\{\psia,\psib\}$, including superposition states.  This gives rise to a great variety of possible relaxometry measurements.  For example, the method can be used to probe different vector components $V_j(t)$ (or commuting and non-commuting signals $\Vzero(t)$ and $\Vone(t)$, respectively) based on the selection of sensing states.  Moreover, different sensing states typically have vastly different transition energies, providing a means to cover a wide frequency spectrum.
If multiple sensing qubits are available, the relaxation of higher-order quantum transitions can be measured, which gives additional freedom to probe different symmetries of the Hamiltonian.

\begin{figure}[t!]
\centering
\includegraphics[width=0.8\figurewidth]{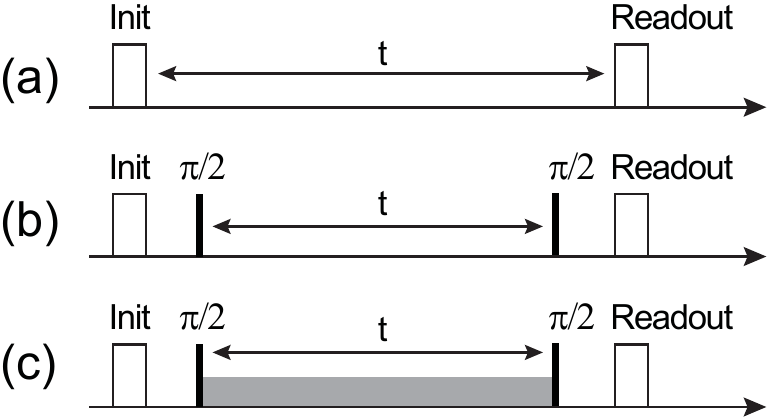}
\caption{Common relaxometry protocols.
(a) $T_1$ relaxometry.
(b) $\Ttwoast$ relaxometry.
(c) Dressed state relaxometry.
Narrow black boxes represent $\pi/2$ pulses and the grey box in (c) represents a resonant or off-resonant drive field.  
}
\label{fig:relaxometrysequences}
\end{figure}

An overview of the most important relaxometry protocols is given in Table \ref{table:relaxometry} and Fig. \ref{fig:relaxometrysequences}.  They are briefly discussed in the following.

\subsubsection{$T_1$ relaxometry}

$T_1$ relaxometry probes the transition rate between states $\ma$ and $\mb$.  This is the canonical example of energy relaxation.
Experimentally, the transition rate is measured by initializing the sensor into $\ma$ at time $t'=0$ and inspecting $p=|\langle 1\malpha|^2$ at time $t'=t$ without any further manipulation of the quantum system (see Fig. \ref{fig:relaxometrysequences}(a)).  The transition rate is
\begin{align}
(T_1)^{-1}
	&= \frac12 \gamma^2 S_{\Vone}(\wo) \ ,
\end{align}
where $T_1$ is the associated relaxation time and $S_{\Vone}=S_{V_x}+S_{V_y}$.  Thus, $T_1$ relaxometry is only sensitive to the transverse component of $\vec V$.  Because $T_1$ can be very long, very high sensitivities are in principle possible, assuming that the resonance is not skewed by low-frequency noise.
By tuning the energy splitting $\wo$ between $\ma$ and $\mb$, for example through the application of a static control field, a frequency spectrum of $S_{\Vone}(\omega)$ can be recorded \cite{Kimmich04}.  For this reason and because it is experimentally simple, $T_1$ relaxometry has found many applications.  For example, single-spin probes have been used to detect the presence of magnetic ions~\cite{Steinert13}, spin waves in magnetic films~\cite{Vandersar15},  high-frequency magnetic fluctuations near surfaces~\cite{Rosskopf14,Myers14,Romach15}, and single molecules \cite{Sushkov14}.  $T_1$ relaxometry has also been applied to perform spectroscopy of electronic and nuclear spins~\cite{Hall16}.  In addition, considerable effort has been invested in mapping the noise spectrum near superconducting flux qubits by combining several relaxometry methods~\cite{Bialczak07,Lanting09,Bylander11,Yan13}.
\begin{figure}[t!]
\centering
\includegraphics[width=\figurewidth]{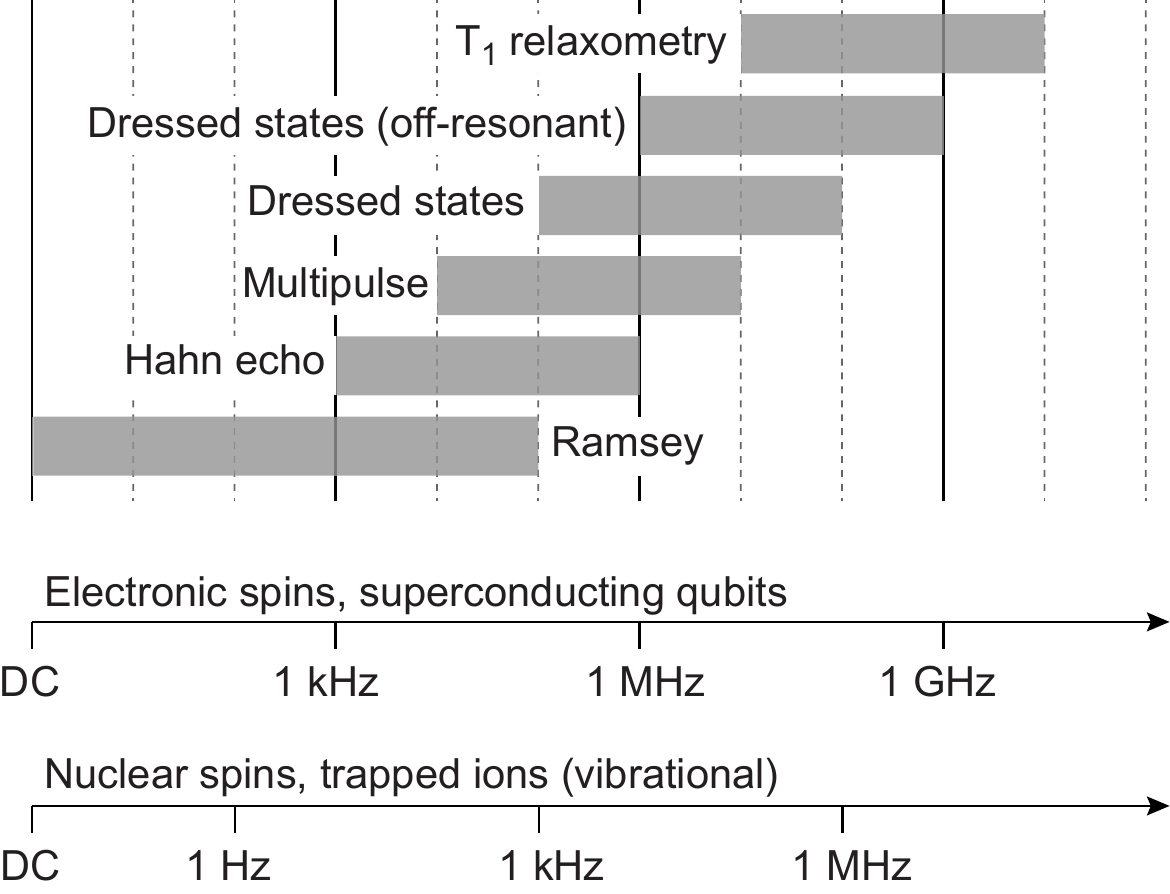}
\caption{Typical spectral range of noise spectroscopy protocols.
Scales refer to the quantum sensors discussed in Section \ref{sec03}.
}
\label{fig:spectroscopy}
\end{figure}

\subsubsection{$T_2^\ast$ and $T_2$ relaxometry}

$T_2^\ast$ relaxometry probes the transition rate between the superposition states $\xab = (\ma \pm e^{-i\wo t}\mb)/\sqrt2$.  This corresponds to the free induction decay observed in a Ramsey experiment (Fig. \ref{fig:relaxometrysequences}(b)).  The associated dephasing time $T_2^\ast$ is given by
\begin{align}
(\Ttwoast)^{-1}
	&= \frac14 \gamma^2 S_{\Vone}(\wo) + \frac12 \gamma^2 S_{\Vzero}(0) \ ,
\label{eq:ttwoast}
\end{align}
where $S_{\Vzero}=S_{V_z}$ (see also Eq. (\ref{eq:chi:ramsey})).  The transverse $S_{\Vone}$ term in Eq. (\ref{eq:ttwoast}) involves a ``bit flip'' and the parallel $S_{\Vzero}$ term involves a ``phase flip''.  Because a phase flip does not require energy, the spectral density is probed at zero frequency.  Since $S_V(\omega)$ is often dominated by low-frequency noise, $S_{\Vzero}(0)$ is typically much larger than $S_{\Vone}(\wo)$ and the high-frequency contribution can often be neglected.  Note that Eq. (\ref{eq:ttwoast}) is exact only when the spectrum $S_{\Vzero}(\omega)$ is flat up to $\omega \sim \pi/t$.

$\Ttwoast$ relaxometry can be extended to include dephasing under dynamical decoupling sequences.  The relevant relaxation time is then usually denoted by $T_2$ rather than $\Ttwoast$.  Dephasing under dynamical decoupling is more rigorously described by using filter functions (see Section \ref{sec07:filterfunction}).

\subsubsection{Dressed state methods}

Relaxation can also be analyzed in the presence of a resonant drive field.  This method is known as ``spin locking'' in magnetic resonance~\cite{Slichter96}.  Due to the presence of the resonant field the degeneracy between $\xab$ is lifted and the states are separated by the energy $\hbar\wone$, where $\wone\ll\wo$ is the Rabi frequency of the drive field.  A phase flip therefore is no longer energy conserving.  The associated relaxation time $T_{1\rho}$ is given by
\begin{align}
(T_{1\rho})^{-1}
	&\approx \frac14 \gamma^2 S_{\Vone}(\wo) + \frac12 \gamma^2 S_{\Vzero}(\wone)
\end{align}
By systematically varying the Rabi frequency $\wone$, the spectrum $S_{\Vzero}(\wone)$ can be recorded~\cite{Loretz13,Yan13}.  Because $\wone\ll\wo$, dressed states provide useful means for covering the medium frequency range of the spectrum (see Fig. \ref{fig:spectroscopy}).  In addition, since dressed state relaxometry does not require sweeping a static control field for adjusting the probe frequency, it is more versatile than standard $T_1$ relaxometry.

Dressed state methods can be extended to include off-resonant drive fields.  Specifically, if the drive field is detuned by $\Dw$ from $\wo$, relaxation is governed by a modified relaxation time
\begin{align}
(T_{1\rho})^{-1}
	&\approx \gamma^2  \frac14 \left[1+\frac{\Dw^2}{\weff^2}\right] S_{\Vone}(\wo)  
	+ \frac12 \frac{\wone^2}{\weff^2} \gamma^2 S_{\Vzero}(\weff) \ ,
\end{align}
where $\weff = \sqrt{\wone^2 + \Dw^2}$ is the effective Rabi frequency.  A detuning therefore increases the accessible spectral range towards higher frequencies.  For a very large detuning the effective frequency becomes similar to the detuning $\weff \approx \Dw$, and the drive field only enters as a scaling factor for the spectral density.  Detuned drive fields have been used to map the $1/f$ noise spectrum of transmon qubits up to the GHz range \cite{Slichter12}.

\section{Dynamic range and adaptive sensing}
\label{sec08}

``Adaptive sensing'' refers to a class of techniques addressing the intrinsic problem of limited dynamic range in quantum sensing:  \textit{The basic quantum sensing protocol cannot simultaneously achieve high sensitivity and measure signals over a large amplitude range}.

The origin of this problem lies in the limited range of values for the probability $p$, which must fall between 0 and 1.  For the example of a Ramsey measurement, $p$ oscillates with the signal amplitude $V$ and phase wrapping occurs once $\gamma V t$ exceeds $\pm \pi/2$, where $t$ is the sensing time.  Given a measured transition probability $p$, there is an infinite number of possible signal amplitudes $V$ that can correspond to this value of $p$ (see top row of Fig.~\ref{fig:adaptive_ramsey}).  A unique assignment hence requires \textit{a priori} knowledge -- that $V$ lies within $\pm\pi/(2\gamma t)$, or within half a Ramsey fringe, of an expected signal amplitude.  This defines a maximum allowed signal range,
\begin{equation}
\Vmax = \frac{\pi}{\gamma t} \ .
\label{eq:vmax}
\end{equation}

The sensitivity of the measurement, on the other hand, is proportional to the slope of the Ramsey fringe and reaches its optimum when $t \approx \Ttwoast$.  The smallest detectable signal is approximately $\Vmin \approx 2/(\gamma C\sqrt{\Ttwoast T})$, where $T$ is the total measurement time and $C$ the readout efficiency parameter (see Eqs. \ref{eq:voptslope} and \ref{eq:vminscaling1}).
The dynamic range is then given by the maximum allowed signal divided by the minimum detectable signal,
\begin{equation}
\DR = \frac{\Vmax}{\Vmin} =   \frac{\pi C \sqrt{T}} {2\sqrt{\Ttwoast}}  \propto \sqrt{T} \ .
\end{equation}
Hence, the basic measurement protocol can be applied only to small changes of a quantity around a fixed known value, frequently zero.  The protocol does not apply to the problem of determining the value of a large and \textit{a priori} unknown quantity.   Moreover, the dynamic range only improves with the square root of the total measurement time $T$.

\subsection{Phase estimation protocols}

\begin{figure}
\centering
\includegraphics[width=\figurewidth]{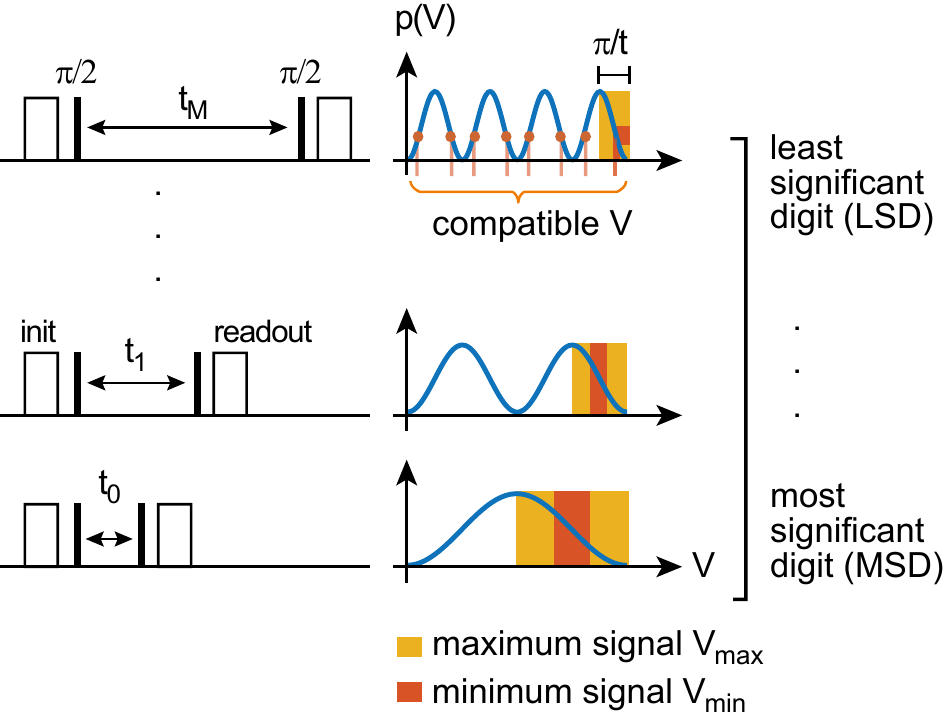}
\caption{High dynamic range sensing. A series of measurements with different interrogation times $t$ is combined to estimate a signal of interest. The shortest measurement (lowest line) has the largest allowed signal $\Vmax$ and provides a coarse estimate of the quantity, which is subsequently refined by longer and more sensitive measurements.  Although the $p(V)$ measured in a sensitive measurement (top line) can correspond to many possible signal values, the coarse estimates allow one to extract a unique signal value $V$. 
}
\label{fig:adaptive_ramsey}
\end{figure}

Interestingly, a family of advanced sensing techniques can efficiently address this problem and achieve a dynamic range that scales close to $\DR \propto T$.  This scaling is sometimes referred to as another instance of the Heisenberg limit, because it can be regarded as a $1/T$ scaling of sensitivity at a fixed $\Vmax$.  The central idea is to combine measurements with different sensing times $t$ such that the least sensitive measurement with the highest $\Vmax$ yields a coarse estimate of the quantity of interest, which is subsequently refined by more sensitive measurements (Fig. \ref{fig:adaptive_ramsey}).

In the following we discuss protocols that use exponentially growing sensing times $t_m = 2^m t_0$, where $m=0,1,\dots,M$ and $t_0$ is the smallest time element (see Fig. \ref{fig:adaptive_ramsey}).  Although other scheduling is possible, this choice allows for an intuitive interpretation: subsequent measurements estimate subsequent digits of a binary expansion of the signal.  The maximum allowed signal is then set by the shortest sensing time, $\Vmax = \pi/(\gamma t_0)$, while the smallest detectable signal is determined by the longest sensing time, $\Vmin \approx 2/(\gamma C\sqrt{t_M T})$.  Because $T \propto t_M$ due to the exponentially growing interrogation times, the dynamic range of the improved protocol scales as
\begin{equation}
\dr \propto \frac{\sqrt{t_M T}}{t_0} \propto T \ .
\end{equation}
This scaling is obvious from an order-of-magnitude estimate: adding an additional measurement step increases both precision and measurement duration $t$ by a factor of two, such that precision scales linearly with total acquisition time $T$.  We will now discuss three specific implementations of this idea. 

\begin{figure*}
\centering
\includegraphics[width=0.9\textwidth]{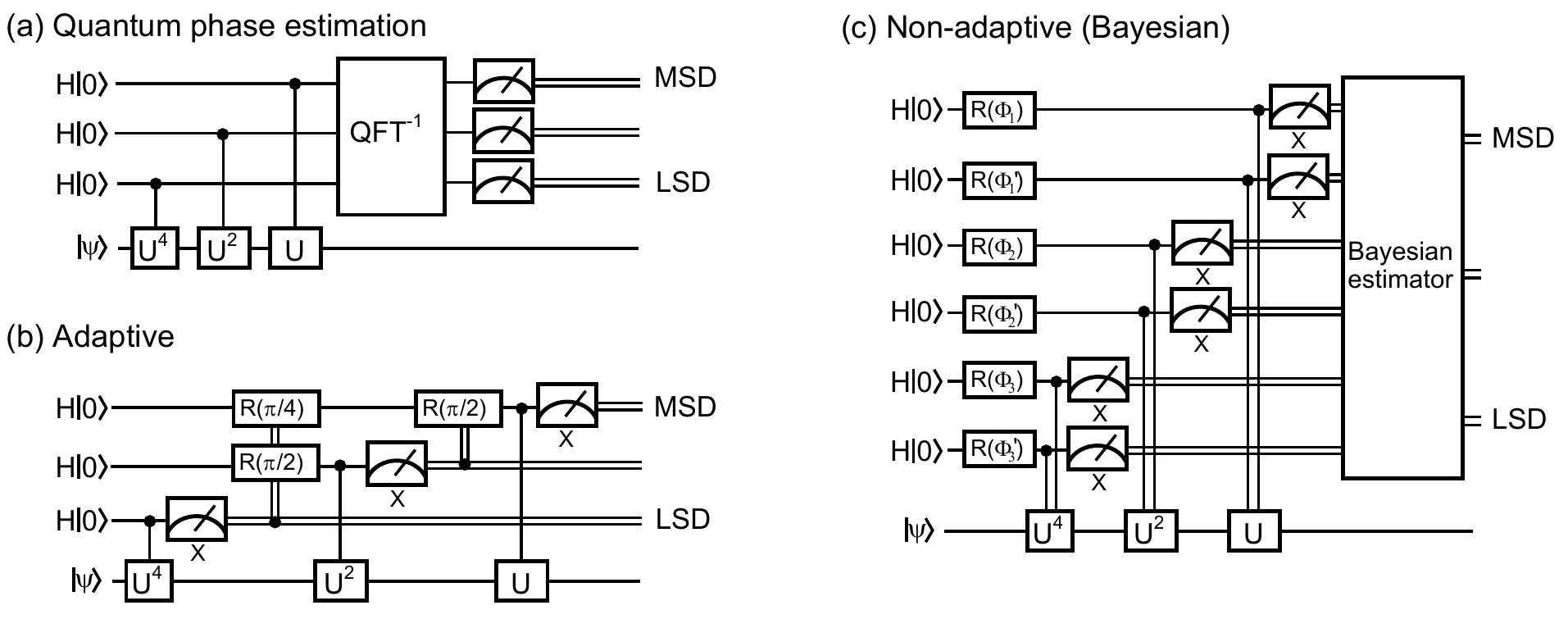}
\caption{Phase estimation algorithms.
(a) Quantum phase estimation by the inverse Quantum Fourier transform, as it is employed in prime factorization algorithms \cite{Shor94,Kitaev95}.
(b) Adaptive phase estimation. Here, the quantum Fourier transform is replaced by measurement and classical feedback. Bits are measured in ascending order, substracting the lower digits from measurements of higher digits by phase gates that are controlled by previous measurement results.
(C) Non-adaptive phase estimation. Measurements of all digits are fed into a Bayesian estimation algorithm to estimate the most likely value of the phase.
$H$ represents a Hadamard gate, $R(\Phi)$ a $Z$-rotation by the angle $\Phi$, and $U$ the propagator for one time element $t_0$.  The box labeled by ``x'' represents a readout.
}
\label{fig:qpe_adaptive_bayesian}
\end{figure*}

\subsubsection{Quantum phase estimation}

All three protocols can be considered variations of the phase estimation scheme depicted in Fig. \ref{fig:qpe_adaptive_bayesian}(a).  The scheme was originally  put forward by~\onlinecite{Shor94}, in the seminal proposal of a quantum algorithm for prime factorization and has been interpreted by~\onlinecite{Kitaev95}, as a phase estimation algorithm. 

The original formulation applies to the problem of finding the phase $\phi$ of the eigenvalue $e^{2\pi i\phi}$ of a unitary operator $\hat U$, given a corresponding eigenvector $\ket \psi$.  This problem can be generalized to estimating the phase shift $\phi$ imparted by passage through an interferometer or exposure to an external field.  The algorithm employs a register of  $N$ auxiliary qubits ($N=3$ in Fig. \ref{fig:qpe_adaptive_bayesian}) and prepares them into a digital representation $\ket \phi= \ket{\phi_1}\ket{\phi_{2}}\dots\ket{\phi_M}$ of a binary expansion of $\phi=\sum_{m=1}^M \phi_m 2^{-m}$ by a sequence of three processing steps:

\begin{enumerate}
\item{State preparation:}
All qubits are prepared in a coherent superposition state $\xa = (\ket 0 + \ket 1)/\sqrt2$ by an initial Hadamard gate.  The resulting state of the full register can then be written as
\begin{equation}
\frac 1 {\sqrt{2^M}} \sum_{j=0}^{2^M-1} \ket j_M
\end{equation}
where $\ket j_M$ denotes the register state in binary expansion $\ket 0_M = \ket{00\dots 0}$, $\ket 1_M = \ket{00\dots 1}$, $\ket 2_M = \ket{00\dots 10}$, etc.

\item{Phase encoding:}
The phase of each qubit is tagged with a multiple of the unknown phase $\phi$.  Specifically, qubit $m$ is placed in state $(\ket 0 + e^{2\pi i 2^m \phi}\ket 1)/\sqrt 2$.  Technically, this can be implemented by exploiting the back-action of a controlled-$\hat U^{2^m}$-gate that is acting on the eigenvector $\psi$ conditional on the state of qubit $m$.  Since $\psi$ is an eigenvector of $\hat U^j$ for arbitrary $j$, this action transforms the joint qubit-eigenvector state as
\begin{align}
\tfrac 1 {\sqrt 2} (\ket 0 + \ket 1)\otimes \ket{\psi} \nonumber \\
\rightarrow \tfrac 1 {\sqrt 2} (\ket 0 + e^{2\pi i 2^m \phi} \ket 1)\otimes \ket\psi
\end{align}
Here, the back-action on the control qubit $m$ creates the required phase tag. The state of the full register evolves to 
\begin{equation}
\frac 1 {\sqrt{2^M}} \sum_{j=0}^{2^M-1} e^{2\pi i \phi j/2^M} \ket j
\label{eq:phi_fourier}
\end{equation}
In quantum sensing, phase tagging by back-action is replaced by the exposure of each qubit to an external field for a time $2^m t_0$ (or passage through an interferometer of length $2^m l_0$).

\item{Quantum Fourier Transform:}
In a last step, an inverse quantum Fourier transform (QFT) \cite{Nielsen00b} is applied to the qubits.  This algorithm can be implemented with polynomial effort (\ie, using $\mathcal O(M^2)$ control gates).  The QFT recovers the phase $\phi$ from the Fourier series (\ref{eq:phi_fourier}) and places the register in state
\begin{equation}
\ket \phi= \ket{\phi_1}\ket{\phi_2}\dots\ket{\phi_M} \ .
\end{equation}
A measurement of the register directly yields a digital representation of the phase $\phi$.  To provide an estimate of $\phi$ with $2^{-M}$ accuracy, $2^M$ applications of the phase shift $\hat U$ are required.  Hence, the algorithm scales linearly with the number of applications of $\hat U$ which in turn is proportional to the total measurement time $T$.
\end{enumerate}

Quantum phase estimation is the core component of Shor's algorithm, where it is used to compute discrete logarithms with polynomial time effort \cite{Shor94}.

\subsubsection{Adaptive phase estimation}

While quantum phase estimation based on the QFT can be performed with polynomial time effort, the algorithm requires two-qubit gates, which are difficult to implement experimentally, and the creation of fragile entangled states. This limitation can be circumvented by an adaptive measurement scheme that reads the qubits sequentially, feeding back the classical measurement result into the quantum circuit \cite{Griffiths96}.  The scheme is illustrated in Fig. \ref{fig:qpe_adaptive_bayesian}(b). 

The key idea of adaptive phase estimation is to first measure the least significant bit of $\phi$, represented by the lowest qubit in Fig. \ref{fig:qpe_adaptive_bayesian}(b).  In the measurement of the next significant bit, this value is subtracted from the applied phase.  The subtraction can be implemented by classical unitary rotations conditioned on the measurement result, for example by controlled $R(\pi/j)$ gates as shown in Fig. \ref{fig:qpe_adaptive_bayesian}(b).  This procedure is then repeated in ascending order of the bits.  The QFT is thus replaced by measurement and classical feedback, which can be performed using a single qubit sensor. 

In practical implementations \cite{Higgins07}, the measurement of each digit is repeated multiple times or performed on multiple parallel qubits.  This is possible because the controlled-$U$ gate does not change the eigenvector $\psi$, so that it can be re-used as often as required.  The Heisenberg limit can only be reached if the number of resources (qubits or repetitions) spent on each bit are carefully optimized \cite{Berry09,Said11,Cappellaro12}.  Clearly, most resources should be allocated to the most significant bit, because errors at this stage are most detrimental to sensitivity.
The implementation by~\onlinecite{Bonato16}, for example, scaled the number of resources $N_m$ linearly according to
\begin{equation}
N_m = G + F (M-1-m).
\end{equation}
with typical values of $G=5$ and $F=2$.

\subsubsection{Non-adaptive phase estimation}

Efficient quantum phase estimation can also be implemented without adaptive feedback, with the advantage of technical simplicity \cite{Higgins09}.
A set of measurements $\{x_j\}_{j=1\dots N}$ (where $N>M$) is used to separately determine each unitary phase $2^m \phi$ with a set of fixed, classical phase shifts before each readout.
This set of measurements still contains all the information required to extract $\phi$, which can be motivated by the following arguments:  given a redundant set of phases, a post-processing algorithm can mimic the adaptive algorithm by postselecting those results that have been measured using the phase most closely resembling the correct adaptive choice.  From a spectroscopic point of view, measurements with different phases correspond to Ramsey fringes with different quadratures.  Hence, at least one qubit of every digit will perform its measurement on the slope of a Ramsey fringe, allowing for a precise measurement of $2^m \phi$ regardless of its value. 

The phase $\phi$ can be recovered by Bayesian estimation.
Every measurement $x_j = \pm 1$ provides information about $\phi$, which is described by the \textit{a posteriori} probability
\begin{equation}
p(\phi|x_j),
\end{equation}
%
the probability that the observed outcome $x_j$ stems from a phase $\phi$.
This probability is related to the inverse conditional probability $p(x_l|\phi)$ -- the excitation probability describing Ramsey fringes -- by Bayes' theorem.  The joint probability distribution of all measurements is obtained from the product
\begin{equation}
p(\phi) \propto \prod_j p(\phi|x_j) \ ,
\end{equation}
from which the most likely value of $\phi$ is picked as the final result \cite{Waldherr12,Nusran12}. Here, too, acquisition time scales with the significance of the bit measured to achieve the Heisenberg limit.

\subsubsection{Comparison of phase estimation protocols}

All of the above variants of phase estimation achieve a $\DR\propto T$ scaling of the dynamic range.  They differ, however, by a constant offset.  Adaptive estimation is slower than quantum phase estimation by the QFT since it trades spatial resources (entanglement) into temporal resources (measurement time).  Bayesian estimation in turn is slower than adaptive estimation due to additional redundant measurements.

Experimentally, Bayesian estimation is usually simple to implement because no real-time feedback is needed and the phase estimation can be performed \textit{a posteriori}.  Adaptive estimation is technically more demanding since real-time feedback is involved, which often requires dedicated hardware (such as field-programmable gate arrays or a central processing unit) for the fast decision making.  Quantum phase estimation by the QFT, finally, requires many entangled qubits.

\subsection{Experimental realizations}

The proposals of Shor \cite{Shor94}, Kitaev \cite{Kitaev95} and Griffiths \cite{Griffiths96} were followed by a decade where research towards Heisenberg-limited measurements has focused mostly on the use of entangled states, such as the N00N state (see Section \ref{sec09}).  These states promise Heisenberg scaling in the spatial dimension (number of qubits) rather than time \cite{Giovannetti04,Giovannetti06,Lee02} and have been studied extensively for both spin qubits \cite{Bollinger96,Leibfried04,Leibfried05,Jones09} and photons \cite{Fonseca99,Edamatsu02,Walther04,Mitchell04,Nagata07,Xiang11}. 

Heisenberg scaling in the temporal dimension has shifted into focus with an experiment published in 2007, where adaptive phase estimation was employed in a single-photon interferometer \cite{Higgins07}.  The experiment has subsequently been extended to a non-adaptive version \cite{Higgins09}.  Shortly after, both variants have been translated into protocols for spin-based quantum sensing \cite{Said11}.  Meanwhile, high-dynamic-range protocols have been demonstrated on NV centers in diamond using both non-adaptive implementations \cite{Nusran12,Waldherr12} and an adaptive protocol based on quantum feedback \cite{Bonato16}. 
As a final remark, we note that a similar performance -- $1/T$ scaling and an increased dynamic range -- may be achieved by weak measurement protocols, which continuously track the evolution of the phase over the sensing sequence \cite{Shiga12, Kohlhaas15}. Weak measurements have been more generally proposed to enhance sensing protocols \cite{}, but their ultimate usefulness is still under debate \cite{}.

\section{Ensemble sensing}
\label{sec09}

Up to this point, we have mainly focused on single qubit sensors.  In the following two sections, quantum sensors consisting of more than one qubit will be discussed.  The use of multiple qubits opens up many additional possibilities that cannot be implemented on a single qubit sensor.

This section considers ensemble sensors where many (usually identical) qubits are operated in parallel.  Apart from an obvious gain in readout sensitivity, multiple qubits allow for the implementation of second-generation quantum techniques, including entanglement and state squeezing, which provide a true ``quantum'' advantage that cannot be realized with classical sensors.
Entanglement-enhanced sensing has been pioneered with atomic systems, especially atomic clocks \cite{Leibfried04,Giovannetti04}.  In parallel, state squeezing is routinely applied in optical systems, such as optical interferometers \cite{Ligo11}.

\subsection{Ensemble sensing}

Before discussing entanglement-enhanced sensing techniques, we briefly consider the simple parallel operation of $M$ identical single-qubit quantum sensors.
This implementation is used, for example, in atomic vapor magnetometers \cite{Budker07} or solid-state spin ensembles \cite{Wolf15}.  The use of $M$ qubits accelerates the measurement by a factor of $M$, because the basic quantum sensing cycle (Steps 1-5 of the protocol, Fig. \ref{fig:sensingprocessB}) can now be operated in parallel rather than sequentially.  Equivalently, $M$ parallel qubits can improve the sensitivity by $\sqrt{M}$ per unit time.

This scaling is equivalent to the situation where $M$ classical sensors are operated in parallel.  The scaling can be seen as arising from the projection noise associated with measuring the quantum system, where it is often called the Standard Quantum Limit (SQL) \cite{Braginskii74,Giovannetti04} or shot noise limit.
In practice, it is sometimes difficult to achieve a $\sqrt{M}$ scaling because instrumental stability is more critical for ensemble sensors \cite{Wolf15}.

For ensemble sensors such as atomic vapor magnetometers or spin arrays, the quantity of interest is more likely the number density of qubits, rather than the absolute number of qubits $M$.  That is, how many qubits can be packed into a certain volume without deteriorating the sensitivity of each qubit.  The sensitivity is then expressed per unit volume ($\propto \textrm{meters}^{-3/2}$).
The maximum density of qubits is limited by internal interactions between the qubits.  Optimal densities have been calculated both for atomic vapor magnetometers \cite{Budker07} and ensembles of NV centers \cite{Taylor08,Wolf15}.

\subsection{Heisenberg limit}

The standard quantum limit can be overcome by using quantum-enhanced sensing strategies to reach a more fundamental limit where the uncertainty $\sigp$ (Eq. ( \ref{eq:sigpq})) scales as $1/M$.  This limit is also known as the \textit{Heisenberg limit}.  Achieving the Heisenberg limit requires reducing the variance of a chosen quantum observable at the expenses of the uncertainty of a conjugated observable. This in turn requires preparing the quantum sensors in an entangled state. In particular, squeezed states \cite{Caves81,Wineland92,Kitagawa93} have been proposed early on to achieve the Heisenberg limit and thanks to experimental advances have recently shown exceptional sensitivity \cite{Hosten16}. 

The fundamental limits of sensitivity (quantum metrology) and strategies to achieve them have been discussed in many reviews \cite{Giovannetti04,Giovannetti06,Giovannetti11,Wiseman09b,Paris09}
and they will not be the subject of our review. 
In the following, we will focus on the most important states and methods that have been used for entanglement-enhanced sensing.

\subsection{Entangled states}

\subsubsection{GHZ and N00N states}
\begin{figure}[t]
\includegraphics[width=0.48\textwidth]{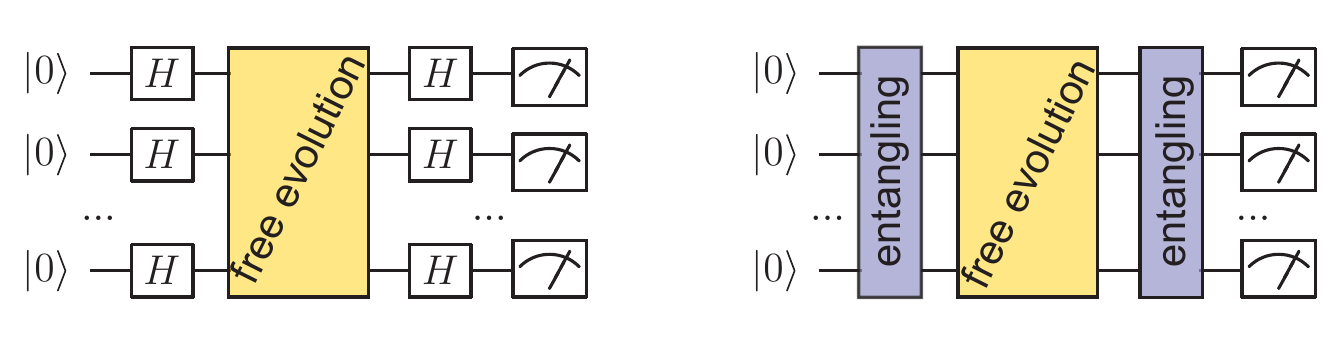}
	\caption{Left: Ramsey scheme. Right: entangled Ramsey scheme for Heisenberg-limited sensitivity}
\label{fig:GHZ}
\end{figure}

To understand the benefits that an entangled state can bring to quantum sensing, the simplest example is given by Greenberger-Horne-Zeilinger (GHZ) states. The sensing scheme is similar to a Ramsey protocol, however, if $M$ qubit probes are available, the preparation and readout pulses are replaced by entangling operations (Fig.~\ref{fig:GHZ}). 

We can thus replace the procedure in Sec.~\ref{sec04:ramsey} with the following:
\begin{enumerate}
\item The quantum sensors are initialized into $\ma\otimes\ma\otimes ... \otimes\ma\otimes\ma\equiv\ket{00\dots0}$.
\item Using entangling gates, the quantum sensors are brought into the  GHZ state $\psia =  (\ket{00\dots0} + \ket{11\dots1})/\sqrt2$. 
\item The superposition state evolves under the Hamiltonian $\hat{H}_0$ for a time $t$.  The superposition state picks up an \textit{enhanced phase} $\phi = M\wo t$, and the state after the evolution is
\begin{equation}
\psit =  \frac{1}{\sqrt{2}}(\ket{00\dots0} +  e^{iM\wo t}\ket{11\dots1}) \ ,
\label{eq:ramseyphaseGHZ}
\end{equation}
\item Using the inverse entangling gates, the state $\psit$ is converted back to an observable state, e.g. $\malpha = [\frac{1}{2}(e^{iM\wo t}+1)\ket{0_1} + \frac{1}{2} (e^{iM\wo t}-1)\ket{1_1}]\ket{0\dots0}_{2,M}$.
\item[5,6.] The final state is read out (only the first quantum probe needs to be measured in the case above).  The transition probability is
\begin{align}
p &= 1-|\langle 0 \malpha|^2 \nonumber \\
  &= \frac{1}{2}[1-\cos(M\wo t)] = \sin^2(M\wo t/2).
\end{align}
\end{enumerate}
Comparing this result with what obtained in Sec.~\ref{sec04:ramsey}, we see that the oscillation frequency of the signal is  enhanced by a factor $M$ by preparing a GHZ state, while the shot noise is unchanged, since we still measure only one qubit. This allows using an $M$-times shorter interrogation time or achieving an improvement of the sensitivity (calculated from the QCRB) by a factor $\sqrt{M}$. While for $M$ uncorrelated quantum probes the QCRB of Eq.~(\ref{eq:QCRB}) becomes
\begin{equation}
\Delta V_{N,M}=\frac{1}{\gamma\sqrt{N \mathcal F}}=\frac{e^\chi}{\gamma t\sqrt{M\,N}},
\label{eq:QCRBM}	
\end{equation}
for the GHZ state, the Fisher information reflects the state entanglement to give
\begin{equation}
\Delta V_{N,\mr{GHZ}}=\frac{1}{\gamma\sqrt{N \mathcal F}_\mr{GHZ}}=\frac{e^\chi}{\gamma Mt\sqrt{N}}
\end{equation}
Heisenberg-limited sensitivity with a GHZ state was demonstrated using three entangled Be ions \cite{Leibfried04}. 

Unfortunately, the $\sqrt{M}$ advantage in sensitivity is usually compensated by the GHZ state's increased decoherence rate \cite{Huelga97}, which is an issue common to most entangled states.  Assuming, for example, that each probe is subjected to uncorrelated dephasing noise, the rate of decoherence of the GHZ state is $M$ time faster than for a product state. Then, the interrogation time also needs to be reduced by a factor $M$ and no net advantage in sensitivity can be obtained. This has led to the quest for different entangled states that could be more resilient to decoherence.

Similar to GHZ states, N00N states have been conceived to improve interferometry \cite{Lee02}. They were first introduced by~\onlinecite{Yurke86l}, in the context of neutron Mach-Zender interferometry as the fermionic ``response'' to the squeezed states proposed by~\onlinecite{Caves81}, for Heisenberg metrology.
Using an $M$-particle interferometer, one can prepare an entangled Fock state,
\begin{equation}
\ket{\psi_{N00N}}=(\ket{M}_a\ket{0}_b+\ket{0}_a\ket{M}_b)/\sqrt{2} \ ,
\end{equation}
where $\ket{N}_a$ indicates the N-particle Fock state in spatial mode $a$. Already for small $M$, it is possible to show sensitivity beyond the standard quantum limit \cite{Kuzmich98a}.
If the phase is applied only to mode 
$a$ of the interferometer, the phase accumulated is then
\begin{equation}
\ket{\psi_{N00N}}{=(e^{iM\phi_a}}\ket{M}{_a}\ket{0}{_b+}\ket{0}{_a}\ket{M}{_b)/\sqrt{2} }\ {,
}\end{equation}
that is, $M$ times larger than for a one-photon state.
Experimental progress has allowed to reach ``high N00N'' (with $M>2$) states \cite{Mitchell04,Walther04,Monz10} by using strong nonlinearities or measurement and feed-forward.  They have been used not only for sensing but also for enhanced lithography \cite{Boto00}. Still, ``N00N'' states are very fragile \cite{Bohmann15} and they are afflicted by a small dynamic range.

\subsubsection{Squeezing}
Squeezed states are promising for quantum-limited sensing as they can reach sensitivity beyond the standard quantum limit. Squeezed states of light have been introduced by~\onlinecite{Caves81}, as a mean to reduce noise in interferometry experiments. 
One of the most impressive application of squeezed states of light \cite{Walls83,Schnabel10,Ligo11} has been the sensitivity enhancement of the LIGO gravitational wave detector \cite{Aasi13short}, obtained by injecting  vacuum squeezed states in  the interferometer.

Squeezing has also been extended to fermionic degrees of freedom (spin squeezing,~\onlinecite{Kitagawa93}) to reduce the uncertainty in spectroscopy measurements of ensemble of qubit probes.  The Heisenberg uncertainty principle bounds the minimum error achievable in the measurement of two conjugate variables. While for typical states the uncertainty in the two observables is on the same order,  it is possible to redistribute the fluctuations in the two conjugate observables. Squeezed states are then characterized by a reduced uncertainty in one observable at the expense of another observable. Thus, these states can help improving the sensitivity of quantum interferometry, as demonstrated by~\onlinecite{Wineland92},\onlinecite{Wineland94}. Similar to GHZ and N00N states, a key ingredient to this sensitivity enhancement is entanglement \cite{Sorensen01,Wang03}. However, the description of squeezed states is simplified by the use of a single collective angular-momentum variable.

The degree of spin squeezing can be measured by several parameters. For example, from the commutation relationship for the collective angular momentum, $\Delta J_\alpha\Delta J_\beta\geq |\langle J_\gamma\rangle|$, one can naturally define a squeezing parameter
\begin{equation}
\xi=\Delta J_\alpha/\sqrt{|\langle J_\gamma\rangle/2} \ ,
\end{equation}
with $\xi<1$ for squeezed states.  To quantify the advantage of squeezed states in sensing, it is advantageous to directly relate squeezing to the improved sensitivity.  This may be done by considering the ratio of the uncertainties on the acquired phase for the squeezed state $\Delta\phi_{sq}$ and for the uncorrelated state $\Delta\phi_0$ in, \eg, a Ramsey measurement.  The metrology squeezing parameter, proposed by~\onlinecite{Wineland92}, is then
\begin{equation}
\xi_R = \left|\frac{|\Delta\phi|_{sq}}{|\Delta\phi|_0}\right| = \frac{\sqrt{N}\Delta J_y(0)}{|\langle J_z(0)\rangle |}. 
\end{equation}

Early demonstrations of spin squeezing were obtained by entangling trapped ions via their shared motional modes \cite{Meyer01}, using repulsive interactions in a Bose-Einstein condensate \cite{Esteve08}, or partial projection by measurement \cite{Appel09}.  More recently, atom-light interactions in high-quality cavities have enabled squeezing of large ensembles atoms \cite{Leroux10,Schleier-Smith10a,Gross10,Louchet10,Bohnet14,Cox16,Hosten16} that can perform as atomic clocks beyond the standard quantum limit.  Spin squeezing can be also implemented in qubit systems \cite{Auccaise15,Cappellaro09b,Bennett13,Sinha03} following the original proposal by~\onlinecite{Kitagawa93}. 

In this context, a simple quantum sensing scheme,  following the procedure in Sec.~\ref{sec04:ramsey}, could be constructed by replacing step 2 with the preparation of a squeezed state, so that $\psia$ is a squeezed state. The state is prepared by evolving a reference (ground) state $|0\rangle$ under a squeezing Hamiltonanian, such as the one-axis $H_1=\chi J_z^2$ or two-axis $H_1=\chi(J_x^2-J_y^2)$ squeezing Hamiltonians. 
Then, during the free evolution (step 3) an enhanced phase can be acquired, similar to what happens for entangled states. The most common sensing protocols with squeezed states forgo step 4, and directly measure the population difference between the state $|0\rangle$ and $|1\rangle$. However, imperfections in this measurement limits the  sensitivity, since  achieving the Heisenberg  limit requires single-particle state detection. While this is difficult to obtain for large qubit numbers, recent advances show great promise~\cite{Zhang12,Hume13} (see also next section on alternative detection methods). A different strategy is to follow more closely the sensing protocol for entangled states, and refocus the squeezing (reintroducing step 4) before readout~\cite{Davis16}.

\subsubsection{Parity measurements}
A challenge in achieving the full potential of multi-qubit enhanced metrology is the widespread inefficiency of quantum state readout.  Metrology schemes often require single qubit readout or the measurement of complex, many-body observables. In both cases, coupling of the quantum system to the detection apparatus is inefficient, often because strong coupling would destroy the very quantum state used in the metrology task. 

To reveal the properties of entangled states and to take advantage of their enhanced sensitivities, an efficient observable is the parity of the state. The parity observable was first introduced in the context of ion qubits \cite{Bollinger96,Leibfried04} and it referred to the excited or ground state populations of the ions.  The parity has become widely adopted for the readout of N00N states, where the parity measures the even/odd number of photons in a state \cite{Gerry10}. Photon parity measurements are as well used in quantum metrology with squeezed states. 
While the simplest method for parity detection would be via single photon counting, and recent advances in superconducting single photon detectors approach the required efficiency \cite{Natarajan12}, photon numbers could also be measured with single-photon resolution using quantum non-demolition (QND) techniques \cite{Imoto85} that exploit nonlinear optical interactions.
Until recently, parity detection for atomic ensembles containing more than a few particles was out of reach. However, recent breakthroughs in spatially resolved  \cite{Bakr09} and cavity-based atom detection \cite{Schleier-Smith10,Hosten16a}  enabled atom counting in mesoscopic ensembles containing $M \gtrsim 100$ atoms.

\subsubsection{Other types of entanglement}

The key difficulty with using entangled states for sensing is that they are less robust against noise.  Thus, the advantage in sensitivity is compensated by a concurrent reduction in coherence time.  In particular, it has been demonstrated that for  frequency estimation, any non-zero value of uncorrelated dephasing noise cancels the advantage of the maximally entangled state over a classically correlated state \cite{Huelga97}.  An analogous result can be proven for magnetometry \cite{Auzinsh04}. 

Despite this limitation, non-maximally entangled states can provide an advantage over product states \cite{UlamOrgikh01,Shaji07}.  Optimal states for quantum interferometry in the presence of photon loss can, for example, be found by numerical searches \cite{Huver08,Lee09}.

Single-mode states have also been considered as a more robust alternative to  two-mode states.  Examples include pure Gaussian states in the presence of phase diffusion \cite{Genoni11}, mixed Gaussian states in the presence of loss \cite{Aspachs09} and single-mode variants of two-mode states \cite{Maccone09}.

Other strategies include the creation of states that are more robust to the particular noise the system is subjected to \cite{Goldstein11,Cappellaro12a} or the use of entangled ancillary qubits that are not quantum sensors themselves \cite{Kessler14,Huang16,Demkowicz14,Dur14}.  These are considered in the next section (Section \ref{sec10}).

\section{Sensing assisted by auxiliary qubits}
\label{sec10}

In the previous section we considered potential improvements in sensitivity derived from the availability of multiple quantum systems operated in parallel. A different scenario arises when only a small number of additional quantum systems is available, or when the additional quantum systems do not directly interact with the signal to be measured.  Even in this situation, however, ``auxiliary qubits'' can aid in the sensing task.  Although auxiliary qubits -- or more generally, additional quantum degrees of freedom -- cannot improve the sensitivity beyond the quantum metrology limits, they can aid in reaching these limits, for example when it is experimentally difficult to optimally initalize or readout the quantum state.  Auxiliary qubits may be used to increase the effective coherence or memory time of a quantum sensor, either by operation as a quantum memory or by enabling quantum error correction. 

In the following we discuss some of the schemes that have been proposed or implemented with auxiliary qubits.

\subsection{Quantum logic clock}

Clocks based on optical transitions of an ion kept in a high-frequency trap exhibit significantly improved accuracy over more common atomic clocks. 
Single-ion atomic clocks currently detain the record for the most accurate optical clocks, with uncertainties of $2.1\times10^{−18}$ for a $^{87}$Sr ensemble clock~\cite{Nicholson15} and $3.2\times10^{−18}$ for a single a single trapped $^{171}$Yb~\cite{Huntemann16}. 

The remaining limitations on optical clocks are related to their long-term stability and isolation from external perturbations such as electromagnetic interference. These limitations are even more critical when such clocks are based on a string of ions in a trap, because of the associated unavoidable electric field gradients. Only some ion species, with no quadrupolar moment, can then be used, but not all of them present a suitable transition for laser cooling and state detection beside the desired, stable clock transition.  To overcome this dilemma, quantum logic spectroscopy has been introduced~\cite{Schmidt05}.  The key idea is to employ two ion species: a clock ion that presents a stable clock transition (and represents the quantum sensor), and a logic ion (acting as auxiliary qubit) that is used to prepare, via a cooling transition, and readout the clock ion.  The resulting ``quantum logic'' ion clock can thus take advantage of the most stable ion clock transitions, even when the ion cannot be efficiently read out, thus achieving impressive clock performance~\cite{Rosenband07,Rosenband08,Hume07}.  Advanced quantum logic clocks may incorporate multi-ion logic~\cite{Tan15} and use quantum algorithms for more efficient readout~\cite{Schulte16}.

\subsection{Storage and retrieval}

The quantum state $\ket \psi$ can be stored and retrieved in the auxiliary qubit.  Storage can be achieved by a SWAP gate (or more simply two consecutive c-NOT gates) on the sensing and auxiliary qubits, respectively \cite{Rosskopf16}.  Retrieval uses the same two c-NOT gates in reverse order.   For the example of an electron-nuclear qubit pair, c-NOT gates have been implemented both by selective pulses \cite{Rosskopf16,Pfender16} and using coherent rotations \cite{Zaiser16}.

Several useful applications of storage and retrieval have been demonstrated.  A first example includes correlation spectroscopy, where two sensing periods are interrupted by a waiting time $t_1$ \cite{Laraoui13,Zaiser16,Rosskopf16}.  A second example includes a repetitive (quantum non-demolition) readout of the final qubit state, which can be used to reduce the classical readout noise \cite{Jiang09}.

\subsection{Quantum error correction}

Quantum error correction~\cite{Shor95,Nielsen00b} aims at counteracting errors during quantum computation by encoding the qubit information into redundant degrees of freedom. The logical qubit is thus encoded in a subspace of the total Hilbert space (the \textit{code}) such that each error considered maps the code to an orthogonal subspace, allowing detection and correction of the error.  Compared to dynamical decoupling schemes, qubit protection covers the entire noise spectrum and is not limited to low-frequency noise.  On the other hand, qubit protection can typically only be applied against errors that are orthogonal to the signal, because otherwise the signal itself would be ``corrected''.  In particular, for vector fields, quantum error correction can be used to protect against noise in one spatial direction while leaving the sensor responsive to signals in the orthogonal spatial direction.  Thus, quantum error correction suppresses noise according to spatial symmetry, and not according to frequency.

The simplest code is the 3-qubit repetition code, which corrects against one-axis noise with depth one (that is, it can correct up to one error acting on one qubit). For example, the code $|0\rangle_L=|000\rangle$ and $|1\rangle_L=|111\rangle$ can correct against a single qubit flip error.  Note that~\cite{Ozeri13x,Dur14} equal superpositions of these two logical basis states are also optimal to achieve Heisenberg-limited sensitivity in estimating a global phase~\cite{Bollinger96,Leibfried04}.  While this seems to indicate that QEC codes could be extremely useful for metrology, the method is hampered by the fact that QEC often cannot discriminate between signal and noise.  In particular, if the signal to be detected couples to the sensor in a similar way as the noise, the QEC code also eliminates the effect of the signal.  This compromise between error suppression and preservation of signal sensitivity is common to other error correction methods.  For example, in dynamical decoupling schemes, a separation in the frequency of noise and signal is required.  Since QEC works independently of noise frequency, distinct operators for the signal and noise interactions are required.  This imposes an additional condition on a QEC code: the quantum Fisher information~\cite{Giovannetti11,Lu15} in the code subspace must be non-zero. 

Several situations for QEC-enhanced sensing have been considered.  One possible scenario is to protect the quantum sensor against a certain type of noise (\eg, single qubit, bit-flip or transverse noise), while measuring the interaction between qubits~\cite{Dur14,Herrera15}.  More generally, one can measure a many-body Hamiltonian term with a strength proportional to the signal to be estimated~\cite{Herrera15}.  Since this can typically only be achieved in a perturbative way, this scheme still leads to a compromise between noise suppression and effective signal strength. 

The simplest scheme for QEC is to use a single \textit{good} qubit (unaffected by noise) to protect the sensor qubit~\cite{Arrad14,Kessler14,Hirose16,Ticozzi06}. In this scheme, which has recently been implemented with NV centers~\cite{Unden16}, the qubit sensor detects a signal along one axis (\eg, a phase) while being protected against noise along a different axis (\eg, against bit flip).  Because the ``good'' ancillary qubit can only protect against one error event (or, equivalently, suppress the error probability for continuous error), the signal acquisition must be periodically interrupted to perform a corrective step.  Since the noise strength is typically much weaker than the noise fluctuation rate, the correction steps can be performed at a much slower rate compared to dynamical decoupling. Beyond single qubits, QEC has also been applied to N00N states~\cite{Bergmann16}.  These recent results hint at the potential of QEC for sensing which has just about begun to being explored.


\section{Outlook}
\label{sec11}

Despite its rich history in atomic spectroscopy and classical interferometry, quantum sensing is an excitingly new and refreshing development advancing rapidly along the sidelines of mainstream quantum engineering research.  Like no other field, quantum sensing has been uniting diverse efforts in science and technology to create fundamental new opportunities and applications in metrology.  Inputs have been coming from traditional high-resolution optical and magnetic resonance spectroscopy, to the mathematical concepts of parameter estimation, to quantum manipulation and entanglement techniques borrowed from quantum information science.
Over the last decade, and especially in the last few years, a comprehensive toolset has been established that can be applied to any type of quantum sensor.
In particular, these allow operation of the sensor over a wide signal frequency range, can be adjusted to maximize sensitivity and dynamic range, and allow discrimination of different types of signals by symmetry or vector orientation.  While many experiments so far made use of single qubit sensors, strategies to implement entangled multi-qubit sensors with enhanced capabilities and higher sensitivity are just beginning to be explored.

One of the biggest attractions of quantum sensors is their immediate potential for practical applications.  This potential is partially due to the immense range of conceived sensor implementations, starting with atomic and solid-state spin systems and continuing to electronic and vibrational degrees of freedom from the atomic to the macroscale.  In fact, quantum sensors based on SQUID magnetometers and atomic vapors are already in everyday use, and have installed themselves as the most sensitive magnetic field detectors currently available.  Likewise, atomic clocks have become the ultimate standard in time keeping and frequency generation.  Many other and more recent implementations of quantum sensors are just starting to make their appearance in many different niches.  Notably, NV centers in diamond have started conquering many applications in nanoscale imaging due to their small size.  

What lies ahead in quantum sensing?  On the one hand, the range of applications will continue to expand as new types and more mature sensor implementations become available.  Taking the impact quantum magnetometers and atomic clocks had in their particular discipline, it can be expected that quantum sensors will penetrate much of the 21st century technology and find their way into both high-end and consumer devices.  Advances with quantum sensors will be strongly driven by the availability of ``better'' materials and more precise control, allowing their operation with longer coherence times, more efficient readout, and thus higher sensitivity.

In parallel, quantum sensing will profit from efforts in quantum technology, especially quantum computing, where many of the fundamental concepts have been developed, such as dynamical decoupling protocols, quantum storage and quantum error correction, as well as quantum phase estimation.  Vice versa, quantum sensing has become an important resource for quantum technologies as it provides much insight into the ``environment'' of qubits, especially through decoherence spectroscopy.  A better understanding of decoherence in a particular implementation of a quantum system can help the adoption of strategies to protect the qubit, and guide the engineering and materials development.  The border region between quantum sensing and quantum simulation, in addition, is becoming a fertile playground for emulating and detecting many-body physics phenomena.   Overall, quantum sensing has the potential to fundamentally transform our measurement capabilities, enabling higher sensitivity and precision, new measurement types, and covering atomic up to macroscopic length scales.

\section*{Acknowledgments}
The authors thank
Jens Boss, 
Dmitry Budker, 
Kevin Chang, 
Kristian Cujia, 
\L{}ukasz Cywi\'nski, 
Simon Gustavsson, 
Sebastian Hofferberth, 
Dominik Irber, 
Fedor Jelezko, 
Renbao Liu, 
Luca Lorenzelli, 
Tobias Rosskopf, 
Daniel Slichter,
J\"org Wrachtrup and
Jonathan Zopes
for helpful comments and discussions.
CLD acknowledges funding from the DIADEMS program 611143 of the European Commission, the Swiss NSF Project Grant $200021\_137520$, the Swiss NSF NCCR QSIT, and ETH Research Grant ETH-03 16-1.
FR acknowledges funding from the Deutsche Forschungsgemeinschaft via Emmy Noether grant RE 3606/1-1.
PC acknowledges funding from the U.S. Army Research Office through MURI grants No. W911NF-11-1-0400 and W911NF-15-1-0548 and by the NSF PHY0551153 and PHY1415345.


\section*{Appendix A: Table of symbols}
\label{appendixA}
\begin{table*}[h!]
\centering
\begin{tabular}{llll}
\hline\hline
Quantity & Symbol & Unit & \  \\
\hline
Readout efficiency									& $C$						& $0\leq C\leq 1$ & \ \\
Dynamic range												& DR						& --- & \ \\
AC signal: Frequency								& $\fac$				& Hz & \ \\
Multipulse sensing: Bandwidth 			& $\Delta f$  	& Hz & \ \\
Hamiltonian													& $\hat{H}(t)$ 	& Hz & \ \\
- internal Hamiltonian							& $\hat{H}_0$		&  & \ \\
- signal Hamiltonian								& $\hat{H}_V(t)$&  & \ \\
\quad ... commuting part								& $\HVzero(t)$	&  & \ \\
\quad ... non-commuting part						& $\HVone(t)$		&  & \ \\
- control Hamiltonian								& $\hat{H}_\mr{control}(t)$	& & \ \\
Number of qubits in ensemble; other uses	& $M$			& --- & \ \\
Multipulse sensing: Filter order		& $k$ 					& --- & \ \\
Multipulse sensing: No. of pulses		& $n$						& --- & \ \\
Number of measurements							& $N$						& --- & \ \\
Transition probability							& $p$						& $p \in[0...1]$ \ \\
- Bias point												& $p_0$					&  & \ \\
- Change in transition probability	& $\dpp=p-p_0$	& \ \\
Signal spectral density							& $S_V(\omega)$ & Signal squared per Hz & \ \\
Sensing time												& $t$ 					& s & \ \\
Signal autocorrelation time					& $\tc$ 				& s & \ \\
Total measurement time							& $T$ 					& s & \ \\
Relaxation or decoherence time 			& $\Tx$					& s & \ \\
- $T_1$ relaxation time							& $T_1$					&  & \ \\
- Dephasing time	      						& $T_2^\ast$		&  & \ \\
- Decoherence time	    						& $T_2$					&  & \ \\
- Rotating frame relaxation time & $T_{1\rho}$			&  & \ \\
Signal 															& $V(t)$ 	    	& \textit{varies} & \  \\
- parallel signal										& $\Vzero(t)=V_z(t)$	& & \ \\
- transverse signal									& $\Vone(t)=[V_x^2(t)+V_y^2(t)]^{1/2}$	& & \ \\
- vector signal											& $\vec{V}(t)=\{V_x,V_y,V_z\}(t)$	& & \ \\
- rms signal amplitude							& $\Vrms$				& & \ \\
- AC signal amplitude								& $\Vpk$				& & \ \\
- minimum detectable signal amplitude & $\Vmin$ 		& & \  \\
\quad ...	per unit time							& $\vmin$ 			& Unit signal per second & \  \\
Multipulse sensing: Weighting function	& $W(\fac,\alpha)$,$\bar{W}(\fac)$, etc. & --- & \ \\
Physical output of quantum sensor   & $x$, $x_j$		& \textit{varies} & \  \\
Multipulse sensing: Modulation function	& $y(t)$ 		& --- & \ \\
Multipulse sensing: Filter function	& $Y(\omega)$ 	& Hz$^{-1}$ & \ \\
AC signal: Phase										& $\alpha$  		& --- & \ \\
Coupling parameter									& $\gamma$  		& Hz per unit signal & \  \\
Decoherence or transition rate			& $\Gamma$  		& s$^{-1}$ & \ \\
Quantum phase	accumulated by sensor	& $\phi$				& --- & \ \\
- rms phase													& $\phirms$			& --- & \ \\
Pauli matrices											& $\hat{\vec\sigma}=\{\hat\sigma_x,\hat\sigma_y,\hat\sigma_z\}$	& & \ \\
Uncertainty of transition probability & $\sigp$ 		& --- & \ \\
- due to quantum projection noise 	& $\sigpq$ 			& & \ \\
- due to readout noise 							& $\sigpr$ 			& & \ \\
Multipulse sequence pulse delay			& $\tau$ 	  		& s & \  \\
Transition frequency								& $\wo$ 	  		& Hz & \  \\
Rabi frequency											& $\wone$ 			& Hz & \  \\
- effective Rabi frequency					& $\weff$ 			& Hz & \  \\
Decoherence function  							& $\chi(t)$ 		& --- & \  \\
Basis states (energy eigenstates)		& $\{\ma,\mb\}$     	& ---& \ \\
Superposition states								& $\{\xa,\xb\}$		 		& ---& \ \\
Sensing states											& $\{\psia,\psib\}$ 	& ---& \ \\
\hline\hline
\end{tabular}
\caption{Frequently used symbols.}
\label{table:symbols}
\end{table*}

%

\newpage


%

\end{document}